\documentclass[12pt]{JHEP3}
\input epsf.tex
\epsfclipon
\usepackage{epsfig}
\usepackage{epstopdf}
\usepackage{epsf}
\input{epsf.sty}
\usepackage{graphicx,amsmath,amssymb}
\usepackage{epsfig,multicol}
\usepackage{subfig}
\allowdisplaybreaks

\usepackage{bbm,bm,amsmath,amssymb}

\def\bg{\begin{eqnarray}}
\def\nd{\end{eqnarray}}
\def\sin{{\rm sin}}
\def\cos{{\rm cos}}
\def\tan{{\rm tan}}
\def\log{{\rm log}}
\def\del{\partial}
\title{Non-K\"ahler Deformed Conifold, Ultra-Violet Completion and Supersymmetric Constraints in the Baryonic Branch}
\author{Keshav Dasgupta, Jake Elituv, Maxim Emelin, Anh-Khoi Trinh\\
\vskip.03in
Ernest Rutherford Physics Department, McGill University \\
3600 rue University, Montr\'{e}al, Qu\'{e}bec, Canada H3A 2T8\\
{\tt dasgupta.keshav8@gmail.com, jake.elituv@mail.mcgill.ca}\\
{\tt anh-khoi.trinh@mail.mcgill.ca, maxim.emelin@mail.mcgill.ca}}

\date{\today}
\maketitle

\abstract{Gravity duals for a class of UV complete minimally supersymmetric non-conformal gauge theories require deformed conifolds with fluxes. However these manifolds do not allow for the standard K\"ahler or conformally 
K\"ahler metrics on them, instead the metrics are fully non-K\"ahler. We take a generic such configuration of a 
non-K\"ahler deformed conifold with fluxes and ask what constraints do supersymmetry impose in the Baryonic branch. We study the supersymmetry conditions and show that for the correct choices of the vielbeins and the complex structure all the equations may be consistently solved. The constraints now lead  not only to the known cases in the literature but also to some new backgrounds. We also show how geometric features of these backgrounds, including the overall warp factor and the resolution parameters, can be seen on the field theory side from perturbative ``probe-brane'' type calculations by Higgsing the theory and studying one-loop 4-point functions of vector and chiral multiplets. Finally we discuss how UV completions of these gauge theories may be seen from our set-up, both from type IIB as well as from the T-dual type IIA brane constructions.}

\begin{document}

\section{Introduction and summary}

Since the introduction of conifolds and their immediate generalizations, the resolved and the 
deformed conifolds \cite{candelas90}, string theory has been enriched not only by their ubiquity but also by the possibility of having a large number of exact solutions for hitherto unsolvable systems. One such unsolvable system is of course gauge theory with running couplings at  strong couplings. It is now known that the dynamics of a class of such gauge theories at large $N$ may be determined by using dual descriptions that involve deformed conifolds with fluxes. Such a  duality allows a one-to-one matching between operators in gauge theory and classical states 
in the deformed conifold background. This matching, or more appropriately, this dictionary, lies at the very heart of the most recent developments in string theory that promises to shed light on the numerous strongly coupled phenomena that were out of reach of analytical studies  so far.   

Examples of deformed conifolds that were presented early on in \cite{candelas90} included mostly, but not exclusively, those that had K\"ahler metrics on them. In other words, the early examples were generically constructed with Calabi-Yau metrics. However it turns out, the examples that are actually useful for solving strongly coupled systems are not the ones with K\"ahler metrics, but the ones with non-K\"ahler metrics. Surprisingly, some of these manifolds may not even support integrable complex structures, yet could give rise to supersymmetric solutions 
in string theory \cite{DRS, anke}. 

Demanding supersymmetry in the presence of non-K\"ahlerity is a subtle affair. For the K\"ahler case, the existence of supersymmetry and the existence of a certain amount of holonomy (we will call it the $\mathbb{G}$-holonomy), go hand in hand \cite{CHSW}.  In fact even the integrability of the complex structure follows from the above identification. 
Once we get rid of K\"ahlerity, and possibly also the integrability of the complex structure by relying only on the presence of an {\it almost} complex structure, the analysis of supersymmetry no longer follows the criteria laid out in \cite{CHSW}. The $\mathbb{G}$-holonomy changes to $\mathbb{G}$-structure and the manifold develops a torsion. Supersymmetry, as well as the integrability of the complex structure, are then analyzed using the so-called torsion classes \cite{torsion}. 

There is however another criterion for demanding supersymmetry, developed mostly for the type IIB theory 
 \cite{GKP}, by studying the flux structure associated with the choice of an almost complex structure. Supersymmetry requires the complexified three-form flux, which we will call ${\bf G}_3$ in this paper, to be a  (2, 1) form and Imaginary Self-Dual (ISD). This criterion is a bit more restrictive compared to the torsion class analysis, but
equivalence between these two approaches may be shown for the IIB case wherever fluxes are present. Interestingly, either of the two approaches do not require the internal manifold to be compact which turns  out to be an added bonus because gauge/gravity dualities are only defined with non-compact internal manifolds. Clearly the non-K\"ahler deformed conifolds that we want to study  fall in this category.  

The solutions that we construct in section \ref{IR} however are more general than the non-K\"ahler deformed conifolds 
and in particular feature a resolution parameter that introduces a relative difference in the curvature of the 2-cycles of the internal manifold. We call them the resolved warped-deformed conifolds. It is interesting to see the origin of this resolution parameter from the field theory side. To do this we perform a field theory calculation that describes the interactions of a probe brane with the brane stack that hosts the field theory. These interactions are studied by higgsing the theory and integrating out the resulting massive fields. The resulting effective action must then match the probe brane action in the dual geometry. 

Such ``probe brane'' calculations can be a useful tool for determining features of the gravity dual entirely from the field theory perspective despite the different regimes of validity of the two descriptions. In particular, in section \ref{navneet}, we will be able to determine the warp factor of the dual geometry and account for the resolution parameters of the gravity duals in terms of expectation values of certain ``baryonic'' operators in the field theory. In the conformal case, where the field theory is $ \mathcal{N} = 1, ~SU(N) \times SU(N)$ with bifundamental matter, we can relate these resolution parameters to the resolution parameters of the original background on which the branes are placed. In the non-conformal case, we can also obtain a resolution parameter by augmenting the theory by some adjoint matter. This opens a new branch in the moduli space of the theory, at least some of which we claim is dual to the non-K\"ahler solutions presented here.

All of the results, and specifically  the new branches in the moduli space of the gauge theories,  should 
in principle also be visible from a T-dual type IIA framework. T-duality in such a set-up is known to convert geometries to branes \cite{angel, DM1, DM2} and therefore the expectation would be to have a complete brane descriptions of the results of sections \ref{IR} and \ref{navneet}. This will indeed turn out to be true, and in section \ref{youvee} we will have detailed elaborations of various such configurations. Interestingly however the IIA side  allows us to 
construct many new brane configurations whose geometrical descriptions in the IIB dual are unclear. Additionally, the brane configurations in IIA suffer from various {\it bendings} because of the non-conformalities of the underlying gauge theories. This makes the  reverse T-duality map highly non-trivial, and we comment on some of the allowed possibilities.  

\subsection{Organization of the paper}

The paper is organized as follows. In section \ref{IR} we will construct our class of supersymmetric non-K\"ahler resolved warped-deformed  conifold backgrounds. The brane side, or equivalently the gauge theory side, of the story is first presented 
in section \ref{rescon}, followed by the gravity dual description in section \ref{defcon}.   
Starting with an ansatze for the K\"ahler structure in section \ref{duck} and the 
complex structure in \ref{bazaar}, we work out the three-form fluxes, first by imposing a specific constraint in section
\ref{a=b}, and then by going beyond it, in section \ref{salonfem}. In this section, 
requiring that the three-form fluxes preserve supersymmetry, we also obtain a consistent set of equations for the warp factors. Finally in section \ref{rebchut}, we analyze the supersymmetry equations again by choosing a different complex structure.

In section \ref{navneet} we study the dynamics of a D3 probe near a brane stack at a conifold singularity. We find warp factors that reproduce with the known gravity duals to the Klebanov-Witten and Klebanov-Strassler theories in sections \ref{alchut} and \ref{marmichut} respectively, 
and show how these geometries can acquire additional resolution factors by turning on expectation values of different ``baryonic'' operators. Our analysis in section \ref{alchut} delves in details on various issues. For example in 
section \ref{meyechut} we lay out the fundamentals of gauge fixing and Higgsing in this set-up. They in turn open up avenues to compute the vector and the chiral multiplets'  4-point functions in sections \ref{machli} and \ref{jesschut} respectively. The former helps us to estimate the warp-factors and the latter helps us to study the resolution parameters. The extensions of these results to the non-conformal cascading theories are non-trivial, and we detail our procedures in section \ref{marmichut}. 

In section \ref{youvee} we discuss the moduli space and UV completion of the Klebanov-Strassler field theory in terms of the brane pictures in both the IIB and IIA descriptions. On the type IIB side there are two avenues to approach the problem. The first one, as discussed in section \ref{dracula}, involves a CFT as our starting point, while the second one, as discussed in section \ref{cvetic}, involves a non-CFT as our starting point. The CFT description tells us how one may Higgs the theory to reach a cascading Klebanov-Strassler model, and we elaborate the story in section 
\ref{phantom}. Our analysis requires the presence of branes and anti-branes, and we study the issue of stability 
in section \ref{stabul}. A similar picture emerges with a non-CFT starting point, and we elaborate the story in section 
\ref{cvetic}. In section \ref{tdula} we show that there exists T-dual type IIA descriptions of all the type IIB results.
For example the UV completions as well as the Higgsing phenomena, discussed on the IIB side, have precise type IIA brane duals. We elaborate these brane configurations in section \ref{bhadush}, and point out the existence of many new ``branches"  in the moduli space in section \ref{branches}. We also speculate on their connections to the known Mesonic branches of the theory. 

Finally, section \ref{konoklota} contains some additional discussion of our results and concluding remarks.

\section{IR physics, dualities and supersymmetry \label{IR}}

The gauge theory that we want to study here is not conformal and therefore we will have to worry about the RG flow 
from the UV to the far IR. We will also want to analyze the strong coupling limit of the theory in terms of a supergravity 
dual so that computations may be performed  at both weak and strong coupling, the former being studied using a perturbative approximation. Existence of a supergravity dual at strong 't Hooft coupling implies large $N$, where $N$ is the number of colors, so that $g^2_{YM}$ may be kept small. 

One well known candidate for such a theory is the Klebanov-Strassler model \cite{KS}, whose IR and intermediate energy physics is well studied. However UV completion is necessary to avoid issues related to the UV divergence of the
Wilson loop and Landau poles in the presence of fundamental matter. A candidate UV completion using an AdS cap was first proposed in \cite{mia1}, which was developed further in \cite{mia2, mia3} and \cite{maxim}.  In this paper we will 
derive similar results but from a completely different point of view using moduli spaces of certain conformal and non-conformal gauge theories. In fact the moduli space dynamics that we will illustrate here would be more generic  and new results may be derived from here. 

We will start by studying the IR physics and the corresponding ${\cal N} = 1$ supersymmetry again using slightly different point of view, namely non-K\"ahler manifolds and torsion classes. 

\subsection{Non-K\"ahler resolved conifold and IR gauge theory \label{rescon}}

The IR gauge theory in question using configuration of type IIA branes has been studied in details since 
\cite{angel, DM1, DM2}. The configuration involves $M$ D4-branes straddling between two orthogonal NS5 branes. 
The orientations of the branes may be described using three complex coordinates:
\bg\label{comcor}
u \equiv x_4 + i x_5, ~~~~~~ v \equiv x_8 + i x_9, ~~~~~~~  w \equiv x_6 + i x_7, \nd
such that one of the NS5-branes is parametrized by $u$, and the other NS5-brane is parametrized by $v$. The $M$
D4-branes between the two NS5-branes are parametrized by $w$ such that $\vert  w \vert$ will be related to the 
YM coupling on the straddling D4-branes. Note that  in \cite{DM1, DM2} the YM coupling was related to 
${\rm Re}~w$ i.e the distance $x_6$ between the two orthogonal NS5-branes.  One may also use ${\rm Im}~w$
to be the YM coupling constant. This leads to a somewhat similar theory as described in \cite{DOT}.

The T-dual along the compact ${\rm Re}~w$ direction turns out to be the IR limit of the Klebanov-Strassler model 
\cite{DM2}, which is $M$ D5-branes wrapping the vanishing 2-cycle of a conifold in type IIB theory. 
The distance $x_6$ on the brane side now relates to the NS B-field in the type IIB side. On the other hand a T-dual 
of the compact $x_7$ direction similarly translates into $M$ D5-branes wrapping the resolution 2-cycle of 
a resolved conifold. This is the model studied by Vafa \cite{vafaGT}. 
The distance $x_7$ on the brane side translates into the size of the 2-cycle. Generically therefore we expect the YM coupling to be related to the complexified K\"ahler parameter:
\bg\label{alurjen}
\omega \equiv J + iB, \nd
where both $J$ and $B$ are restricted to the 2-cycle in the type IIB side.  This configuration arises from the two 
NS5-branes separated in the full $w$ plane with $M$ D4-branes straddling between them. The T-dual of this configuration however is more non-trivial as it involves both the NS B-field as well as non-zero size of the 
2-cycle. Together they lead to $M$ D5-branes wrapping the 2-cycle of a non-K\"ahler resolved conifold first 
discussed in \cite{anke} and more recently elaborated in \cite{DEM}. 

The actual T-dual configuration, following the Buscher's rule of \cite{buscher}, is highly non-trivial because, although the two NS5-branes T-dualize to orthogonal Taub-NUT spaces, the T-dual of the straddling $M$ D4-branes is 
unknown. However in \cite{DM2} enough evidence was provided to suggest that the T-dual of this configuration 
goes to fractional D3-branes in the type IIB set-up. Using this, we claimed in \cite{DEM} that the T-dual configuration leads to the 
following background in the type IIB side:
\bg\label{kalbinder}
&&ds^2 = {1\over \sqrt{h}} ~ds^2_{0123} + \sqrt{h}~ds^2_6 , \nonumber\\
&& {\bf F}_3 = {\rm cosh}~\gamma e^{-2\phi} \ast_6 d\left(e^{2\phi} J\right), ~~~~~~ {\bf H}_3 = -{\rm sinh}~\gamma ~d\left(e^{2\phi} J\right) , \nonumber\\
&&{\widetilde{\bf F}}_5 = -{\rm sinh}~\gamma~{\rm cosh}~\gamma\left(1 + \ast_{10}\right) {\cal C}_5(r)~d\psi \wedge \prod_{i=1}^2~\sin~\theta_i~ d\theta_i \wedge d\phi_i, \nd
where $\phi$ is the type IIB dilaton, ${\cal C}_5$ is some specific function of $r$ \cite{DEM}, 
and $\gamma$ is the so-called boosting angle (see \cite{MM, DEM} for details). 
The background has both RR and NS three-forms switched on in such a way that the backreaction of these forms 
convert the six-dimensional internal space $ds_6$ to a non-K\"ahler warped resolved conifold:
\bg\label{ryaconn} 
ds^2_6 = {\cal G}_1~ dr^2 + {\cal G}_2 (d\psi + {\rm cos}~\theta_1 d\phi_1 + {\rm cos}~\theta_2 d\phi_2)^2  + \sum_{i = 1}^2 {\cal G}_{2+i}
(d\theta_i^2 + {\rm sin}^2\theta_i d\phi_i^2) ,\nonumber\\
\nd
where ${\cal G}_i$ are the warp-factors which for simplicity would be taken as a function of the radial coordinate $r$ 
only. This will be generalized later. The Hodge star in \eqref{kalbinder} is with respect to the metric \eqref{ryaconn}. 
Finally, the warp-factor $h$ appearing in \eqref{kalbinder} is defined as:
\bg\label{snowman}
h =  e^{-2\phi}\left(1 + {\rm sinh}^2\gamma\right) - {\rm sinh}^2\gamma. \nd
Let us now discuss ths issue of supersymmetry here. It turns out, and as discussed in \cite{DEM}, there are two ways 
to verify supersymmetry of the background \eqref{kalbinder}. The first one is to define a complexified three-form
${\bf G}_3$ in the following way: 
\bg\label{harhole}
{\bf G}_3 \equiv {\bf F}_3 - i e^{-\phi} {\bf H}_3, \nd
such that it is an ISD (2, 1) form and {\it not} an ISD (1, 2) form \cite{GKP}. Supersymmetry is clearly broken if the 
three-form flux \eqref{harhole} is either a mixture of (2, 1) and (1, 2) forms,  or is a (3, 0) or (0, 3) form. The second 
way of verifying supersummetry appears from the torsion class constraints  \cite{torsion}:
\bg\label{torcons}
{\cal W}_1 = {\cal W}_2 = 0, ~~~~~ {\cal W}_3 \ne 0, ~~~~~ 2{\cal W}_4 + {\rm Re}~{\cal W}_5 = 0, \nd
where the first equation allows an integrable complex structure, the second equation with non-vanishing ${\cal W}_3$ 
allows a non-zero torsion and the third equation allows a supersymmetric $SU(3)$ structure. Together, \eqref{torcons}
gives rise to four-dimensional ${\cal N} = 1$ supersymmetry. 

We expect both the above conditions should be satisfied simultaneously, and indeed in \cite{DEM}  a 
rigorous proof of the existence of supersymmetry for the background \eqref{kalbinder} was given, provided the 
warp-factors ${\cal G}_i$ satisfy the following constraint \cite{DEM}:
\bg\label{virgop}
{{\cal G}_{3r} - \sqrt{{\cal G}_1 {\cal G}_2} \over {\cal G}_3} + {{\cal G}_{4r} - \sqrt{{\cal G}_1 {\cal G}_2} \over {\cal G}_4} = 0, \nd
with a constant dilaton profile. The above constraint appears from demanding the vanishing of the (1, 2) form 
of the three-form flux \eqref{harhole}. On the other hand, demanding the closure of the fundamental (3, 0) form 
$\Omega$ gives rise to the following additional constraint:
\bg\label{zeroeffect} 
{{\cal G}_{3r}\over {\cal G}_3} + {{\cal G}_{4r}\over {\cal G}_4} + {{\cal G}_{2r} - 2 \sqrt{{\cal G}_1 {\cal G}_2}\over {\cal G}_2} ~ = ~ 0 . \nd
Interestingly \eqref{virgop} is related to the vanishing of ${\cal W}_4$ whereas \eqref{zeroeffect} is related to the 
vanishing of ${\rm Re}~{\cal W}_5$. 
Mathematically such an internal manifold is termed as a non-K\"ahler special-Hermitian
manifold, which is a complex manifold with a closed (3, 0) form and ${\cal W}_3 \ne 0$. As we saw above, all other torsion class vanish.

\subsection{Non-K\"ahler deformed conifold and gravity dual \label{defcon}}

The gauge theory side of the story appears from either the type IIA brane configuration or from type IIB D5-branes wrapped on 2-cycle of a warped resolved conifold. For the latter case one looks for the {\it decoupling} limit to 
concentrate only on the gauge theory on the wrapped  D5-branes. 

How does one determine the gravity dual of the IR gauge theory? There are two ways to proceed here. The first one is 
from the type IIA brane construction, as discussed in \cite{DOT}, and the second one is from the type IIB wrapped 
brane picture, as discussed in \cite{vafaGT, anke}. Both these viewpoints lead to one conclusion: the 
gravity duals of the ${\cal N} = 1$ theories discussed in section \ref{IR} are generically given, in the type IIB theory, by resolved warped-deformed conifolds with background fluxes. In the past  a specific gravity dual involving 
warped-deformed conifold was first discussed in \cite{KS} followed by \cite{vafaGT} and \cite{malnun}. In \cite{KS} 
an explicit solution involving conformally Calabi-Yau metric, i.e the metric of a deformed conifold, with fluxes was 
presented. In \cite{malnun} a solution involving a specific non-K\"ahler metric on the deformed confiold, along with 
type IIB fluxes, was presented. This was elaborated later in \cite{MM} to show how one may interpolate between the IR picture 
of \cite{malnun} and the baryonic branch description of \cite{butti}. Again a specific non-K\"ahler metric showed up 
in the analysis. A more generic picture, involving a resolved warped-deformed conifold, was discussed in 
\cite{anke}, starting with local descriptions and then going to generic global set-ups. However the detailed fluxes 
were not spelled out completely in \cite{anke}, 
part of the reason being the underlying technicalities of the analysis involved. 
In the following we will generalize the story in a way as to make transparent the issue of the interplay between supersymmetry and non-K\"ahlerity, much in the vein of \cite{DEM}. As far as we know, this was not attempted before.

\subsubsection{The fundamental (1, 1) form and non-K\"ahlerity \label{duck}}

The type IIB background interestingly takes the {\it same} form as we had in \eqref{kalbinder}, except now $h, J, \phi, {\cal C}_5$ as well as the internal metric $ds^2_6$ are all different.  We start by defining $ds^2_6$ in the following way:
\bg\label{crooked}
ds^2_6 & = &  {\cal G}_1~ dr^2 + {\cal G}_2 (d\psi + {\rm cos}~\theta_1 d\phi_1 + {\rm cos}~\theta_2 d\phi_2)^2  + \sum_{i = 1}^2 {\cal G}_{2+i}
(d\theta_i^2 + {\rm sin}^2\theta_i d\phi_i^2) \nonumber\\
& + & {\cal G}_5 ~\cos~\psi\left(d\theta_1d\theta_2 - \sin~\theta_1 \sin~\theta_2 d\phi_1 d\phi_2\right)  + 
{\cal G}_6~  \sin~\psi\left(\sin~\theta_1d\phi_1 d\theta_2  + \sin~\theta_2d\phi_2 d\theta_1\right), \nonumber\\
\nd 
where we have used similar coefficients to express the internal metric \eqref{crooked}  as in \eqref{ryaconn}. This 
is intentional as we want to relate the gravity dual framework to the wrapped brane scenario. Clearly the internal 
metric \eqref{crooked} is a non-K\"ahler space and we want to see how the various warp-factors are related to 
each other once we demand supersymmetry and EOM constraints.  

The following analysis will be more involved than the one we encountered earlier in \cite{DEM} because of the 
complicated nature of the internal metric \eqref{crooked}. To proceed, we first define the left-invariant Maurer-Cartan forms in the usual way:
\bg\label{leftinv}
&&\hskip-2in \left(\begin{matrix} \sigma_1 \\ \sigma_2 \end{matrix}\right) 
= \left(\begin{matrix}{~~\cos}~\psi_1 & ~~~{\sin}~\psi_1 \\ -\sin~\psi_1 & ~~~{\cos}~\psi_1 \end{matrix}\right) 
\left(\begin{matrix} d\theta_1 \\ \sin~\theta_1 ~d\phi_1 \end{matrix}\right)\nonumber\\
&& \hskip-2in \left(\begin{matrix} \Sigma_1 \\ \Sigma_2 \end{matrix}\right) 
= \left(\begin{matrix}{~~\cos}~\psi_2 & ~~~{\sin}~\psi_2 \\ -\sin~\psi_2 & ~~~{\cos}~\psi_2 \end{matrix}\right) 
\left(\begin{matrix} d\theta_2 \\ \sin~\theta_2 ~d\phi_2 \end{matrix}\right)\nonumber\\
&& \hskip-2in \sigma_3 =  d\psi_1 + \cos~\theta_1~d\phi_1, ~~~  \Sigma_3 =  d\psi_2 + \cos~\theta_2 ~d\phi_2, 
\nd
where note that we took two angles $\psi_1$ and $\psi_2$. These two angles are related to $\psi$ as $\psi \equiv \psi_1 + \psi_2$. We could also take $\psi_1 = \psi_2 = {\psi\over 2}$, but then this leads to some subtlety 
in extending the internal space to a $G_2$ structure manifold in M-theory as the eleventh 
direction is related to $\psi_1 - \psi_2$ \cite{cvetic, ankegwyn}.  It is also interesting to note that, if we allow:
\bg\label{g5=g6}
{\cal G}_5(r) = {\cal G}_6(r), \nd
then the  second line of \eqref{crooked} may be simplified further under certain conditions. In fact new isometry 
arises under such consideration as discussed in the first reference of \cite{anke}. 

The choice of the vielbeins for our case is more subtle now as we need to guarantee the K\"ahler, Calabi-Yau nature 
of the internal space in the absence of any background fluxes. For the resolved conifold case studied in 
\eqref{ryaconn}, the choice of the vielbein with $\psi_1 = \psi_2 = {\psi\over 2}$  in \cite{DEM} was:
\bg\label{macbeth}
&& e_1 = \sqrt{{\cal G}_1\sqrt{h}}e_r, ~~~~~~ e_2 = \sqrt{{\cal G}_2\sqrt{h}} e_\psi\\
&& e_3 = \sqrt{{\cal G}_3\sqrt{h}}\left(-\sin~{\psi\over 2} ~e_{\phi_1} + \cos~{\psi\over 2}~e_{\theta_1}\right), 
~~e_4 = \sqrt{{\cal G}_3\sqrt{h}}\left(\cos~{\psi\over 2} ~e_{\phi_1} + \sin~{\psi\over 2}~e_{\theta_1}\right) , \nonumber\\   
&& e_5 = \sqrt{{\cal G}_4\sqrt{h}}\left(-\sin~{\psi\over 2} ~e_{\phi_2} + \cos~{\psi\over 2}~e_{\theta_2}\right), 
~~e_6 = \sqrt{{\cal G}_4\sqrt{h}}\left(\cos~{\psi\over 2} ~e_{\phi_2} + \sin~{\psi\over 2}~e_{\theta_2}\right) .
\nonumber\nd
For the present case the choice of the vielbeins considered in \cite{KS, minpis} do not lead to the right Calabi-Yau 
deformed conifold limit in the absence of fluxes as shown in \cite{ankegwyn}. A more non-trivial choice of the 
vielbeins is called for here. 

A choice of vielbeins that work for a Calabi-Yau deformed conifold has been proposed in \cite{ankegwyn}. However 
our internal manifold \eqref{crooked} is non-K\"ahler, so we will have to modify them further to get the correct vielbeins. 
We will also have to make sure that we reproduce the correct set of vielbeins for the non-K\"ahler resolved confiold 
from our choice. It turns out there exist another set of vielbeins for the resolved case that is related to \eqref{macbeth} 
simply by $\psi \to -\psi$ for ($e_3, e_4, e_5, e_6$). This means we can express $e_i$ in \eqref{macbeth}  by the Maurer-Cartan forms \eqref{leftinv} with one additional change: the first complex vielbein for $h \equiv 1$ becomes:
\bg\label{jesica}
E_1 \equiv  \sqrt{{\cal G}_2} e_\psi + i \sqrt{{\cal G}_1} e_r. \nd
Of course we could resort to the old veilbeins choice in \cite{DEM}, but the present set of changes will allow us to express the new veilbeins for the non-K\"ahler deformed conifold in a more consistent set-up. 
Furthermore, to
simplify the problem a bit, we will consider the case \eqref{g5=g6}. For such a scenario, one may express the vielbeins in the following way:
\bg\label{IT}
&& e_1 = \sqrt{{\cal G}_1\sqrt{h}}dr, ~~~~~~ e_2 = \sqrt{{\cal G}_2\sqrt{h}} e_\psi =\sqrt{{\cal G}_2\sqrt{h}} \left(\sigma_3 + \Sigma_3\right)\nonumber\\ 
&& e_3 = \sqrt{{\cal G}_3\sqrt{h}}\left(\alpha_1 \sigma_1 + {\beta}_3 \Sigma_1\right), ~~~~
e_4 = \sqrt{{\cal G}_4\sqrt{h}}\left(\alpha_2 \sigma_2 - {\beta}_4 \Sigma_2\right) \nonumber\\
&& e_5 = \sqrt{{\cal G}_3\sqrt{h}}\left(\beta_1 \sigma_1 + {\alpha}_3 \Sigma_1\right), ~~~~
e_6 = \sqrt{{\cal G}_4\sqrt{h}}\left(-\beta_2 \sigma_2 + {\alpha}_4 \Sigma_2\right), \nd
where $\alpha_i$ and $\beta_i$ are functions of $r$, the radial coordinate. With a Calabi-Yau metric on the deformed
conifold, we expect $\alpha \equiv \alpha_1 = \alpha_2 = \alpha_3 = \alpha_4$ and $\beta \equiv \beta_1 = \beta_2 = \beta_3 = \beta_4$, although they will still be non-trivial functions of 
the radial coordinate. 
For the present case we need:
\bg\label{lathi}
&&\alpha_1^2 + \beta_1^2 = 1 = \alpha_4^2 + \beta_4^2\nonumber\\
&& \alpha_2^2 + \beta_2^2 = {{\cal G}_3\over {\cal G}_4}, ~~~ \alpha_3^2 + \beta_3^2 = 
{{\cal G}_4\over {\cal G}_3}\nonumber\\
&& \alpha_1 \beta_3 + \alpha_3 \beta_1 = {{\cal G}_6\over 2{\cal G}_3}, ~~~~ 
\alpha_2 \beta_4 + \alpha_4 \beta_2 = {{\cal G}_6\over 2 {\cal G}_4}, \nd
which, as one may check, reproduces the background resolved warped-deformed conifold metric 
\eqref{crooked}. However the choice of the vielbeins 
 is more complicated than the ones considered in \cite{ankegwyn}. Simplification could happen if we further impose
 ${\cal G}_3 =  {\cal G}_4$, i.e the two base spheres in the metric \eqref{crooked} are resolved in identical fashion. In
 this limit the values of $\alpha$ and $\beta$ may be determined as:
\bg\label{alphabeta}
&& \alpha^2 = {1\over 2}\left(1 \pm {\sqrt{4{\cal G}_3^2 - {\cal G}_6^2}\over 2{\cal G}_3}\right), ~~~ 
\beta = {{\cal G}_6\over 4 \alpha {\cal G}_3}, \nd
where the inequalities between the $\alpha_i$ rest solely on the inequality between ${\cal G}_3$ and 
${\cal G}_4$; and we will assume that both of them, at any point in $r$, are bigger than ${\cal G}_6$ at that point 
in $r$. Furthermore we will take the plus sign for  $\alpha$ to avoid putting further constraints on the 
warp-factors.  Note that when:
\bg\label{edenlac} 
\alpha_1 = \alpha_4 = 1, ~~~~ \alpha_2 = \sqrt{{\cal G}_3 \over {\cal G}_4}, ~~~~ \alpha_3 = \sqrt{{\cal G}_4 \over {\cal G}_3}, ~~~~\beta_i = 0, \nd 
we recover the veilbeins for the resolved conifold 
case \cite{ankegwyn, DEM}.  This also gives a reason for choosing the relative plus sign  in \eqref{alphabeta}.

Our next step would be to define an almost complex structure for the internal space \eqref{crooked}. If this almost 
complex structure is integrable, then the internal space will be a complex manifold. We follow the standard procedure to define our complex vielbeins in the following way:
\bg\label{machine}
E_1 = e_2 + i \sigma e_1 , ~~~~~ E_2 = e_3 + i e_4, ~~~~~ E_3 = e_5 + i e_6, \nd
with an almost complex structure ($i\sigma, i, i$). This is similar to what we had earlier in \cite{DEM}. On the other hand, we also require the fundamental ($1, 1$) form $J$. However the $J$ appearing in \eqref{kalbinder} is {\it not}
associated with the $J$ that we expect from the metric in \eqref{kalbinder}. For our case, the fundamental form $J$ appears from a pre Maldacena-Martelli \cite{MM} dual metric and flux configuration given by:
\bg\label{preMM}
ds^2 = ds^2_{0123} + e^{-2\phi} ds^2_6, ~~~~~ {\bf H} = e^{-2\phi} \ast_6 d\left(e^{2\phi} J\right), \nd
where $ds^2_6$ is the deformed conifold metric \eqref{crooked}, ${\bf H}$ is the NS three-form and the Hodge star is 
with respect to the above metric.  To proceed further we will need the two torsion classes ${\cal W}_1$ and 
${\cal W}_2$ to argue for the integrability of the complex structure. The torsion classes may be expressed as:
\bg\label{torclass}
{\cal W}_1 J \wedge J \wedge J = d\Omega \wedge J, ~~~~ 
\left[d\Omega\right]^{(2, 2)} = {\cal W}_1 J \wedge J + {\cal W}_2 \wedge J.\nd
We expect $\Omega$ to be a (3, 0) form, so $d\Omega$ is unlikely to have a (2, 2) piece. This means ${\cal W}_2$ 
vanishes if ${\cal W}_1$ vanishes.  For simplicity, let us first start by putting an integrable complex structure 
on the manifold \eqref{crooked}, and define the {\it pre-dual} vielbeins in the following way:
\bg\label{predual} 
&& {\cal E}_1 = e^{-\phi} \left[i\sqrt{{\cal G}_1} dr +  \sqrt{{\cal G}_2} \left(d\psi + \cos~\theta_1 d\phi_1 
+ \cos~\theta_2 d\phi_2\right)\right] \nonumber\\
&& {\cal E}_2 = e^{-\phi}\left[\sqrt{{\cal G}_3}\left(\alpha_1 \sigma_1 + \beta_3 \Sigma_1\right) 
+ i\sqrt{{\cal G}_4} \left(\alpha_2 \sigma_2 - \beta_4 \Sigma_2\right)\right] \nonumber\\
&& {\cal E}_3 = e^{-\phi}\left[\sqrt{{\cal G}_3}\left(\beta_1 \sigma_1 + \alpha_3 \Sigma_1\right) 
- i\sqrt{{\cal G}_4}\left(\beta_2 \sigma_2 - \alpha_4 \Sigma_2\right)\right], \nd  
where $\phi$ remains the dilaton in the {\it post-dual} scenario whose generic picture is given in \eqref{kalbinder}. The
fundamental two-form now may be constructed in the usual way as: 
\bg\label{funform}
J  &= &   -{i\over 2}\left({\cal E}_1 \wedge \overline{\cal E}_1  + {\cal E}_2 \wedge \overline{\cal E}_2 + {\cal E}_3 \wedge \overline{\cal E}_3\right) \nonumber\\
&=&  e^{-2\phi}\sqrt{{\cal G}_1 {\cal G}_2}~dr \wedge \left(d\psi + \cos~\theta_1 d\phi_1 + \cos~\theta_2 d\phi_2\right)\nonumber\\
&-& e^{-2\phi}\sqrt{{\cal G}_3{\cal G}_4}\left[\left(\alpha_1 \alpha_2 -\beta_1 \beta_2\right) \sigma_1 \wedge \sigma_2 
+ \left(\alpha_3 \alpha_4 -\beta_3 \beta_4\right)  \Sigma_1 \wedge \Sigma_2\right] \nonumber\\
&-& e^{-2\phi}\sqrt{{\cal G}_3{\cal G}_4}\left[\left(\beta_1 \alpha_4 -\alpha_1 \beta_4\right) \sigma_1 \wedge \Sigma_2   +   \left(\beta_3 \alpha_2 -\alpha_3 \beta_2\right)   \Sigma_1 \wedge \sigma_2\right], \nd  
where we see that if $\alpha_i = \alpha$ and $\beta_i = \beta$ as in \eqref{alphabeta} then the fundamental form simplifies to take 
the somewhat usual form for a deformed conifold as given in \cite{ankegwyn}. Such an equality may only be assumed 
for our generic construction if:
\bg\label{jungle}
{\cal G}_3(r) = {\cal G}_4(r), \nd
which is the limit in which the manifold \eqref{crooked} is a non-K\"ahler warped deformed conifold and {\it not} a
non-K\"ahler warped resolved-deformed conifold. It is therefore the additional resolution parameter that makes the fundamental form more complicated. This would mean that the resulting fluxes will also get highly non-trivial as we shall see soon. The fundamental form \eqref{funform} becomes:
\bg\label{funform2} 
e^{2\phi} J & = & \sqrt{{\cal G}_1{\cal G}_2} dr \wedge \left(d\psi + \cos~\theta_1 d\phi_1 + \cos~\theta_2 d\phi_2\right)\\
&-& {1\over 2} \sqrt{{\cal G}_3{\cal G}_4} \left({\bf A} - {\bf B}\right) \sin~\psi
\left(e_{\phi_1} \wedge e_{\phi_2} - e_{\theta_1} \wedge e_{\theta_2}\right) \nonumber\\
&-& \sqrt{{\cal G}_3{\cal G}_4}\left[\left(\alpha_1 \alpha_2 -\beta_1 \beta_2\right) e_{\theta_1} \wedge e_{\phi_1} 
+ \left(\alpha_3 \alpha_4 -\beta_3 \beta_4\right)  e_{\theta_2} \wedge e_{\phi_2} \right] \nonumber\\
&-& \sqrt{{\cal G}_3{\cal G}_4}\left[\left({\bf A}~\cos^2 {\psi\over 2}  + {\bf B}~\sin^2 {\psi\over 2}\right) e_{\theta_1} \wedge e_{\phi_2}  + 
\left({\bf B}~\cos^2 {\psi\over 2}  + {\bf A}~\sin^2 {\psi\over 2}\right) e_{\theta_2} \wedge e_{\phi_1}\right], \nonumber \nd 
where $\psi \equiv \psi_1 + \psi_2$, and we see that the complications come from the cross terms that proportional to
$\cos~{\psi\over 2}$ and $\sin~ {\psi\over 2}$. One simplification is possible at this stage without 
incorporating \eqref{jungle}: this is the case where ${\bf A} = {\bf B}$ where we have 
defined ${\bf A}$ and ${\bf B}$ as:
\bg\label{pendant}
{\bf A} \equiv \beta_1 \alpha_4 - \alpha_1 \beta_4, ~~~~~~
{\bf B} \equiv \beta_3 \alpha_2 - \alpha_3 \beta_2. \nd
Looking at \eqref{lathi}, we see that there are not enough equations to fix all the $\alpha_i$ and $\beta_i$. 
Thus \eqref{pendant} doesn't seem to over-constrain the system instead helps in simplifying $J$  by
killing off the second term in 
\eqref{funform2} and removing the unnecessary $\psi$ dependence in the 
last term of \eqref{funform2}.  Note however the imbalance factors for the two spheres still remain.

The resulting manifold is clearly non-K\"ahler as one may check from the fundamental form $J$. Checking the closure of $J$ gives us the following expression:
\bg\label{warning}
d\left(e^{2\phi}J\right) & = & dr \wedge \left[e_{\theta_1} \wedge e_{\phi_1}  \left(\sqrt{{\cal G}_1 {\cal G}_2} 
- {\del {\bf C}\over \del r}\right) 
+ e_{\theta_2} \wedge e_{\phi_2}  \left(\sqrt{{\cal G}_1 {\cal G}_2} 
- {\del {\bf D}\over \del r}\right)\right]\\
&-& {\del {\cal F}\over \del r} ~dr \wedge  \left(e_{\theta_1} \wedge e_{\phi_2} + e_{\theta_2} \wedge e_{\phi_1}\right)
- {\cal F} ~d\theta_1 \wedge d\theta_2 \wedge \left(\cos~\theta_1 d\phi_1 - \cos~\theta_2 d\phi_2\right), \nonumber  \nd
where $e_i$ are the standard one-forms and we have identified ${\bf A}$ with ${\bf B}$ to simplify $J$ 
as mentioned above. Generalization of this is possible by keeping ${\bf A}$ and ${\bf B}$ but doesn't lead to any new physics, so we shall stick with this simplified form here. 
The functions ${\bf C}, {\bf D}$ and  ${\cal F}(r)$ may be defined in the following way:
\bg\label{IT2}
{\bf C} \equiv  \sqrt{{\cal G}_3 {\cal G}_4}\ \left(\alpha_1 \alpha_2 - \beta_1 \beta_2\right), ~~~
{\bf D} \equiv  \sqrt{{\cal G}_3 {\cal G}_4}\ \left(\alpha_3 \alpha_4 - \beta_3 \beta_4\right), ~~~
{\cal F}(r) \equiv  {\bf A} \sqrt{{\cal G}_3 {\cal G}_4}, \nonumber\\
\nd
where ${\bf A}$ is defined in \eqref{pendant}. All these 
which would vanish if we impose \eqref{jungle}. On the other hand, \eqref{warning} still remains non-zero because 
generically we do not expect $\alpha_i$ to equal $\beta_i$. The manifold \eqref{crooked} becomes K\"ahler if and only 
if the warp-factors ${\cal G}_i$ satisfy the following constraints:
\bg\label{gold}
&& {\del {\cal F}\over \del r} = {\cal F} = 0 \nonumber\\
&&\sqrt{{\cal G}_1 {\cal G}_2}  ~ = ~  {\del {\bf C} \over \del r} ~ = ~ {\del {\bf D} \over \del r}. \nd
The first condition of vanishing ${\cal F}$ is easily achieved by making $\alpha_i = \alpha$ and $\beta_i = \beta$. This is of course the expected constraint \eqref{jungle}. Thus the non-trivial constraint is the second one 
involving all the warp-factors. Clearly imposing \eqref{jungle} simplifies this but there is still a non-trivial equation relating the warp factors ${\cal G}_1, {\cal G}_2, {\cal G}_3$ and ${\cal G}_6$ that need to be satisfied. 

To verify the set of constraints \eqref{gold}, let us take the specific case of a deformed conifold. A deformed conifold allows a K\"ahler metric for the following choice of the warp-factors ${\cal G}_i$ with vanishing dilaton $\phi$:
\bg\label{hush}
&& {\cal G}_3 = {\cal G}_4 = {{\gamma}\over 4}, ~~~~ {\cal G}_6 = {\mu^2 {\gamma} \over 2 r^2} \nonumber\\
&& {\cal G}_1 =   {{\gamma} + \left(r^2 {\gamma}' - {\gamma}\right)\left(1 -
{\mu^4\over r^4}\right) \over r^2 \left(1 - {\mu^4\over r^4}\right)}, ~~~ {\cal G}_2 =  {1\over 4} \left[{\gamma} + \left(r^2 {\gamma}' - {\gamma}\right)\left(1 -
{\mu^4\over r^4}\right)\right], \nd 
where $\mu^2$ is associated with the deformation parameter, i.e the size of the three-cycle at $r = 0$,  of the deformed conifold and ${\gamma}$ is associated with the K\"ahler potential ${\bf F}$ by the relation \cite{candelas90}:
\bg\label{crisis}
{\gamma} \equiv r^2 {\bf F}, \nd
with the derivative on ${\gamma}$ is defined with respect to $r^2$ and not $r$. This is of course a matter of convention, but is widely followed in the literature. It is also easy to infer that the Ricci flatness of the deformed conifold 
implies  the following differential equation for $\gamma$  \cite{candelas90}:
\bg\label{brycenunu}
r^2\left(r^4 - \mu^4\right) {d{\gamma}^3\over dr^2} + 3 \mu^4 {\gamma}^3 = 2 r^8, \nd
which basically fixes the functional form of ${\gamma}$.  We will not need the explicit solution for $\gamma$, and one
may plug in the values of the warp-factors \eqref{hush} in \eqref{alphabeta} to determine the functional form for 
$\alpha_i$ and $\beta_i$ in the following way (see also \cite{ankegwyn}):
\bg\label{carlas}
&& \alpha_i  \equiv \alpha = {1\over 2} \sqrt{1 + {\mu^2\over r^2}} + {1\over 2} \sqrt{1 - {\mu^2\over r^2}},~~~~ 
\beta_i \equiv \beta = {\mu^2\over 2 r^2 \alpha},
 \nd
implying the vanishing of ${\cal F}(r)$ from \eqref{IT2}. This is of course the expected simplification. One may now verify 
that the set of warp-factors \eqref{hush} along with the ($\alpha, \beta$) values from \eqref{carlas} satisfy the 
constraint equations for K\"ahlerity, namely  \eqref{gold}.  

\subsubsection{The holomorphic (3, 0) form and complex structures \label{bazaar}}

The pre-dual set-up that we studied in the previous section assumes an integrable complex structure as evident from 
our choice of the vielbeins in \eqref{predual}. It is therefore essential for us to check the closure of the holomorphic 
(3, 0) form $\Omega$. The most generic three-form can be expressed as:
\bg\label{threeform} 
\Omega &  = &  e_r \wedge \left({\bf A}_{11}~e_{\theta_2} \wedge e_{\phi_2}  - {\bf A}_{12}~e_{\theta_1} \wedge e_{\phi_1}\right) 
+  e_r \wedge \left({\bf A}_{21}~e_{\phi_1} \wedge e_{\theta_2}  + {\bf A}_{22}~e_{\theta_1} \wedge e_{\phi_2}\right) \nonumber\\
&+& e_r \wedge \left({\bf A}_{31}~e_{\theta_1} \wedge e_{\theta_2}  - {\bf A}_{32}~e_{\phi_1} \wedge e_{\phi_2}\right) 
+ e_\psi \wedge \left({\bf A}_{41}~e_{\theta_2} \wedge e_{\phi_2}  - {\bf A}_{42}~e_{\theta_1} \wedge e_{\phi_1}\right)\nonumber\\
&+&  e_\psi \wedge \left({\bf A}_{51}~e_{\phi_1} \wedge e_{\theta_2}  + {\bf A}_{52}~e_{\theta_1} \wedge e_{\phi_2}\right) 
+ e_\psi \wedge \left({\bf A}_{61}~e_{\theta_1} \wedge e_{\theta_2}  - {\bf A}_{62}~e_{\phi_1} \wedge e_{\phi_2}\right), 
\nonumber\\ \nd
with $e_k$ being the standard vielbeins defined earlier without using the warp-factors and ${\cal G}_k$ are the 
warp-factors in the metric  \eqref{crooked}. The coefficients ${\bf A}_{ij}$ are defined using the warp-factors in the 
following way:
\bg\label{planetarium} 
{\bf A}_{11} &=& - \sqrt{{\cal G}_1{\cal G}_3{\cal G}_4} \left(\alpha_4 \beta_3 + \alpha_3 \beta_4\right), ~~~~
{\bf A}_{12} =  - \sqrt{{\cal G}_1{\cal G}_3{\cal G}_4} \left(\alpha_1 \beta_2 + \alpha_2 \beta_1\right)\nonumber\\
 {\bf A}_{21}  &= & i ~\sin~\psi_1~\cos~\psi_2~{\cal G}_3 \sqrt{{\cal G}_1} \left(\alpha_1 \alpha_3 - \beta_1 \beta_3\right) + 
 i ~\cos~\psi_1~\sin~\psi_2~{\cal G}_4 \sqrt{{\cal G}_1} \left(\alpha_2 \alpha_4 - \beta_2 \beta_4\right) \nonumber\\
 &+&  \sin~\psi_1~\sin~\psi_2~\sqrt{{\cal G}_1{\cal G}_3 {\cal G}_4} \left(\alpha_1 \alpha_4 + \beta_1 \beta_4\right) -
  \cos~\psi_1~\cos~\psi_2~\sqrt{{\cal G}_1{\cal G}_3 {\cal G}_4} \left(\alpha_2 \alpha_3 + \beta_2 \beta_3 \right) 
  \nonumber\\ 
{\bf A}_{22}  &= & i ~\cos~\psi_1~\sin~\psi_2~{\cal G}_3 \sqrt{{\cal G}_1} \left(\alpha_1 \alpha_3 - \beta_1 \beta_3\right) + 
 i ~\sin~\psi_1~\cos~\psi_2~{\cal G}_4 \sqrt{{\cal G}_1} \left(\alpha_2 \alpha_4 - \beta_2 \beta_4\right) \nonumber\\
 &-&  \cos~\psi_1~\cos~\psi_2~\sqrt{{\cal G}_1{\cal G}_3 {\cal G}_4} \left(\alpha_1 \alpha_4 + \beta_1 \beta_4\right) +
  \sin~\psi_2~\sin~\psi_1~\sqrt{{\cal G}_1{\cal G}_3 {\cal G}_4} \left(\alpha_2 \alpha_3 + \beta_2 \beta_3 \right) 
  \nonumber\\ 
{\bf A}_{31}  &= & i~ \cos~\psi_1~\cos~\psi_2~{\cal G}_3 \sqrt{{\cal G}_1} \left(\alpha_1 \alpha_3 - \beta_1 \beta_3\right) -
 i~ \sin~\psi_1~\sin~\psi_2~{\cal G}_4 \sqrt{{\cal G}_1} \left(\alpha_2 \alpha_4 - \beta_2 \beta_4\right) \nonumber\\
 &+&  \cos~\psi_1~\sin~\psi_2~\sqrt{{\cal G}_1{\cal G}_3 {\cal G}_4} \left(\alpha_1 \alpha_4 + \beta_1 \beta_4\right) +
  \cos~\psi_2~\sin~\psi_1~\sqrt{{\cal G}_1{\cal G}_3 {\cal G}_4} \left(\alpha_2 \alpha_3 + \beta_2 \beta_3 \right) 
  \nonumber\\ 
{\bf A}_{32}  &= & -i ~\sin~\psi_1~\sin~\psi_2~{\cal G}_3 \sqrt{{\cal G}_1} \left(\alpha_1 \alpha_3 - \beta_1 \beta_3 \right) +
 i ~\cos~\psi_1~\cos~\psi_2~{\cal G}_4 \sqrt{{\cal G}_1} \left(\alpha_2 \alpha_4 - \beta_2 \beta_4\right) \nonumber\\
 &+&  \sin~\psi_1~\cos~\psi_2~\sqrt{{\cal G}_1{\cal G}_3 {\cal G}_4} \left(\alpha_1 \alpha_4 + \beta_1 \beta_4\right) +
  \cos~\psi_1~\sin~\psi_2~\sqrt{{\cal G}_1{\cal G}_3 {\cal G}_4} \left(\alpha_2 \alpha_3 + \beta_2 \beta_3 \right), 
  \nonumber\\  \nd
where $\alpha_i$ and $\beta_i$ are defined in \eqref{alphabeta} and note the appearances of ${\psi_1}$ as well as $\psi_2$ instead of the identification  $\psi_1 = \psi_2 \equiv {\psi\over 2}$ as used earlier. Our choice here is motivated
to encompass generalizations that will be useful soon. Note also that the other coefficients appearing in 
\eqref{planetarium} are given by:
\bg\label{paquett}
{\bf A}_{nk} \equiv - i~{\bf A}_{n-3, k} \sqrt{{\cal G}_2 \over {\cal G}_1}, ~~~~ n \ge 4, \nd
which are related to the wedge product with $e_\psi$. The above three-form \eqref{threeform} is rather non-trivial and 
one may simplify this by going to limit where ${\cal G}_3   = {\cal G}_4$. This will immediately make $\alpha_i  \equiv \alpha$ and $\beta_i \equiv \beta$ and one may easily infer from \eqref{planetarium} 
that in this limit $\Omega$ becomes:
\bg\label{logan}
\Omega & = & 2i ~{\cal G}_3 ~ \alpha \beta \left(e_{\theta_2} \wedge e_{\phi_2}  - e_{\theta_1} \wedge e_{\phi_1}\right) 
\wedge \left(i\sqrt{{\cal G}_1}~e_r +  \sqrt{{\cal G}_2}~e_\psi\right) \nonumber\\
&+&i \left[\left(\alpha^2 - \beta^2\right) \sin~\psi + i ~\cos~\psi\right] {\cal G}_3   \sqrt{{\cal G}_1}
~e_r \wedge \left(e_{\phi_1} \wedge e_{\theta_2}  + e_{\theta_1} \wedge e_{\phi_2}\right) \nonumber\\
&+& i\left[\left(\alpha^2 - \beta^2\right) \cos~\psi - i ~\sin~\psi\right] {\cal G}_3   \sqrt{{\cal G}_1}
~e_r \wedge \left(e_{\theta_1} \wedge e_{\theta_2}  - e_{\phi_1} \wedge e_{\phi_2}\right) \nonumber\\
&-&i \left[i\left(\alpha^2 - \beta^2\right) \sin~\psi - \cos~\psi\right] {\cal G}_3   \sqrt{{\cal G}_2}
~e_\psi \wedge \left(e_{\phi_1} \wedge e_{\theta_2}  + e_{\theta_1} \wedge e_{\phi_2}\right) \nonumber\\
&-&i \left[i\left(\alpha^2 - \beta^2\right) \cos~\psi + \sin~\psi\right] {\cal G}_3   \sqrt{{\cal G}_2}
~e_\psi \wedge \left(e_{\theta_1} \wedge e_{\theta_2}  - e_{\phi_1} \wedge e_{\phi_2}\right), 
\nd
where we have defined $\psi$ as the combination $\psi_1 + \psi_2$, and 
used $\alpha^2 + \beta^2 = 1$ to simplify further the above expression. At this stage we can use the explicit form for the ${\cal G}_i$ warp-factors given in \eqref{hush} as well as the values of $\alpha$ and $\beta$ given in \eqref{carlas} to express $\Omega$ for a K\"ahler manifold as:
\bg\label{kongkiz} 
\Omega & = & {i\mu^2 {\bf T} \over r^2} \left(e_{\theta_2} \wedge e_{\phi_2}  - e_{\theta_1} \wedge e_{\phi_1}\right) 
\wedge \left({2i\over r{\bf S}}~e_r + e_\psi\right)\nonumber\\
&-& i{\bf T} \left(i{\bf S}~\sin~\psi - \cos~\psi\right) 
~e_\psi \wedge \left(e_{\phi_1} \wedge e_{\theta_2}  + e_{\theta_1} \wedge e_{\phi_2}\right) \nonumber\\
&-& i{\bf T} \left(i{\bf S}~\cos~\psi + \sin~\psi\right) 
~e_\psi \wedge \left(e_{\theta_1} \wedge e_{\theta_2}  - e_{\phi_1} \wedge e_{\phi_2}\right) \nonumber\\
&+& {2i{\bf T} \over r {\bf S}}\left({\bf S}~ \sin~\psi + i ~\cos~\psi\right)
~e_r \wedge \left(e_{\phi_1} \wedge e_{\theta_2}  + e_{\theta_1} \wedge e_{\phi_2}\right) \nonumber\\
&+& {2i{\bf T}\over r {\bf S}}\left({\bf S}~ \cos~\psi - i ~\sin~\psi\right) 
~e_r \wedge \left(e_{\theta_1} \wedge e_{\theta_2}  - e_{\phi_1} \wedge e_{\phi_2}\right), 
\nd
where $\mu^2$ is the deformation parameter of a warped-deformed conifold. This expression matches well with the 
three-form derived in \cite{ankegwyn}, the difference arising from some redefinition of the $\psi_i$ coordinates of  
\cite{ankegwyn} with the ones used here. We have also defined ${\bf S}$ and ${\bf T}$ in the following way:
\bg\label{dumpling} 
{\bf S} \equiv \sqrt{1 - {\mu^4\over r^4}}, ~~~~~ 
{\bf T} \equiv {\gamma\over 8} \sqrt{\gamma + \left(r^2 \gamma' - \gamma\right)\left(1 - {\mu^4\over r^4}\right)}. \nd
We now have all the ingredients to compute the closure of the fundamental three-form $\Omega$. We will start with the simple case of K\"ahler manifold with $\Omega$ given by \eqref{kongkiz} above. We find:
\bg\label{belblond}
d\Omega & = & {\bf E}_1 ~e_r \wedge d\psi \wedge \left(e_{\theta_2} \wedge e_{\phi_2} - e_{\theta_1} \wedge e_{\phi_1}\right) \\
& + & \left({\bf E}_2~\cos~\psi - i {\bf E}_3~\sin~\psi\right)~e_r\wedge d\psi \wedge \left(e_{\theta_1} \wedge e_{\theta_2}
- e_{\phi_1} \wedge e_{\phi_2}\right) \nonumber\\
& + & \left({\bf E}_2~\sin~\psi + i {\bf E}_3~\cos~\psi\right)~e_r\wedge d\psi \wedge \left(e_{\phi_1} \wedge e_{\theta_2}
+ e_{\theta_1} \wedge e_{\phi_2} \right)\nonumber\\
& + & \left({\bf E}_2~\cos~\psi - i {\bf E}_3~\sin~\psi\right)
e_r \wedge e_{\theta_1} \wedge e_{\theta_2}\wedge \left(\cot~\theta_1 ~e_{\phi_1} + \cot~\theta_2~e_{\phi_2}\right) 
\nonumber\\
&+& \left[{\bf E}_1 ~\cot~\theta_1 + \cot~\theta_2\left({\bf E}_2 ~\sin~\psi + i {\bf E}_3 ~\cos~\psi\right)\right] ~e_r \wedge
e_{\phi_1} \wedge e_{\theta_2} \wedge e_{\phi_2} \nonumber\\  
&+& \left[{\bf E}_1 ~\cot~\theta_2 + \cot~\theta_1\left({\bf E}_2 ~\sin~\psi + i {\bf E}_3 ~\cos~\psi\right)\right] ~e_r \wedge
e_{\phi_1} \wedge e_{\theta_1} \wedge e_{\phi_2} , \nonumber
\nd
where interestingly, other components vanish because of the anti-symmetric nature of them. We have also defined 
${\bf E}_i$ in the following way:
\bg\label{atomicb}
{\bf E}_1 \equiv i\mu^2 \del_r \left({{\bf T}\over r^2}\right), ~~~~~ {\bf E}_2 \equiv \del_r\left({\bf S T}\right) - {2{\bf T} \over r {\bf S}}, ~~~~~ {\bf E}_3 \equiv \del_r {\bf T} - {2{\bf T} \over r},  \nd 
where ${\bf S}$ and ${\bf T}$ are defined earlier in \eqref{dumpling}. 
Demanding an integrable complex structure would imply the closure of $\Omega$, which would further imply the vanishing of the three ${\bf E}_i$ coefficients. Vanishing of ${\bf E}_1$ implies that ${\bf T}$ given in \eqref{dumpling} is a much simpler function and should be proportional to $r^2$. In fact, since ${\bf T}$ is constructed out of $\gamma$ and $\gamma$ satisfies the differential equation \eqref{brycenunu}, one may easily check that ${\bf T}$ may indeed be 
expressed alternatively as:
\bg\label{nascar}
{\bf T} ~ = ~ {r^2 \over 4\sqrt{6}}. \nd
Plugging in the value of ${\bf T}$ from \eqref{nascar} (or \eqref{dumpling}) and ${\bf S}$  from \eqref{dumpling} in 
\eqref{atomicb}, one may verify that both ${\bf E}_2$ and ${\bf E}_3$ also vanish. Thus our choice of the pre-dual complex vielbeins indeed leads to a complex manifold once we choose the warp-factors to take the form \eqref{hush}. This of course confirms the Calabi-Yau nature of the background \eqref{hush}. 

What happens in case when the warp-factors are different from the Calabi-Yau choice \eqref{hush}? Clearly, as we saw earlier, with generic choice of the ${\cal G}_i$ warp-factors, the manifold cannot be a K\"ahler manifold.  The question is whether the manifold can still be a complex manifold\footnote{Complexity is defined with respect to the simplest complex structure.}. 
To see this let us ask what constraints do the warp-factors have to satisfy to allow for an integrable complex structure.  Computing $d\Omega$ gives us the following form:
\bg\label{highway}
d\Omega & = & e_r \wedge e_{\theta_1} \wedge e_{\theta_2} \wedge \left({\bf C}_{11} ~\cot~\theta_1~e_{\phi_1}  + 
{\bf C}_{12} ~\cot~\theta_2~e_{\phi_2}\right) \\    
&-& i~d\psi  \wedge e_{\theta_1} \wedge e_{\theta_2} \wedge \left({\bf C}_{21} ~\cot~\theta_1~e_{\phi_1}  + 
{\bf C}_{22} ~\cot~\theta_2~e_{\phi_2}\right) \nonumber\\  
& + & e_r \wedge d\psi \wedge \left({\bf B}_{31}~e_{\theta_1} \wedge e_{\theta_2}  +  {\bf B}_{32}~e_{\phi_1} \wedge e_{\phi_2}\right) + i~{\bf E} ~e_{\theta_1} \wedge e_{\phi_1} \wedge e_{\theta_2} \wedge e_{\phi_2} \nonumber\\   
&+& e_r \wedge  d\psi \wedge \left({\bf B}_{11}~e_{\theta_2} \wedge e_{\phi_2}  +  {\bf B}_{12}~e_{\theta_1} \wedge e_{\phi_1} 
+  {\bf B}_{21}~e_{\phi_1} \wedge e_{\theta_2}  + {\bf B}_{22}~e_{\theta_1} \wedge e_{\phi_2}\right) \nonumber\\
&+& e_{\phi_1} \wedge e_{\phi_2} \wedge \left[e_r \wedge \left({\bf D}_{11}~e_{\theta_2}  + {\bf D}_{12}~
e_{\theta_1}\right)
+ i ~d\psi \wedge  \left({\bf D}_{21}~\cot~\theta_2 ~e_{\theta_2}  + {\bf D}_{22}~\cot~\theta_1 ~e_{\theta_1}\right)\right],\nonumber \nd
where, as expected, contains more terms than the simplified version that we studied earlier in \eqref{belblond}. The various coefficients are now defined in the following way:
\bg\label{syrendel}
&& {\bf D}_{1k} = \cot~\theta_{k+s}\left[{\bf A}_{32} + i~\del_r\left({\bf A}_{2k} \sqrt{{\cal G}_2\over {\cal G}_1}\right) \right] + 
i~\cot~\theta_k~\del_r\left({\bf A}_{1k} \sqrt{{\cal G}_2\over {\cal G}_1}\right) \\
&& {\bf B}_{3k} = (-1)^k\left[\del_\psi {\bf A}_{3k} + i \del_r\left({\bf A}_{3k} \sqrt{{\cal G}_2\over {\cal G}_1}\right)\right],~~
{\bf C}_{1k} = {\bf A}_{2k} - i \del_r\left({\bf A}_{31} \sqrt{{\cal G}_2 \over {\cal G}_1}\right), \nonumber\\
&& {\bf D}_{2k} = \sqrt{{\cal G}_2 \over {\cal G}_1}\left(\del_\psi {\bf A}_{2k} - {\bf A}_{32}\right), ~~
{\bf E} = \sqrt{{\cal G}_2 \over {\cal G}_1} \left[{\bf A}_{11} - {\bf A}_{12} + \cot~\theta_1 ~\cot~\theta_2 \left({\bf A}_{22} - {\bf A}_{21}\right)\right] \nonumber\\
&& {\bf B}_{1k} = i(-1)^k \del_r\left({\bf A}_{1k} \sqrt{{\cal G}_2\over {\cal G}_1}\right), ~
{\bf B}_{2k} = -\del_\psi {\bf A}_{2k} -i \del_r\left({\bf A}_{2k} \sqrt{{\cal G}_2\over {\cal G}_1}\right), ~
{\bf C}_{2k} = {{\bf A}_{2k} + \del_\psi {\bf A}_{31}\over \sqrt{{\cal G}_1/{\cal G}_2}},   \nonumber \nd
where $k = 1, 2$ and $s \equiv (-1)^{k+1}$. Note that, in defining the coefficients we have taken $\psi$ derivatives 
instead of resorting to more generic $\psi_1$ and $\psi_2$ derivatives. The latter could be easily implemented, at the cost of making the analysis more cumbersome. To avoid this, we have taken $\psi_1 = \psi_2 = {\psi\over 2}$, 
although the generalization that we indulged in \eqref{planetarium} will become useful later. 

Vanishing of $d\Omega$ now implies the vanishing of {\it all} the coefficients 
in \eqref{syrendel}. This will lead to multiple constraints, so let us tread carefully here. First, the vanishing of ${\bf E}$ 
and ${\bf B}_{1k}$ in \eqref{syrendel} immediately tells us that:
\bg\label{kissof}
{\bf A}_{11} = {\bf A}_{12} = \alpha \sqrt{{\cal G}_1\over {\cal G}_2}, ~~~~~ {\bf A}_{22} = {\bf A}_{21}, \nd
where the ${\bf A}_{ij}$ coefficients may be read from \eqref{planetarium}, and $\alpha$ is a constant whose value 
will be determined later. The other constraints may be determined from the vanishing of ${\bf C}_{1k}, {\bf C}_{2k}, 
{\bf D}_{1k}$ and ${\bf D}_{2k}$ in the following way:
\bg\label{femfat}
&& {\bf A}_{21} - i~ \del_r\left({\bf A}_{31} \sqrt{{\cal G}_2\over {\cal G}_1}\right) = 0, ~~~ {\bf A}_{32} 
- \del_\psi {\bf A}_{21} = 0
\nonumber\\
&& {\bf A}_{32} + i ~\del_r\left({\bf A}_{21} \sqrt{{\cal G}_2\over {\cal G}_1}\right) = 0, ~~~ {\bf A}_{21} 
+ \del_\psi {\bf A}_{31} = 0,
\nd
where the vanishing of the RHS of the equations does not imply simple separation of variables because each of the 
${\bf A}_{ij}$ coefficients appearing above are complex functions as should be evident from \eqref{planetarium}.
The above constraints relate many of ${\bf A}_{ij}$ coefficients and we can combine them to generate other relations. A useful way to manipulate \eqref{femfat} is to get the following three relations:
\bg\label{xmen3}
&&\del_\psi {\bf A}_{21} + i~\del_r\left({\bf A}_{21} \sqrt{{\cal G}_2\over {\cal G}_1}\right) = 0 \nonumber\\
&&\del_\psi {\bf A}_{31} + i~\del_r\left({\bf A}_{31} \sqrt{{\cal G}_2\over {\cal G}_1}\right) = 0 \nonumber\\
&&\del_\psi {\bf A}_{32} + i~\del_r\left({\bf A}_{32} \sqrt{{\cal G}_2\over {\cal G}_1}\right) = 0, \nd
which separate the coefficients in a meaningful way. Amazingly these equations are exactly the equations one may get by demanding the vanishing of ${\bf B}_{2k}$  and ${\bf B}_{3k}$ coefficients! Thus \eqref{femfat} and \eqref{xmen3} 
are equivalent sets of equations\footnote{Note that one of the equation appearing in \eqref{femfat} is redundant.}.
This shows the consistency of the analysis once we demand an integrable complex structure for the pre-dual framework.   

One may also verify the consistency of the set of equations \eqref{xmen3} by going to the warped-resolved conifold
limit. This is the limit where the $\alpha_i$ and $\beta_i$ coefficients in \eqref{alphabeta} satisfy the condition
\eqref{edenlac}. In this limit, one may easily verify that the relevant ${\bf A}_{ij}$ coefficients in \eqref{planetarium} 
satisfy:
\bg\label{xmen1} 
{\bf A}_{11} = {\bf A}_{12} = 0, ~~~~ {\bf A}_{31} = {\bf A}_{32} = - i~{\bf A}_{21} = 
\sqrt{{\cal G}_1{\cal G}_3{\cal G}_4} \left(i~\cos~\psi + \sin~\psi\right), \nd
with all other coefficients related to this via \eqref{paquett}. Now plugging \eqref{xmen1} in any of the three equations given in \eqref{xmen3} immediately gives us the condition \eqref{zeroeffect}.

\subsubsection{Background three-form fluxes in the ${\bf A} = {\bf B}$ limit \label{a=b}}

It is time now to work out the consistent three-form fluxes in the resolved warped-deformed conifold background that 
would solve all the type IIB EOMs. The NS three form flux ${\bf H}_3$ is easy to determine. Comparing 
\eqref{kalbinder} and \eqref{warning} one may express the NS flux is the following expected way:
\bg\label{phonemarie}
{{\bf H}_3\over {\rm sinh}~\gamma} & = & -dr \wedge \left[e_{\theta_1} \wedge e_{\phi_1}  \left(\sqrt{{\cal G}_1 
{\cal G}_2} 
- {\del {\bf C}\over \del r}\right) 
+ e_{\theta_2} \wedge e_{\phi_2}  \left(\sqrt{{\cal G}_1 {\cal G}_2} 
- {\del {\bf D}\over \del r}\right)\right]\\
&+& {\del {\cal F}\over \del r} ~dr \wedge  \left(e_{\theta_1} \wedge e_{\phi_2} + e_{\theta_2} \wedge e_{\phi_1}\right)
+ {\cal F} ~d\theta_1 \wedge d\theta_2 \wedge \left(\cos~\theta_1 d\phi_1 - \cos~\theta_2 d\phi_2\right), \nonumber  \nd
where the various coefficients appearing above have already been described in \eqref{IT2}, and they all depend on 
the coefficients in the metric \eqref{crooked} as well as our choice of the vielbeins \eqref{IT}. 

On the other hand the RR three-form flux ${\bf F}_3$ is more non-trivial as it relies on the Hodge star operation 
in \eqref{kalbinder}. To determine this we will have to perform a series of manipulations that will express all the relevant variables in terms of the vielbeins $e_i$ in \eqref{IT}. First note that:
\bg\label{coupleHF}
&&e_{\theta_k} = \cos~\psi_k\left(\delta_{k1}\sigma_1 + \delta_{k2}\Sigma_1\right) 
- \sin~\psi_k\left(\delta_{k1}\sigma_2 + \delta_{k2}\Sigma_2\right) \nonumber\\
&& e_{\phi_k} = \sin~\psi_k\left(\delta_{k1}\sigma_1 + \delta_{k2}\Sigma_1\right) 
+ \cos~\psi_k\left(\delta_{k1}\sigma_2 + \delta_{k2}\Sigma_2\right), \nd
for $k = 1, 2$. The above redefinition is useful because we can use the vielbeins in \eqref{IT} to rewrite the $\sigma_i$
and $\Sigma_i$ in the following suggestive way:
\bg\label{serling}
&&\sigma_1 \equiv  {\alpha_3 e_3 - \beta_3 e_5\over \left(\alpha_1 \alpha_3 - \beta_1 \beta_3\right)\sqrt{{\cal G}_3}}, 
~~~~~~ 
\sigma_2 \equiv  {\alpha_4 e_4 + \beta_4 e_6\over \left(\alpha_2 \alpha_4 - \beta_2 \beta_4\right)\sqrt{{\cal G}_4}}
\nonumber\\ 
&&\Sigma_1 \equiv  {\alpha_1 e_5 - \beta_1 e_3\over \left(\alpha_1 \alpha_3 - \beta_1 \beta_3\right)\sqrt{{\cal G}_3}}, 
~~~~~~ 
\Sigma_2 \equiv  {\alpha_2 e_6 + \beta_2 e_4\over \left(\alpha_2 \alpha_4 - \beta_2 \beta_4\right)\sqrt{{\cal G}_4}}.\nd
Another quantity that will be useful from our earlier setting is the functional form for ${\bf A}_{21}$ in
\eqref{planetarium}. We can use the $\psi_1, \psi_2$ dependence of ${\bf A}_{21}$ to write it in the ordered form 
as ${\bf A}_{21}(\psi_1, \psi_2, r)$. This will be useful to express the following three functions that we will need soon:
\bg\label{thicmil}
&&{\bf H}_0 \equiv {{\cal G}_3{\cal G}_4\sqrt{{\cal G}_1{\cal G}_2} {\cal F}_r \over 
{\bf A}_{21}\left(\pi/2, 0, r\right) {\bf A}_{21}\left(0, \pi/2, r\right)} \nonumber\\
&&{\bf H}_1 \equiv {{\cal G}_3{\cal G}_4\sqrt{{\cal G}_1{\cal G}_2}\left({\bf C}_r - \sqrt{{\cal G}_1{\cal G}_2}\right)\over 
{\bf A}_{21}\left(\pi/2, 0, r\right) {\bf A}_{21}\left(0, \pi/2, r\right)} \nonumber\\
&& {\bf H}_2 \equiv {{\cal G}_3{\cal G}_4\sqrt{{\cal G}_1{\cal G}_2}\left({\bf D}_r - \sqrt{{\cal G}_1{\cal G}_2}\right)\over 
{\bf A}_{21}\left(\pi/2, 0, r\right) {\bf A}_{21}\left(0, \pi/2, r\right)}, \nd
where the subscript $r$ denotes derivative with respect to $r$, and the functional form for ${\bf C}, {\bf D}$ and ${\cal F}$ have been defined 
in \eqref{IT2}. At this point we will also resort to a more specific setting of $\psi_1 = \psi_2 = {\psi\over 2}$ in 
\eqref{coupleHF}, and define the following four additional functions:
\bg\label{fantdes}
&&{\bf H}_{4a} \equiv {{\cal F} {\cal G}_1 {\cal G}_3 \sqrt{{\cal G}_1 {\cal G}_2} ~\cot~\theta_1~\sin~{\psi\over 2} \over 
{\bf A}_{21}(\pi/2, 0, r) {\bf A}_{21}(0, \pi/2, r)}, ~~~~
{\bf H}_{4b} \equiv {{\cal F} {\cal G}_1 {\cal G}_4 \sqrt{{\cal G}_1 {\cal G}_2} ~\cot~\theta_1~\cos~{\psi\over 2} \over 
{\bf A}_{21}(\pi/2, 0, r) {\bf A}_{21}(0, \pi/2, r)} \nonumber\\
&&{\bf H}_{4c} \equiv {{\cal F} {\cal G}_1 {\cal G}_4 \sqrt{{\cal G}_1 {\cal G}_2} ~\cot~\theta_2~\cos~{\psi\over 2} \over 
{\bf A}_{21}(\pi/2, 0, r) {\bf A}_{21}(0, \pi/2, r)}, ~~~~
{\bf H}_{4d} \equiv {{\cal F} {\cal G}_1 {\cal G}_3 \sqrt{{\cal G}_1 {\cal G}_2} ~\cot~\theta_2~\sin~{\psi\over 2} \over 
{\bf A}_{21}(\pi/2, 0, r) {\bf A}_{21}(0, \pi/2, r)}, \nonumber\\ \nd
where note the appearance of ${\cal F}(r)$, defined in \eqref{IT2}, in all of the four functions above as well as in 
${\bf H}_0$, defined in \eqref{thicmil}. Since ${\cal F}(r)$ vanishes in the non-K\"ahler resolved conifold setting
\eqref{edenlac}, this would imply the vanishing of ${\bf H}_0$ as well as all the four functions in \eqref{fantdes}. 
With these in hand, we are now ready to express the RR three-form flux ${\bf F}_3$ in the following 
way:
\bg\label{sacredeer}
{{\bf F}_3 \over  e^{-2\phi} {\rm cosh}~\gamma} & = & 
\left({\bf H}_1 {\bf J}_{11} + {\bf H}_2 {\bf J}_{12} + {\bf H}_0 {\bf J}_{13}\right)~ e_\psi \wedge \sigma_1 \wedge \sigma_2 \\
&+& \left({\bf H}_1 {\bf J}_{21} + {\bf H}_2 {\bf J}_{22} + {\bf H}_0 {\bf J}_{23}\right)~ e_\psi \wedge \sigma_1 \wedge \Sigma_2 \nonumber\\
&+& \left({\bf H}_1 {\bf J}_{31} + {\bf H}_2 {\bf J}_{32} + {\bf H}_0 {\bf J}_{33}\right)~ e_\psi \wedge \Sigma_1 \wedge \sigma_2 \nonumber\\
&+& \left({\bf H}_1 {\bf J}_{41} + {\bf H}_2 {\bf J}_{42} + {\bf H}_0 {\bf J}_{43}\right)~ e_\psi \wedge \Sigma_1 \wedge \Sigma_2 \nonumber\\
&+& \left[{\bf H}_{4a} \left({\bf K}_{11} ~\sigma_1  + {\bf K}_{12} ~\Sigma_1\right)   + 
{\bf H}_{4d} \left({\bf K}_{21} ~\sigma_1  + {\bf K}_{22} ~\Sigma_1\right)\right] \wedge e_r \wedge e_\psi \nonumber\\   
&+& \left[{\bf H}_{4b} \left({\bf K}_{31} ~\sigma_2  + {\bf K}_{32} ~\Sigma_2\right)   + 
{\bf H}_{4c} \left({\bf K}_{41} ~\sigma_2  + {\bf K}_{42} ~\Sigma_2\right)\right] \wedge e_r \wedge e_\psi, \nonumber \nd
where note that we have used the Maurer-Cartan forms to express the RR three-form flux. We can go back to the 
conifold coordinates also to write ${\bf F}_3$ equivalently. The former choice is dictated by the brevity of the expression, but reveals no interesting physics. Resorting to the conifold coordinates is useful to see how the fluxes are distributed over the internal cycles $-$ an exercise that we will indulge in later. 

Note also the appearances of ${\bf J}_{ab}$ and ${\bf K}_{ab}$ coefficients. These solely depend on the $\alpha_i$ and $\beta_i$ coefficients that we used to define our vielbeins \eqref{IT}. For example, the ${\bf J}_{1n}$ coefficients are defined in the following way:
\bg\label{infinite}
{\bf J}_{12} &\equiv & \alpha_1^2 \alpha_2^2 + \alpha_1^2 \beta_2^2 + \alpha_2^2 \beta_1^2 
+ \beta_1^2 \beta_2^2 \nonumber\\
 {\bf J}_{13}& \equiv&  \left(\alpha_1 \alpha_4 - \beta_2 \beta_3\right) \alpha_1 \beta_2 + 
\left(\alpha_1 \beta_4 - \beta_3 \alpha_2\right) \alpha_1 \alpha_2  \nonumber\\
&-& \left(\alpha_3 \alpha_2 - \beta_1 \beta_4\right) \alpha_2 \beta_1
- \left(\alpha_3 \beta_2 - \beta_1 \alpha_4\right) \beta_1 \beta_2\nonumber\\
 {\bf J}_{11} &\equiv &   -\beta_1 \beta_2 \alpha_3 \alpha_4 -\beta_1 \beta_4 \alpha_3 \alpha_2 -\beta_3 \beta_2 \alpha_1 \alpha_4 -\beta_3 \beta_4 \alpha_1 \alpha_2, \nd
which are in general non-trivial functions of the radial coordinate $r$.   Note that the $\beta_i$ dependences of the above expressions will tell us that, in the limit when $\beta_i = 0$,  the only non-zero coefficient is ${\bf J}_{12}$ and 
is given by:
\bg\label{chucca}
{\bf J}_{12} \equiv \alpha_1^2 \alpha_2^2, \nd
with the other coefficients vanishing. Furthermore, since ${\bf H}_0$ also vanishes in this limit, at this stage the only non-zero contribution comes from the ${\bf H}_2 {\bf J}_{12}$ piece in the first line of \eqref{sacredeer}. To see the contributions from the other pieces, let us elaborate the next 
three coefficients ${\bf J}_{2n}$ in the following way:
\bg\label{52omnibus}
{\bf J}_{21} &\equiv &  \alpha_3 \alpha^2_4 \beta_1 + \beta_1 \beta_4^2 \alpha_3 + \alpha_1 \alpha_4^2 \beta_3 
+ \alpha_1 \beta_4^2 \beta_3 \nonumber\\
 {\bf J}_{22} &\equiv & -\alpha_2 \alpha_1^2 \beta_4 - \alpha_4 \alpha_1^2 \beta_2 - \alpha_2 \beta_1^2 \beta_4
- \alpha_4 \beta_1^2 \beta_2  \nonumber\\
{\bf J}_{23}& \equiv&  -\left(\alpha_1 \alpha_4 - \beta_2 \beta_3\right) \alpha_1 \alpha_4 - 
\left(\alpha_1 \beta_4 - \beta_3 \alpha_2\right) \alpha_1 \beta_4  \nonumber\\
&&+ \left(\alpha_3 \alpha_2 - \beta_1 \beta_4\right) \beta_1 \beta_4
+ \left(\alpha_3 \beta_2 - \beta_1 \alpha_4\right) \beta_1 \alpha_4, \nd
which take the form somewhat similar to what we had in \eqref{infinite}. In fact the similarity is more prominent for the 
${\bf J}_{13}$ and ${\bf J}_{23}$ components. The $\beta_i$ dependences of the above expressions however tell us that the only non-zero coefficient now, in the limit when $\beta_i = 0$, is ${\bf J}_{23}$ given by:
\bg\label{bendis}
{\bf J}_{23} =  - \alpha_1^2 \alpha_4^2, \nd
with others vanishing. We can compare this with \eqref{chucca}  where the non-vanishing component was 
${\bf J}_{12}$. However compared to \eqref{chucca}, \eqref{bendis} contributes nothing as it couples to ${\bf H}_0$ which in fact vanishes when $\beta_i$ vanishes. A similar story also emerges for the ${\bf J}_{3n}$ coefficients, which
take the following form:
  \bg\label{52villians}
{\bf J}_{32} &\equiv &  \alpha_1 \alpha^2_2 \beta_3 + \alpha_1 \beta_2^2 \beta_3 + \alpha_3 \alpha_2^2 \beta_1 
+ \alpha_3 \beta_2^2 \beta_1 \nonumber\\
 {\bf J}_{31} &\equiv & -\alpha_4 \alpha_3^2 \beta_2 - \alpha_2 \alpha_3^2 \beta_4 - \alpha_4 \beta_3^2 \beta_2
- \alpha_2 \beta_3^2 \beta_4  \nonumber\\
{\bf J}_{33}& \equiv&  \left(\alpha_1 \alpha_4 - \beta_2 \beta_3\right) \beta_2 \beta_3  + 
\left(\alpha_1 \beta_4 - \beta_3 \alpha_2\right) \alpha_2 \beta_3  \nonumber\\
& - &\left(\alpha_3 \alpha_2 - \beta_1 \beta_4\right) \alpha_2 \alpha_3
- \left(\alpha_3 \beta_2 - \beta_1 \alpha_4\right) \beta_2 \alpha_3 , \nd
with the $\beta_i$ dependences as in \eqref{52omnibus}. As before, both ${\bf J}_{31}$ and ${\bf J}_{32}$ vanish
in the limit of vanishing $\beta_i$, and the non-zero coefficient is ${\bf J}_{33}$ with the following value:
\bg\label{microwave}
{\bf J}_{33} =  -\alpha_2^2 \alpha_3^2. \nd
This again couples to ${\bf H}_0$ so doesn't contribute in the resolved conifold case. Finally, we can write the 
${\bf J}_{4n}$ series in the following way:
\bg\label{hepafilter}
{\bf J}_{41} & \equiv & \alpha_3^2 \alpha_4^2  + \alpha_3^2 \beta_4^2 + \beta_3^2 \beta_4^2 
+ \beta_3^2 \alpha_4^2 \nonumber\\
 {\bf J}_{43}& \equiv&  -\left(\alpha_1 \alpha_4 - \beta_2 \beta_3\right) \beta_3 \alpha_4  - 
\left(\alpha_1 \beta_4 - \beta_3 \alpha_2\right) \beta_3 \beta_4  \nonumber\\
&& + \left(\alpha_3 \alpha_2 - \beta_1 \beta_4\right) \alpha_3 \beta_4
+ \left(\alpha_3 \beta_2 - \beta_1 \alpha_4\right) \alpha_3 \alpha_4 \nonumber\\
{\bf J}_{42} & \equiv & -\alpha_1 \alpha_2 \beta_3 \beta_4    -\alpha_1 \alpha_4 \beta_2 \beta_3 
-\alpha_2 \alpha_3 \beta_1 \beta_4 -\alpha_3 \alpha_4 \beta_1 \beta_2, \nd        
whose forms are similar to the ${\bf J}_{1n}$ series in \eqref{infinite}. In fact the similarity even extends to the fact that 
the only non-zero component for vanishing $\beta_i$ is ${\bf J}_{41}$ and takes the value:
\bg\label{libofam}
{\bf J}_{41} = \alpha_3^2 \alpha^2_4. \nd 
 Both \eqref{chucca} and \eqref{libofam} are non-zero and couple to ${\bf H}_2$ and 
 ${\bf H}_1$ respectively in \eqref{thicmil}. Therefore they do contribute to the 
 RR three-form flux in the expression \eqref{sacredeer} for vanishing $\beta_i$. 
  
The only other remaining terms are the ${\bf K}_{ij}$ coefficients appearing in the expression for ${\bf F}_3$ in 
\eqref{sacredeer}. They can also be written in terms of the $\alpha_i$ and $\beta_i$ factors used in the 
vielbeins \eqref{IT}. Interestingly, the constraints \eqref{lathi} help us to express the ${\bf K}_{ij}$ coefficients 
completely in terms of the metric factors ${\cal G}_i$ in the following way:
\bg\label{chukk}
&& {\bf K}_{11} = {\bf K}_{22} = - {{\cal G}_6 \over 2 {\cal G}_3}, ~~~~
{\bf K}_{31} = {\bf K}_{42} = {{\cal G}_6 \over 2 {\cal G}_4}\nonumber\\
&&{\bf K}_{21} =  {\bf K}_{32} = -1, ~~~~~ {\bf K}_{12} =  {1\over {\bf K}_{41}} = -{{\cal G}_4 \over {\cal G}_3}. \nd
The second line above would be non-vanishing when $\beta_i$ vanishes, but since they all couple to ${\bf H}_{4m}$, 
they do not contribute in the resolved conifold set-up. 

This completes our analysis of the three-form fluxes for the non-K\"ahler deformed conifold background \eqref{crooked}. Before moving further it is worthwhile to verify whether we reproduce the correct three-form 
backgrounds for the non-K\"ahler resolved conifold set-up studied in \cite{DEM}.  The constraints on $\alpha_i$ and
$\beta_i$ are clearly \eqref{edenlac}, and imposing them gives us:
\bg\label{bic77}
{\bf F}_3 &=& e^{-2\phi} ~{\rm cosh}~\gamma \left({\bf H}_2 {\bf J}_{12} ~ e_\psi \wedge e_{\theta_1} \wedge e_{\phi_1} 
+ {\bf H}_1 {\bf J}_{41} ~ e_\psi \wedge e_{\theta_2} \wedge e_{\phi_2}\right)\\
& = & {e^{-2\phi} ~{\rm cosh}~\gamma \over {\cal G}_3 {\cal G}_4}
 \sqrt{{\cal G}_2\over {\cal G}_1}\left[ {\cal G}_3^2 \left(\sqrt{{\cal G}_1{\cal G}_2} - {\cal G}_{4r}\right) 
 e_{\theta_1} \wedge e_{\phi_1} + 
  {\cal G}_4^2 \left(\sqrt{{\cal G}_1{\cal G}_2} - {\cal G}_{3r}\right) 
 e_{\theta_2} \wedge e_{\phi_2}\right] \wedge e_\psi, \nonumber \nd
 which matches exactly with the one in \cite{DEM} upto an overall minus sign. The sign difference is because of the definition of $J$ in \eqref{funform} has an overall minus sign compared to the one used in \cite{DEM}. Clearly this 
 also influences the ${\bf H}_3$ flux whose functional form matches exactly with \eqref{phonemarie} again upto 
 an overall minus sign\footnote{The dilaton used in \cite{DEM} is related to $-\phi$ here.}. 
 
 The above consistency is encouraging, although not surprising. The background used in \eqref{crooked} is 
 {\it generic} enough to interpolate between the 
 non-K\"ahler versions of the regular, resolved and the deformed conifolds
 simply by an appropriate choice of the warp-factors ${\cal G}_i$. 
 
 \subsubsection{General study of supersymmetry in the Baryonic branch \label{salonfem}}
 
To analyze the supersymmetry associated with the metric \eqref{crooked} and the three-form fluxes, we will perform
a more generic approach by going away from the constraint ${\bf A} = {\bf B}$, where ${\bf A}$ and ${\bf B}$ are 
defined in \eqref{pendant}. This means we will take the full fundamental form $J$ defined in \eqref{funform2}. 
Let us also define three functions in the following way:
\bg\label{computer}
&&{\bf K} \equiv -{1\over 2} \sqrt{{\cal G}_3 {\cal G}_4} \left({\bf A} - {\bf B}\right) \sin~\psi \\
&& {\cal F}_1 \equiv \left({\bf A}~\cos^2 {\psi\over 2} + {\bf B}~\sin^2 {\psi\over 2}\right) \sqrt{{\cal G}_3 {\cal G}_4}, ~~~~
{\cal F}_2 \equiv \left({\bf B}~\cos^2 {\psi\over 2} + {\bf A}~\sin^2 {\psi\over 2}\right) \sqrt{{\cal G}_3 {\cal G}_4}, \nonumber \nd
where ${\cal F}_1$ and ${\cal F}_2$ modify the original definition of ${\cal F}$ in \eqref{IT2}. When ${\bf A} = {\bf B}$, both ${\cal F}_1$ and ${\cal F}_2$ get identified to ${\cal F}$ and ${\bf K}$ vanishes. One should also check the closure 
of $J$ as in \eqref{warning}, and now it takes the following form:
\bg\label{babywhere}
d\left(e^{2\phi} J\right) & = & \left(e_{\phi_1} \wedge e_{\phi_2} - e_{\theta_1}  \wedge e_{\theta_2}\right) 
\wedge \left({\bf K}_r e_r  + {\bf K}_\psi e_\psi\right) \\
& + & e_{\theta_1}  \wedge e_{\theta_2} \wedge \left({\bf K}_3 e_{\phi_1}   + {\bf K}_4 e_{\phi_2}\right)   
+ e_{\phi_1}  \wedge e_{\phi_2} \wedge \left({\bf K}_5 e_{\theta_1}   + {\bf K}_6 e_{\theta_2}\right) \nonumber\\ 
& + & e_r \wedge \left[ e_{\theta_1}  \wedge e_{\phi_1} \left(\sqrt{{\cal G}_1 {\cal G}_2} - {\bf C}_r\right)  + 
e_{\theta_2}  \wedge e_{\phi_2} \left(\sqrt{{\cal G}_1 {\cal G}_2} - {\bf D}_r\right) \right] \nonumber\\
& - & {\cal F}_{1r} e_r \wedge \left(e_{\theta_1} \wedge e_{\phi_2} + {{\cal F}_{2r} \over {\cal F}_{1r}} 
e_{\theta_2} \wedge e_{\phi_1}\right) 
 -  {\cal F}_{1\psi} e_\psi \wedge \left(e_{\theta_1} \wedge e_{\phi_2} + {{\cal F}_{2\psi} \over {\cal F}_{1\psi}} 
e_{\theta_2} \wedge e_{\phi_1}\right), \nonumber \nd
where the $r $ and $\psi$ subscript denote derivatives with respect to $r$ and $\psi$ respectively, and ${\bf C}$ and 
${\bf D}$ are defined in \eqref{IT2}. The other variables appearing above are defined in the following way:
\bg\label{elisfe}
&&{\bf K}_3 \equiv \left({\bf K}_\psi - {\cal F}_2\right) \cot~\theta_1, ~~~
{\bf K}_4 \equiv \left({\bf K}_\psi + {\cal F}_1\right) \cot~\theta_2 \nonumber\\
&& {\bf K}_5 \equiv \left({\bf K}  - {\cal F}_{1\psi}\right) \cot~\theta_1, ~~~
{\bf K}_6 \equiv \left({\bf K}  +  {\cal F}_{2\psi}\right) \cot~\theta_2 . \nd
With these definitions, \eqref{babywhere}  is clearly 
non-vanishing as before and for the Calabi-Yau case this vanishes in the usual way. 

For supersymmetry we will now require the complexified three-form flux ${\bf G}_3$ defined as the combination 
${\bf F}_3 - i e^{-\phi} {\bf H}_3$. This may be expressed in terms of the original vielbeins $e_1, e_2, ..., e_6$ given in \eqref{macbeth} with $h = 1$, in the following way:
\bg\label{winnie}
{\bf G}_3 & = & e_2 \wedge e_3 \wedge e_4 \left({\bf M}_4 e^{-2\phi} {\rm cosh}~\gamma + i {\bf P}_3 e^{-\phi}
{\rm sinh}~\gamma\right) \nonumber\\
 & + & e_2 \wedge e_5 \wedge e_6 \left({\bf M}_1 e^{-2\phi} {\rm cosh}~\gamma + i {\bf P}_4 e^{-\phi}
{\rm sinh}~\gamma\right)\nonumber\\
& + & e_2 \wedge e_4 \wedge e_5 \left({\bf M}_2 e^{-2\phi} {\rm cosh}~\gamma + i {\bf P}_1 e^{-\phi}
{\rm sinh}~\gamma\right) \nonumber\\
 & + & e_2 \wedge e_3 \wedge e_6 \left({\bf M}_3 e^{-2\phi} {\rm cosh}~\gamma + i {\bf P}_2 e^{-\phi}
{\rm sinh}~\gamma\right) \nonumber\\
& + & e_1 \wedge e_4 \wedge e_6 \left({\bf Q}_3 e^{-2\phi} {\rm cosh}~\gamma + i {\bf Q}_2 e^{-\phi}
{\rm sinh}~\gamma\right) \nonumber\\ 
& + & e_1 \wedge e_3 \wedge e_5 \left({\bf Q}_4 e^{-2\phi} {\rm cosh}~\gamma + i {\bf Q}_1 e^{-\phi}
{\rm sinh}~\gamma\right) \nonumber\\
& - & e_1 \wedge e_3 \wedge e_6 \left({\bf P}_1 e^{-2\phi} {\rm cosh}~\gamma - i {\bf M}_2 e^{-\phi}
{\rm sinh}~\gamma\right) \nonumber\\ 
 & - & e_1 \wedge e_4 \wedge e_5 \left({\bf P}_2 e^{-2\phi} {\rm cosh}~\gamma - i {\bf M}_3 e^{-\phi}
{\rm sinh}~\gamma\right) \nonumber\\
& - & e_1 \wedge e_5 \wedge e_6 \left({\bf P}_3 e^{-2\phi} {\rm cosh}~\gamma - i {\bf M}_4 e^{-\phi}
{\rm sinh}~\gamma\right) \nonumber\\ 
& - & e_1 \wedge e_3 \wedge e_4 \left({\bf P}_4 e^{-2\phi} {\rm cosh}~\gamma - i {\bf M}_1 e^{-\phi}
{\rm sinh}~\gamma\right) \nonumber\\ 
& - & e_2 \wedge e_4 \wedge e_6 \left({\bf Q}_1 e^{-2\phi} {\rm cosh}~\gamma - i {\bf Q}_4 e^{-\phi}
{\rm sinh}~\gamma\right) \nonumber\\ 
& - & e_2 \wedge e_3 \wedge e_5 \left({\bf Q}_2 e^{-2\phi} {\rm cosh}~\gamma - i {\bf Q}_3 e^{-\phi}
{\rm sinh}~\gamma\right) \nonumber\\ 
& + & e_6 \wedge e_3 \wedge e_4 \left(i{\bf N}_1 e^{-\phi} {\rm sinh}~\gamma\right)
+  e_4 \wedge e_5 \wedge e_6 \left(i{\bf N}_2 e^{-\phi} {\rm sinh}~\gamma\right) \nonumber\\
& + & e_5 \wedge e_3 \wedge e_4 \left(i{\bf N}_3 e^{-\phi} {\rm sinh}~\gamma\right)
+  e_3 \wedge e_5 \wedge e_6 \left(i{\bf N}_4 e^{-\phi} {\rm sinh}~\gamma\right) \nonumber\\
& - & e_1 \wedge e_2 \wedge e_5 \left({\bf N}_1 e^{-2\phi} {\rm cosh}~\gamma\right)
-  e_1 \wedge e_2 \wedge e_3 \left({\bf N}_2 e^{-2\phi} {\rm cosh}~\gamma\right) \nonumber\\
& + & e_1 \wedge e_2 \wedge e_6 \left({\bf N}_3 e^{-2\phi} {\rm cosh}~\gamma\right)
+  e_1 \wedge e_2 \wedge e_4 \left({\bf N}_4 e^{-2\phi} {\rm cosh}~\gamma\right), \nd
where note the distribution of the dilaton factor over each individual components. This will be useful soon. We have also used four set of variables: ${\bf M}_i, {\bf Q}_i, {\bf P}_i$ and ${\bf N}_i$ to express the complexified three-form 
${\bf G}_3$. All these variables will be further defined in terms of other variables in the problem, so let us go in steps. 
We first define the following two variables:  
\bg\label{kingi}
{\bf H}^{(0)}_1 \equiv {{\sqrt{{\cal G}_1 {\cal G}_2}} - {\bf C}_r \over 
\sqrt{{\cal G}_1 {\cal G}_3 {\cal G}_4}\left(\alpha_1 \alpha_3 - \beta_1 \beta_3\right) \left(\alpha_2 \alpha_4 - \beta_2 
\beta_4\right)}, ~~~
{\bf H}^{(0)}_2 \equiv \left({{\sqrt{{\cal G}_1 {\cal G}_2}} - {\bf D}_r \over \sqrt{{\cal G}_1 {\cal G}_2} - {\bf C}_r}\right) 
{\bf H}^{(0)}_1, \nonumber\\ \nd
with the subscript $r$ denoting derivatives with respect to $r$ for the two functions ${\bf C}$ and ${\bf D}$ defined 
in \eqref{IT2}. The denominators of \eqref{kingi} may be expressed alternatively using ${\bf A}_{21}$ functions, but we
will not do so here to avoid clutter.  There are {\it four} other sets of $H$ variables that we can define now. The first set being:
\bg\label{thor}
&&{\bf H}_{3c}^{(0)} \equiv {{\bf K}_r~\sin~\psi \over 
\sqrt{{\cal G}_1 {\cal G}_3 {\cal G}_4}\left(\alpha_1 \alpha_3 - \beta_1 \beta_3\right) \left(\alpha_2 \alpha_4 - \beta_2 
\beta_4\right)} \nonumber\\
&& {\bf H}_{3a}^{(0)} \equiv - {{\bf K}_r~\cos~\psi \over 
\left(\alpha_1 \alpha_3 - \beta_1 \beta_3\right) {\cal G}_3 \sqrt{{\cal G}_1}}, ~~~
{\bf H}_{3b}^{(0)} \equiv  {{\bf K}_r~\cos~\psi \over 
\left(\alpha_2 \alpha_4 - \beta_2 \beta_4\right) {\cal G}_4 \sqrt{{\cal G}_1}}, \nd
where ${\bf K}_r$ is the derivative of ${\bf K}$, defined in \eqref{computer}, with respect to the $r$ parameter. The 
second set, on the other hand, may be defined as:
\bg\label{sorenson}
&&{\bf H}_{3c}^{(1)} \equiv {{\cal F}_{1r}~\sin^2 {\psi\over 2} + {\cal F}_{2r}~\cos^2 {\psi\over 2}
\over 
\sqrt{{\cal G}_1 {\cal G}_3 {\cal G}_4}\left(\alpha_1 \alpha_3 - \beta_1 \beta_3\right) \left(\alpha_2 \alpha_4 - \beta_2 
\beta_4\right)} \nonumber\\
&& {\bf H}_{3a}^{(1)} \equiv {\left({\cal F}_{2r}  - {\cal F}_{1r}\right) \sin~\psi \over 2\left(\alpha_1 \alpha_3 - 
\beta_1 \beta_3\right) {\cal G}_3 \sqrt{{\cal G}_1}}, ~~~~
{\bf H}_{3b}^{(1)} \equiv -{\left({\cal F}_{2r}  - {\cal F}_{1r}\right) \sin~\psi \over 2\left(\alpha_2 \alpha_4 - 
\beta_2 \beta_4\right) {\cal G}_4 \sqrt{{\cal G}_1}}, \nd 
which differ from \eqref{thor} by the appearance of ${\cal F}_{ir}$ instead of ${\bf K}_r$. All these are defined with respect to $r$ derivative, so the natural question is to ask whether there are similar definitions with $\psi$ derivatives. The answer is in the affirmative, and appears as the other two sets of $H$ variables, which we will denote as the prime 
variables:
\bg\label{docsavage} 
&&{\bf H}_{3c}^{'(1)} \equiv {{\cal F}_{1\psi}~\sin^2 {\psi\over 2} + {\cal F}_{2\psi}~\cos^2 {\psi\over 2}
\over 
\sqrt{{\cal G}_2 {\cal G}_3 {\cal G}_4}\left(\alpha_1 \alpha_3 - \beta_1 \beta_3\right) \left(\alpha_2 \alpha_4 - \beta_2 
\beta_4\right)} \nonumber\\
&& {\bf H}_{3n}^{'(0)}  ~ = ~ \left({{\bf K}_\psi \over {\bf K}_r}\sqrt{{\cal G}_1 \over {\cal G}_2}\right) {\bf H}_{3n}^{(0)}, ~~~
{\bf H}_{3k}^{'(1)} ~ = ~ \left({{\cal F}_{2\psi} - {\cal F}_{1\psi} \over 
{\cal F}_{2r} - {\cal F}_{1r}}\sqrt{{\cal G}_1 \over {\cal G}_2}\right) {\bf H}_{3k}^{(1)}
\nd
with $n = a, b, c$ and $k = a, b$  
as they appear in \eqref{thor} and \eqref{sorenson} above. 
We now only need two more parameters $b_o$ and $c_o$, defined as:
\bg\label{thepooh}
b_o \equiv {{\cal F}_{1r} ~\cos^2 {\psi\over 2} + {\cal F}_{2r} ~\sin^2 {\psi\over 2} \over 
{\cal F}_{1r} ~\sin^2 {\psi\over 2} + {\cal F}_{2r} ~\cos^2 {\psi\over 2}}, ~~~
c_o \equiv {{\cal F}_{1\psi} ~\cos^2 {\psi\over 2} + {\cal F}_{2\psi} ~\sin^2 {\psi\over 2} \over 
{\cal F}_{1\psi} ~\sin^2 {\psi\over 2} + {\cal F}_{2\psi} ~\cos^2 {\psi\over 2}},
\nd
which become identity when ${\cal F}_1 = {\cal F}_2$. This is of course the case when ${\bf A} = {\bf B}$, but since we are not considering this case, $b_o$ and $c_o$ precisely measure the deviations from the ${\bf A} = {\bf B}$ constraint. With this set of definitions, we are now ready to express ${\bf M}_i$ as the following linear combinations: 
\bg\label{blake}
&& {\bf M}_1 \equiv {\bf H}^{(0)}_1 \alpha_3 \alpha_4  - {\bf H}^{(0)}_2 \beta_1 \beta_2 
+ {\bf H}_{3c}^{(0)} \left(\alpha_3 \beta_2 + \alpha_4 \beta_1\right)   
+ {\bf H}_{3c}^{(1)} \left(\alpha_4 \beta_1 - b_o \alpha_3 \beta_2 \right) \\ 
&& {\bf M}_2 \equiv {\bf H}^{(0)}_1 \alpha_3 \beta_4  - {\bf H}^{(0)}_2 \beta_1 \alpha_2 
+ {\bf H}_{3c}^{(0)} \left(\alpha_2 \alpha_3 + \beta_1 \beta_4\right)   
+ {\bf H}_{3c}^{(1)} \left(\beta_1 \beta_4 - b_o \alpha_2 \alpha_3 \right) \nonumber\\ 
&& {\bf M}_3 \equiv {\bf H}^{(0)}_1 \alpha_4 \beta_3  - {\bf H}^{(0)}_2 \beta_2 \alpha_1 
+ {\bf H}_{3c}^{(0)} \left(\alpha_1 \alpha_4 + \beta_2 \beta_3\right)   
+ {\bf H}_{3c}^{(1)} \left(\alpha_1 \alpha_4 - b_o \beta_2 \beta_3 \right) \nonumber\\ 
&& {\bf M}_4 \equiv - {\bf H}^{(0)}_1 \beta_3 \beta_4  + {\bf H}^{(0)}_2 \alpha_1 \alpha_2 
- {\bf H}_{3c}^{(0)} \left(\alpha_1 \beta_4 + \alpha_2 \beta_3\right)   - 
 {\bf H}_{3c}^{(1)} \left(\alpha_1 \beta_4 - b_o \beta_3 \alpha_2 \right), \nonumber \nd 
which are completely expressed in terms of the un-primed $H$'s defined in \eqref{thor} and \eqref{sorenson} and the deviation factor $b_o$. The 
primed $H$'s, defined in \eqref{docsavage} first appear in defining the variables ${\bf Q}_i$ in the following way:
\bg\label{noworkshop} 
&&{\bf Q}_1 \equiv {\bf H}_{3a}^{(0)}  + {\bf H}_{3a}^{(1)}, ~~~~~ {\bf Q}_2 \equiv {\bf H}_{3b}^{(0)}  + {\bf H}_{3b}^{(1)}
\nonumber\\ 
&& {\bf Q}_3 \equiv {\bf H}_{3a}^{'(0)}  + {\bf H}_{3a}^{'(1)}, ~~~~~ {\bf Q}_4 \equiv {\bf H}_{3b}^{'(0)}  
+ {\bf H}_{3b}^{'(1)}, \nd
which are in some sense a mixture of both primed and un-primed $H$'s. We can however also construct variables that only depend on the primed $H$'s:  these are our ${\bf P}_i$ variables. They take the following form:
\bg\label{altuspress}
&&{\bf P}_1 \equiv {\bf H}_{3c}^{'(0)} \left(\alpha_1 \alpha_4 + \beta_2 \beta_3\right) + 
{\bf H}_{3c}^{'(1)} \left(\alpha_1 \alpha_4 - c_o \beta_2 \beta_3\right) \nonumber\\
&& {\bf P}_2 \equiv {\bf H}_{3c}^{'(0)} \left(\alpha_2 \alpha_3 + \beta_1 \beta_4\right) + 
{\bf H}_{3c}^{'(1)} \left(\beta_1 \beta_4 - c_o \alpha_2 \alpha_3\right) \nonumber\\
&& {\bf P}_3 \equiv {\bf H}_{3c}^{'(0)} \left(\alpha_3 \beta_2 + \alpha_4 \beta_1\right) + 
{\bf H}_{3c}^{'(1)} \left(\beta_1 \alpha_4 - c_o \alpha_3 \beta_2\right) \nonumber\\
&& {\bf P}_4 \equiv -{\bf H}_{3c}^{'(0)} \left(\alpha_1 \beta_4 + \alpha_2 \beta_3\right) -
{\bf H}_{3c}^{'(1)} \left(\alpha_1 \beta_4 - c_o \alpha_2 \beta_3\right), \nd
where $c_o$ is defined in \eqref{thepooh}.
These are somewhat reminiscent of the ${\bf M}_i$ variables in \eqref{blake} both in terms of the appearance of 
${\bf H}^{'(a)}_{3c}$ as well as the deviation parameter $c_o$. The final set of variables appearing in the 
${\bf G}_3$ flux are the ${\bf N}_i$ variables. They take the following form:
\bg\label{pikelia}
&&{\bf N}_1 \equiv \alpha_3 \left({\bf H}^{(0)}_{4a}  + {\bf H}^{'(0)}_{4a}\right)  + 
\beta_1 \left({\bf H}^{(0)}_{4d}  + {\bf H}^{'(0)}_{4d}\right) \nonumber\\  
&& {\bf N}_2 \equiv \beta_3 \left({\bf H}^{(0)}_{4a}  + {\bf H}'^{(0)}_{4a}\right)  + 
\alpha_1 \left({\bf H}^{(0)}_{4d}  + {\bf H}^{'(0)}_{4d}\right) \nonumber\\
&& {\bf N}_3 \equiv \beta_2 \left({\bf H}^{(0)}_{4c}  + {\bf H}^{'(0)}_{4c}\right)
- \alpha_4 \left({\bf H}^{(0)}_{4b}  + {\bf H}^{'(0)}_{4b}\right)  \nonumber\\
&& {\bf N}_4 \equiv \beta_4 \left({\bf H}^{(0)}_{4b}  + {\bf H}^{'(0)}_{4b}\right)
- \alpha_2 \left({\bf H}^{(0)}_{4c}  + {\bf H}^{'(0)}_{4c}\right), \nd
which are now defined in terms of yet another set of $H$-variables which we called ${\bf H}^{(0)}_{4n}$ 
and ${\bf H}^{'(0)}_{4n}$, where $n = a, b, c, d$. To define these, let us express the functional forms for 
${\bf H}^{(0)}_{4a}$
and ${\bf H}^{(0)}_{4b}$ in the following way:
\bg\label{hangman}
&&{\bf H}^{(0)}_{4a} \equiv {{\bf K}_3~\sin~{\psi\over 2} \over \left(\alpha_1 \alpha_3 - \beta_1 \beta_3\right)
\left(\alpha_2 \alpha_4 - \beta_2 \beta_4\right) {\cal G}_4 \sqrt{{\cal G}_3}} \nonumber\\
&& {\bf H}^{(0)}_{4b} \equiv {{\bf K}_3~\cos~{\psi\over 2} \over \left(\alpha_1 \alpha_3 - \beta_1 \beta_3\right)
\left(\alpha_2 \alpha_4 - \beta_2 \beta_4\right) {\cal G}_3 \sqrt{{\cal G}_4}}, \nd
with ${\bf K}_3$ defined as in \eqref{elisfe}.  These two definitions suffice to express the other variables appearing 
in \eqref{pikelia} in the following way:
\bg\label{alpac}
&&{\bf H}^{(0)}_{4c} \equiv - {\bf H}^{(0)}_{4b} \left({{\bf K}_4 \over {\bf K}_3}\right), ~~~~~
{\bf H}^{(0)}_{4d} \equiv - {\bf H}^{(0)}_{4a} \left({{\bf K}_4 \over {\bf K}_3}\right) \nonumber\\
&& {\bf H}^{'(0)}_{4a} \equiv  {\bf H}^{(0)}_{4a}~\cot~{\psi\over 2} \left({{\bf K}_5 \over {\bf K}_3}\right), ~~~~~ 
{\bf H}'^{(0)}_{4b} \equiv  - {\bf H}^{(0)}_{4b}~\tan~{\psi\over 2} \left({{\bf K}_5 \over {\bf K}_3}\right)\nonumber\\
&& {\bf H}'^{(0)}_{4c} \equiv  {\bf H}^{(0)}_{4b}~\tan~{\psi\over 2} \left({{\bf K}_6 \over {\bf K}_3}\right), ~~~~~ 
{\bf H}'^{(0)}_{4d} \equiv  - {\bf H}^{(0)}_{4a}~\cot~{\psi\over 2} \left({{\bf K}_6 \over {\bf K}_3}\right), \nd
where ${\bf K}_3$ is assumed to be a non-zero function. Clearly there are alternative ways to express the 
${\bf H}^{(0)}_{4n}$ and ${\bf H}'^{(0)}_{4n}$ functions, so our descriptions in \eqref{alpac} are {\it not} unique. However since 
the various ways of expressing ${\bf H}^{(0)}_{4n}$ and ${\bf H}'^{(0)}_{4n}$ have no bearing on the outcome of our analysis, we shall stick with \eqref{alpac} here.

We are now ready to analyze supersymmetry for our background. Earlier attempts to study supersymmetry 
have mostly concentrated on specific backgrounds coming from the Maldacena-Nunez \cite{malnun} or the Klebanov-Strassler \cite{KS} solutions, with techniques relying on the study of torsion classes (see for example \cite{butti} and the subsequent follow-up citations). A more direct approach of connecting the MN background with KS is in 
\cite{MM}.  Our approach here will be to analyze the structure of the ${\bf G}_3$ flux in \eqref{winnie}. The subtlety however is that the ${\bf G}_3$ flux is {\it not} ISD in the standard sense. Part of the reason being the presence of a non-trivial almost-complex structure ($i\sigma, i, i$)  as it appears in \eqref{machine}, whose integrability will be the subject of a discussion later. This means we will have to analyze the flux structure \eqref{winnie} with respect to this almost-complex structure.  
We start by studying the ($3, 0$) form:
\bg\label{thesis}
\mathbb{Z}_1~E_1 \wedge E_2 \wedge E_3, \nd
whose presence itself should be a sign of alarm as it breaks supersymmetry. 
The coefficient $\mathbb{Z}_1$ is a complex function that has both real and imaginary pieces that take the following form:
\vskip-.2in
{{\footnotesize
\bg\label{kamran}
&&{\rm Re}~\mathbb{Z}_1 = {{\bf P}_1  + {\bf P}_2 \over 8}\left(e^{-\phi} {\rm sinh}~\gamma 
+ {e^{-2\phi} {\rm cosh}~\gamma \over \sigma}\right) + 
{{\bf Q}_1  -  {\bf Q}_2\over 8} \left({e^{-\phi} {\rm sinh}~\gamma \over \sigma}
+ e^{-2\phi} {\rm cosh}~\gamma \right) \nonumber\\
&& {\rm Im}~\mathbb{Z}_1 = {{\bf Q}_3  -  {\bf Q}_4 \over 8}\left(e^{-\phi} {\rm sinh}~\gamma 
+ {e^{-2\phi} {\rm cosh}~\gamma \over \sigma}\right) -
{{\bf M}_2  +  {\bf M}_3\over 8} \left({e^{-\phi} {\rm sinh}~\gamma \over \sigma}
+ e^{-2\phi} {\rm cosh}~\gamma \right), \nonumber\\ \nd}}
\noindent where the ${\bf Q}_i$, ${\bf P}_i$ and ${\bf M}_i$ variables are defined in \eqref{noworkshop}, 
\eqref{altuspress} and \eqref{blake} respectively. Note the dependence 
on $\gamma$, the parameter defining the Baryonic branch, as well as $\sigma$, the almost-complex structure, for
both the real and the 
imaginary pieces in \eqref{kamran}.  They should vanish individually to preserve supersymmetry. From 
\eqref{kamran} clearly this could happen in multiple ways, but we will not make a specific choice now. Instead we will analyze the other flux components and in the end tabulate all the results to see what generic conditions may emerge from these sets of equations. One of the other flux components is the ($0, 3$) piece that takes the form:
\bg\label{saranun}
\mathbb{Z}_2 ~\overline{E}_1 \wedge \overline{E}_2 \wedge \overline{E}_3, \nd
which should not be thought of as the complex-conjugate of  \eqref{kamran}. Instead the real and the imaginary 
pieces of $\mathbb{Z}_2$ take the following form:
\vskip-.2in
{\footnotesize
\bg\label{shangchi}
&&{\rm Re}~\mathbb{Z}_2 = {{\bf P}_1  + {\bf P}_2\over 8} \left({e^{-2\phi} {\rm cosh}~\gamma \over \sigma}
-e^{-\phi} {\rm sinh}~\gamma \right) -
{{\bf Q}_1  -  {\bf Q}_2\over 8} \left({e^{-\phi} {\rm sinh}~\gamma \over \sigma}
- e^{-2\phi} {\rm cosh}~\gamma \right) \nonumber\\
&& {\rm Im}~\mathbb{Z}_2 = {{\bf Q}_4  -  {\bf Q}_3 \over 8} \left({e^{-2\phi} {\rm cosh}~\gamma \over \sigma}
-e^{-\phi} {\rm sinh}~\gamma \right) +
{{\bf M}_2  +  {\bf M}_3\over 8} \left(e^{-2\phi} {\rm cosh}~\gamma 
-{e^{-\phi} {\rm sinh}~\gamma \over \sigma}\right). \nonumber\\ \nd} 
These have a structure similar to \eqref{kamran}, but there are relative signs that differ. Clearly both the real and the imaginary pieces in \eqref{shangchi} could also
vanish individually in multiple ways, but again we will not make any choices right away. Instead we look at the next 
term that takes the form:
\bg\label{kungfu}
\mathbb{Z}_{31} ~{E}_1 \wedge \overline{E}_2 \wedge \overline{E}_3, \nd
 which is a ($1, 2$) form and is ISD with respect to the almost complex structure \eqref{machine}. This signals a less 
 prominent breaking of supersymmetry than the (3, 0) and the (0, 3) pieces, but is nevertheless still a source of concern. 
 The real and the imaginary pieces of $\mathbb{Z}_{31}$ take the following form:
 \vskip-.2in
 {\footnotesize
 \bg\label{jeantalon}
 &&{\rm Re}~\mathbb{Z}_{31} = {{\bf Q}_1  -  {\bf Q}_2\over 8} \left({e^{-\phi} {\rm sinh}~\gamma \over \sigma}+  
 e^{-2\phi} {\rm cosh}~\gamma \right)  - 
 {{\bf P}_1  + {\bf P}_2\over 8} \left({e^{-2\phi} {\rm cosh}~\gamma \over \sigma}
+e^{-\phi} {\rm sinh}~\gamma \right)\nonumber\\
&& {\rm Im}~\mathbb{Z}_{31} = {{\bf Q}_3  -  {\bf Q}_4 \over 8} \left({e^{-2\phi} {\rm cosh}~\gamma \over \sigma}
+ e^{-\phi} {\rm sinh}~\gamma \right) + 
{{\bf M}_2  +  {\bf M}_3\over 8} \left(e^{-2\phi} {\rm cosh}~\gamma 
+ {e^{-\phi} {\rm sinh}~\gamma \over \sigma}\right),  \nonumber\\ \nd}  
 where the ($\gamma, \sigma$) dependences are similar to the ones in \eqref{kamran}, although the relative signs differ. These small differences will become important later when we will try to put these coefficients to zero. The next supersymmetry breaking term is again an ISD ($1, 2$) form that may be written as: 
 \bg\label{ragnorak}
\mathbb{Z}_{32} ~{E}_2 \wedge \overline{E}_1 \wedge \overline{E}_3, \nd 
 that should not be regarded as anyway related to \eqref{kungfu} by redefinitions of the corresponding veilbeins. 
 This is because the real and the imaginary pieces of $\mathbb{Z}_{32}$ differ from \eqref{jeantalon} in the following interesting way:
 \vskip-.2in
 {\footnotesize
 \bg\label{itonya}
 &&{\rm Re}~\mathbb{Z}_{32} = {{\bf Q}_1  +  {\bf Q}_2\over 8} \left({e^{-\phi} {\rm sinh}~\gamma \over \sigma}+  
 e^{-2\phi} {\rm cosh}~\gamma \right)  + 
 {{\bf P}_2  - {\bf P}_1 \over 8} \left({e^{-2\phi} {\rm cosh}~\gamma \over \sigma}
+e^{-\phi} {\rm sinh}~\gamma \right)\nonumber\\
&& {\rm Im}~\mathbb{Z}_{32} = {{\bf M}_2  -  {\bf M}_3\over 8} \left(e^{-2\phi} {\rm cosh}~\gamma + {e^{-\phi} {\rm sinh}~\gamma \over \sigma}\right) -
{{\bf Q}_3  +  {\bf Q}_4 \over 8} \left({e^{-2\phi} {\rm cosh}~\gamma \over \sigma}
+ e^{-\phi} {\rm sinh}~\gamma \right). 
 \nonumber\\ \nd}  
 The differences arise from the relative signs of the various ${\bf M}_i, {\bf P}_i$ and ${\bf Q}_i$ terms compared to their appearances in \eqref{kamran}, \eqref{shangchi} and \eqref{jeantalon}.  On the other hand, the ($\gamma, \sigma$) dependences remain unchanged from what we had in \eqref{kamran} and \eqref{jeantalon}, and differ from the ones in \eqref{shangchi}. As we will see later, these subtle changes of the forms will help us to fix the supersymmetry conditions uniquely. The last such ($1, 2$) form has the following structure:
\bg\label{ittefaq}
\mathbb{Z}_{33} ~{E}_3 \wedge \overline{E}_1 \wedge \overline{E}_2, \nd 
which is allowed simply by the permutation choices of the veilbeins. The fact that it actually exists for our case comes 
from looking at the real and imaginary values of $\mathbb{Z}_{33}$, namely:
\vskip-.2in
{\footnotesize
 \bg\label{bluebox}
 &&{\rm Re}~\mathbb{Z}_{33} =  {{\bf P}_2  - {\bf P}_1 \over 8} \left({e^{-2\phi} {\rm cosh}~\gamma \over \sigma}+e^{-\phi} {\rm sinh}~\gamma \right) -
  {{\bf Q}_1  +  {\bf Q}_2\over 8} \left({e^{-\phi} {\rm sinh}~\gamma \over \sigma}+  
 e^{-2\phi} {\rm cosh}~\gamma \right)\nonumber\\
&& {\rm Im}~\mathbb{Z}_{33} = {{\bf M}_2  -  {\bf M}_3\over 8} \left(e^{-2\phi} {\rm cosh}~\gamma + {e^{-\phi} {\rm sinh}~\gamma \over \sigma}\right) +
{{\bf Q}_3  +  {\bf Q}_4 \over 8} \left({e^{-2\phi} {\rm cosh}~\gamma \over \sigma}
+ e^{-\phi} {\rm sinh}~\gamma \right). 
 \nonumber\\ \nd}  
 whose only point of difference from \eqref{itonya} appears from the relative signs of the two terms in the real and the imaginary pieces. This difference is of course how it distinguishes itself from \eqref{itonya}, but more crucially, this puts a stronger constraint on the vanishing of certain terms as we shall see soon.
 
The next category of terms are again non-supersymmetric ISD ($1, 2$) forms that appear from a different set of permutations of the three complex vielbeins.  In fact what we now need are the permutations of a set of {\it two} 
complex vielbeins. Our first example is:
\bg\label{hela} 
\mathbb{Z}_{41} ~{E}_2 \wedge \overline{E}_1 \wedge \overline{E}_2, \nd 
that is constructed out of $E_1$ and $E_2$ and their complex conjugates in a suitable way. The real and the imaginary pieces of the coefficient $\mathbb{Z}_{41}$ take the following values:
\bg\label{planethulk}
&&{\rm Re}~\mathbb{Z}_{41} = {1\over 4}\left({\bf P}_3 e^{-\phi} {\rm sinh}~\gamma 
- {{\bf P}_4 e^{-2\phi} {\rm cosh}~\gamma \over \sigma}\right) \nonumber\\
&&{\rm Im}~\mathbb{Z}_{41} = {1\over 4}\left({{\bf M}_1 e^{-\phi} {\rm sinh}~\gamma \over \sigma} 
- {\bf M}_4 e^{-2\phi} {\rm cosh}~\gamma \right), \nd
whose forms are very different from what we had earlier for \eqref{kamran}, \eqref{shangchi}, \eqref{jeantalon}, 
\eqref{itonya} and \eqref{bluebox}. Clearly vanishing of these terms would put further constraints but, as before, we will not make them zero right away. Instead we will look for the next ($1, 2$) form in the same category as before. Such a form can appear by replacing $\overline{E}_1$ in \eqref{hela} with $\overline{E}_3$. In other words, we 
expect:
 \bg\label{sopdee} 
\mathbb{Z}_{42} ~{E}_2 \wedge \overline{E}_3 \wedge \overline{E}_2, \nd 
to be in the same category as \eqref{hela} above. The real and the imaginary pieces of $\mathbb{Z}_{42}$ 
however tell a different story as they take the following form:
\bg\label{atyler}
{\rm Re}~\mathbb{Z}_{42} = - {1\over 4} {\bf N}_3 e^{-\phi} {\rm sinh}~\gamma, ~~~~~
{\rm Im}~\mathbb{Z}_{42} =  {1\over 4} {\bf N}_1 e^{-\phi} {\rm sinh}~\gamma, \nd
which is not only much simpler than \eqref{planethulk}, but also differs from it in the absence of 
$\sigma$. A somewhat similar story presents itself for the next form:
\bg\label{mellmon} 
\mathbb{Z}_{51} ~{E}_3 \wedge \overline{E}_1 \wedge \overline{E}_3, \nd 
for which we expect the real and the imaginary pieces to repeat the same structure as \eqref{planethulk}. This is 
almost true, except that the relative placement of the ${\bf P}_i$ and the ${\bf M}_i$ factors in the real and the 
imaginary pieces:
\bg\label{frogthor}
&&{\rm Re}~\mathbb{Z}_{51} = {1\over 4}\left({\bf P}_4 e^{-\phi} {\rm sinh}~\gamma 
- {{\bf P}_3 e^{-2\phi} {\rm cosh}~\gamma \over \sigma}\right) \nonumber\\
&&{\rm Im}~\mathbb{Z}_{51} = {1\over 4}\left({{\bf M}_4 e^{-\phi} {\rm sinh}~\gamma \over \sigma} 
- {\bf M}_1 e^{-2\phi} {\rm cosh}~\gamma \right), \nd
can be seen to differ from \eqref{planethulk}, although the relative signs of the terms do not change from 
\eqref{planethulk}. Expectedly:
\bg\label{lesterdent} 
\mathbb{Z}_{52} ~{E}_3 \wedge \overline{E}_2 \wedge \overline{E}_3, \nd 
which would be the next form in the same category as \eqref{mellmon}, has the real and the imaginary pieces of 
$\mathbb{Z}_{52}$ to repeat the same structure as in \eqref{atyler}, namely:
\bg\label{julcash}
{\rm Re}~\mathbb{Z}_{52} = - {1\over 4} {\bf N}_4 e^{-\phi} {\rm sinh}~\gamma, ~~~~~
{\rm Im}~\mathbb{Z}_{52} =  {1\over 4} {\bf N}_2 e^{-\phi} {\rm sinh}~\gamma, \nd
in the appearance of ${\rm sinh}~\gamma$ with no $\sigma$ dependence.  Finally the last two ($1, 2$) forms 
appear from $E_1$ and $E_3$ veilbeins, with their complex conjugates, in the following way:
\bg\label{babyback}
\mathbb{Z}_{61} ~{E}_1 \wedge \overline{E}_3 \wedge \overline{E}_1, ~~~~~
\mathbb{Z}_{62} ~{E}_1 \wedge \overline{E}_2 \wedge \overline{E}_1, \nd
that one might expect to follow similar structure as \eqref{planethulk} and \eqref{atyler} or \eqref{frogthor} and
\eqref{julcash} respectively. However it turns out that the real and the imaginary pieces of both 
$\mathbb{Z}_{61}$ and $\mathbb{Z}_{62}$  are not only
much simpler than any of the above set, but also differs from \eqref{atyler} and \eqref{julcash} in the appearance of ${\rm cosh}~\gamma$ as well of $\sigma$.
This is evident from the following form:
\bg\label{pmaalish}
&&{\rm Re}~\mathbb{Z}_{61} = {1\over 4\sigma} ~{\bf N}_3 e^{-2\phi} {\rm cosh}~\gamma, ~~~~
{\rm Im}~\mathbb{Z}_{61} = {1\over 4\sigma} ~{\bf N}_1 e^{-2\phi} {\rm cosh}~\gamma \nonumber\\
&&{\rm Re}~\mathbb{Z}_{62} = {1\over 4\sigma} ~{\bf N}_4 e^{-2\phi} {\rm cosh}~\gamma, ~~~~
{\rm Im}~\mathbb{Z}_{62} = {1\over 4\sigma} ~{\bf N}_2 e^{-2\phi} {\rm cosh}~\gamma . \nd
The above set of terms in \eqref{kamran}, \eqref{shangchi}, \eqref{jeantalon}, \eqref{itonya}, \eqref{bluebox}, 
\eqref{planethulk}, \eqref{atyler}, \eqref{frogthor}, \eqref{julcash}, and \eqref{pmaalish} tabulate {\it all} the supersymmetry equations that may be allowed for our set-up. Once we equate them to zero, they lead to 22 equations that need to be satisfied for supersymmetric consistency of our background. This is a delicate system, so we will have to tread very carefully to find consistent solutions to the system. 

Let us start with the simplest set of equations, namely the vanishing of \eqref{julcash} and \eqref{pmaalish}. Since 
$\gamma$ and $e^{-\phi}$ do not vanish, the only way to satisfy these equations would be to make all ${\bf N}_i$ 
vanish, i.e:
\bg\label{agatha}
{\bf N}_1 = {\bf N}_2 = {\bf N}_3 = {\bf N}_4 = 0, \nd
where ${\bf N}_i$ are defined in \eqref{pikelia}. From the definition it seems \eqref{agatha} can be satisfied if we make  either or both ($\alpha_i, \beta_i$)  and (${\bf H}^{(0)}_{4n}, {\bf H}'^{(0)}_{4n}$) zero, where ${\bf H}^{(0)}_{4n}$ and 
${\bf H}'^{(0)}_{4n}$ are defined in \eqref{hangman} and \eqref{alpac} respectively. We do not expect ($\alpha_i, \beta_i$) to vanish as this would make the background trivial, so the only way to satisfy \eqref{agatha} will be to make 
${\bf H}^{(0)}_{4n}$ and ${\bf H}'^{(0)}_{4n}$ zero simultaneously. Since ${\bf H}^{(0)}_{4n}$ and ${\bf H}'^{(0)}_{4n}$ are in-turn expressed using ${\bf K}_n$ functions, as in \eqref{elisfe}, it amounts to making all the ${\bf K}_n$ functions to zero.  The ${\bf K}_n$ functions vanish when:
\bg\label{poirot}
{\bf K}_\psi = {\cal F}_2 = -{\cal F}_1, ~~~~~ {\bf K} = {\cal F}_{1\psi} = -{\cal F}_{2\psi}, \nd
where ${\bf K}, {\cal F}_1$ and ${\cal F}_2$ are defined in \eqref{computer}. The question then is whether 
\eqref{poirot} can be satisfied without trivializing the background. The answer turns out to be in the affirmative by the following remarkable ansatze\footnote{This ansatze was also anticipated by Veronica Errasti Diez by demanding a hidden $G_2$ structure in the system. More details on this will be presented elsewhere.}:
\bg\label{nileriver}
{\bf A} = - {\bf B}, \nd 
where the form for ${\bf A}$ and ${\bf B}$ appear in \eqref{pendant}. Note that this is {\it not} the ${\bf A} = {\bf B}$ background that we discussed in  section \ref{a=b}, so the careful reader might be a bit alarmed at this stage. However we shall reconcile the conundrum a little later. First, let us see how the functions in \eqref{computer}, and their derivatives,  
behave under 
the constraint \eqref{nileriver}. The result is:
\bg\label{sinn}
&&{\bf K} =  - {\bf A} \sqrt{{\cal G}_3{\cal G}_4}~\sin~\psi, 
~~ {\cal F}_1 = {\bf A} \sqrt{{\cal G}_3{\cal G}_4}~\cos~\psi, ~~ {\cal F}_2 = - {\bf A} \sqrt{{\cal G}_3{\cal G}_4}~\cos~\psi
\\
&& {\bf K}_\psi =  - {\bf A} \sqrt{{\cal G}_3{\cal G}_4}~\cos~\psi, 
~~ {\cal F}_{1\psi} = - {\bf A} \sqrt{{\cal G}_3{\cal G}_4}~\sin~\psi, ~~ {\cal F}_{2\psi} =  {\bf A} \sqrt{{\cal G}_3{\cal G}_4}~\sin~\psi
\nonumber\\
&& {\bf K}_r =  - \left({\bf A} \sqrt{{\cal G}_3{\cal G}_4}\right)_r \sin~\psi, 
~~ {\cal F}_{1r} = \left({\bf A} \sqrt{{\cal G}_3{\cal G}_4}\right)_r \cos~\psi, 
~~ {\cal F}_{2r} = - \left({\bf A} \sqrt{{\cal G}_3{\cal G}_4}\right)_r \cos~\psi, \nonumber \nd
where the subscript $r$ and $\psi$ denote derivatives with respect to $r$  and $\psi$ respectively. Plugging this in the definitions of ${\bf K}_n$ in \eqref{elisfe} immediately tells us that all ${\bf H}^{(0)}_{4n}$ and ${\bf H}'^{(0)}_{4n}$ 
in \eqref{hangman} and \eqref{alpac} respectively vanish. This further implies the vanishing of all ${\bf N}_i$ in 
\eqref{pikelia} as discussed above.

To study the other supersymmetry equations we will have to work out all the $H$-functions using the ansatze 
\eqref{nileriver}. We start with the ${\bf H}_{3n}^{(0)}$ functions given in \eqref{thor}. They can be expressed as:
\bg\label{thor2}
&&{\bf H}_{3c}^{(0)} \equiv - { \left({\bf A} \sqrt{{\cal G}_3{\cal G}_4}\right)_r  \sin^2 \psi \over 
\sqrt{{\cal G}_1 {\cal G}_3 {\cal G}_4}\left(\alpha_1 \alpha_3 - \beta_1 \beta_3\right) \left(\alpha_2 \alpha_4 - \beta_2 
\beta_4\right)} \nonumber\\
&& {\bf H}_{3a}^{(0)} \equiv  { \left({\bf A} \sqrt{{\cal G}_3{\cal G}_4}\right)_r \sin~2\psi \over 
2\left(\alpha_1 \alpha_3 - \beta_1 \beta_3\right) {\cal G}_3 \sqrt{{\cal G}_1}}, ~~~
{\bf H}_{3b}^{(0)} \equiv  - { \left({\bf A} \sqrt{{\cal G}_3{\cal G}_4}\right)_r \sin~2\psi \over 
2 \left(\alpha_2 \alpha_4 - \beta_2 \beta_4\right) {\cal G}_4 \sqrt{{\cal G}_1}}, \nd
by substituting the value of ${\bf K}_r$ from \eqref{sinn} in \eqref{thor}. In the same vein  the ${\bf H}^{(1)}_{3n}$ 
functions from \eqref{sorenson} may be expressed in the following way:
\bg\label{sorenson2}
&&{\bf H}_{3c}^{(1)} \equiv  - {\left({\bf A} \sqrt{{\cal G}_3{\cal G}_4}\right)_r   \cos^2 {\psi}
\over 
\sqrt{{\cal G}_1 {\cal G}_3 {\cal G}_4}\left(\alpha_1 \alpha_3 - \beta_1 \beta_3\right) \left(\alpha_2 \alpha_4 - \beta_2 
\beta_4\right)} \nonumber\\
&& {\bf H}_{3a}^{(1)} \equiv - {\left({\bf A} \sqrt{{\cal G}_3{\cal G}_4}\right)_r  \sin~2\psi \over 2\left(\alpha_1 \alpha_3 - 
\beta_1 \beta_3\right) {\cal G}_3 \sqrt{{\cal G}_1}}, ~~~~
{\bf H}_{3b}^{(1)} \equiv {\left({\bf A} \sqrt{{\cal G}_3{\cal G}_4}\right)_r \sin~2\psi \over 2\left(\alpha_2 \alpha_4 - 
\beta_2 \beta_4\right) {\cal G}_4 \sqrt{{\cal G}_1}}, \nd 
where we again substituted ${\bf K}_r$ from \eqref{sinn} in \eqref{sorenson}. Combining ${\bf H}^{(0)}_{3a}$ and 
${\bf H}^{(1)}_{3a}$ as well as ${\bf H}^{(0)}_{3b}$ and ${\bf H}^{(1)}_{3b}$ as in \eqref{noworkshop}, one can easily 
infer:
\bg\label{hercule}
{\bf Q}_1 = {\bf Q}_ 2 = 0.  \nd
The above equation fixes the values for ${\bf Q}_1$ and ${\bf Q}_2$, but doesn't fix ${\bf Q}_3$ and ${\bf Q}_4$, because the latter are defined in terms of ${\bf H}'^{(0)}_{3n}$ and ${\bf H}'^{(1)}_{3n}$ as may be seen from 
\eqref{noworkshop}. To proceed, let us define the ${\bf H}'^{(0)}_{3n}$ functions using the constraint \eqref{nileriver} as:
\bg\label{thor3}
&&{\bf H}_{3c}^{'(0)} \equiv -{{\bf A}~\sin~2\psi \over 
2\sqrt{{\cal G}_2}\left(\alpha_1 \alpha_3 - \beta_1 \beta_3\right) \left(\alpha_2 \alpha_4 - \beta_2 
\beta_4\right)} \nonumber\\
&& {\bf H}_{3a}^{'(0)} \equiv  {{\bf A}\sqrt{{\cal G}_4}~\cos^2 \psi \over 
\left(\alpha_1 \alpha_3 - \beta_1 \beta_3\right) \sqrt{{\cal G}_2 {\cal G}_3}}, ~~~
{\bf H}_{3b}^{'(0)} \equiv - {{\bf A} \sqrt{{\cal G}_3}~\cos^2 \psi \over 
\left(\alpha_2 \alpha_4 - \beta_2 \beta_4\right) \sqrt{{\cal G}_2 {\cal G}_4}}, \nd
where we used the ${\bf K}_\psi$ value from \eqref{sinn} in \eqref{docsavage}.  Compared to \eqref{thor2} and 
\eqref{sorenson2}, all the functions in \eqref{thor3} are proportional to ${\bf A}$. A similar story emerges for the 
${\bf H}^{'(1)}_{3n}$ functions, which may now be written as:
\bg\label{thor4}
&&{\bf H}_{3c}^{'(1)} \equiv {{\bf A}~\sin~2\psi \over 
2\sqrt{{\cal G}_2}\left(\alpha_1 \alpha_3 - \beta_1 \beta_3\right) \left(\alpha_2 \alpha_4 - \beta_2 
\beta_4\right)} \nonumber\\
&& {\bf H}_{3a}^{'(1)} \equiv  {{\bf A}\sqrt{{\cal G}_4}~\sin^2 \psi \over 
\left(\alpha_1 \alpha_3 - \beta_1 \beta_3\right) \sqrt{{\cal G}_2 {\cal G}_3}}, ~~~
{\bf H}_{3b}^{'(1)} \equiv - {{\bf A} \sqrt{{\cal G}_3}~\sin^2 \psi \over 
\left(\alpha_2 \alpha_4 - \beta_2 \beta_4\right) \sqrt{{\cal G}_2 {\cal G}_4}}, \nd
where the only difference in the last two functions compared to the last two functions in \eqref{thor3}, is the appearance of $\sin^2\psi$ instead of $\cos^2\psi$. This means we can add up ${\bf H}^{'(0)}_{3a}$ and 
${\bf H}^{'(1)}_{3a}$ as well as ${\bf H}^{'(0)}_{3b}$ and ${\bf H}^{'(1)}_{3b}$ to get the following values for 
${\bf Q}_3$  and ${\bf Q}_4$ respectively using \eqref{noworkshop}:
\bg\label{paques}
{\bf Q}_3 =   {{\bf A}\sqrt{{\cal G}_4} \over \left(\alpha_1 \alpha_3 - \beta_1 \beta_3\right) \sqrt{{\cal G}_2 {\cal G}_3}}, ~~~~ {\bf Q}_4 =   -{{\bf A}\sqrt{{\cal G}_3} \over \left(\alpha_2 \alpha_4 - \beta_2 \beta_4\right) \sqrt{{\cal G}_2 {\cal G}_4}}, \nd
which are non-zero but independent of $\psi$. The non-vanishing of ${\bf Q}_3$ and ${\bf Q}_4$ will have important consequence as we shall soon see. On the other hand, the functions ${\bf P}_1$ and ${\bf P}_2$ appear along with 
${\bf Q}_1$ and ${\bf Q}_2$ in \eqref{kamran}, \eqref{shangchi}, \eqref{jeantalon}, \eqref{itonya} and \eqref{bluebox}
for the real parts of the $\mathbb{Z}_{mn}$ coefficients of the (3, 0), (0, 3) as well as some of the (1, 2) forms. Looking at the definitions of the ${\bf P}_i$ functions in \eqref{altuspress}, we see that they are defined with respect to the deviation factors $b_o$ and $c_o$ from \eqref{thepooh}. For the choice of the ansatze \eqref{nileriver}, it is easy to infer that:
\bg\label{marple}
b_o = c_o = -1, \nd
by plugging in the functional forms for ${\cal F}_{nr}$ and ${\cal F}_{n\psi}$ from \eqref{sinn}.  This immediately tells us that all the ${\bf P}_i$ functions in \eqref{altuspress} are now defined in terms of the combination ${\bf H}^{'(0)}_{3c}$  and ${\bf H}^{'(1)}_{3c}$. This combination vanishes, as may be seem from \eqref{thor3} and \eqref{thor4}. Therefore:
\bg\label{mirrorcracked}
{\bf P}_1  = {\bf P}_2 = {\bf P}_3  = {\bf P}_4 = 0, \nd 
which immediately implies that all the real parts of the $\mathbb{Z}_{mn}$ coefficients of the (3, 0), (0, 3) as well as the 
(1, 2) forms vanish. In other words:
\bg\label{cyanide}
{\rm Re} ~\mathbb{Z}_1 = {\rm Re} ~\mathbb{Z}_2 = {\rm Re} ~\mathbb{Z}_{3n} = {\rm Re} ~\mathbb{Z}_{5p} =
{\rm Re} ~\mathbb{Z}_{6p} = 0, \nd 
where $n = 1, 2, 3$ and $p = 1, 2$. The vanishing of ${\rm Im}~\mathbb{Z}_{m2}$ and ${\rm Im}~\mathbb{Z}_{61}$, 
where $m = 4, 5, 6,$ have already been shown from the vanishing of ${\bf N}_i$ functions in \eqref{agatha}. 

Let us now discuss the non-zero pieces. One contribution would appear from the non-vanishing ${\bf Q}_3$ 
and ${\bf Q}_4$ 
functions \eqref{paques}. The remaining contributions appear from the ${\bf M}_i$ functions in \eqref{blake}. 
The fact that they are non-zero may be easily seen by plugging in the ($b_o$,  
$c_o$) values from \eqref{marple}  as well as the ${\bf H}^{(0)}_{3c}$ and ${\bf H}^{(1)}_{3c}$ values  
from \eqref{thor2} and \eqref{sorenson2} respectively  in  \eqref{blake}.  The result is:
\bg\label{blake2}
&& {\bf M}_1 \equiv {\bf H}^{(0)}_1 \alpha_3 \alpha_4  - {\bf H}^{(0)}_2 \beta_1 \beta_2 
- {\left(\alpha_3 \beta_2 + \alpha_4 \beta_1\right) \left({\bf A} \sqrt{{\cal G}_3{\cal G}_4}\right)_r  \over 
\sqrt{{\cal G}_1 {\cal G}_3 {\cal G}_4}\left(\alpha_1 \alpha_3 - \beta_1 \beta_3\right) \left(\alpha_2 \alpha_4 - \beta_2 
\beta_4\right)}
\nonumber\\  
&& {\bf M}_2 \equiv {\bf H}^{(0)}_1 \alpha_3 \beta_4  - {\bf H}^{(0)}_2 \beta_1 \alpha_2 
- { \left(\alpha_2 \alpha_3 + \beta_1 \beta_4\right) \left({\bf A} \sqrt{{\cal G}_3{\cal G}_4}\right)_r  \over 
\sqrt{{\cal G}_1 {\cal G}_3 {\cal G}_4}\left(\alpha_1 \alpha_3 - \beta_1 \beta_3\right) \left(\alpha_2 \alpha_4 - \beta_2 
\beta_4\right)}
 \nonumber\\  
&& {\bf M}_3 \equiv {\bf H}^{(0)}_1 \alpha_4 \beta_3  - {\bf H}^{(0)}_2 \beta_2 \alpha_1 
- {\left(\alpha_1 \alpha_4 + \beta_2 \beta_3\right) \left({\bf A} \sqrt{{\cal G}_3{\cal G}_4}\right)_r  \over 
\sqrt{{\cal G}_1 {\cal G}_3 {\cal G}_4}\left(\alpha_1 \alpha_3 - \beta_1 \beta_3\right) \left(\alpha_2 \alpha_4 - \beta_2 
\beta_4\right)}
\nonumber\\  
&& {\bf M}_4 \equiv - {\bf H}^{(0)}_1 \beta_3 \beta_4  + {\bf H}^{(0)}_2 \alpha_1 \alpha_2 
+ { \left(\alpha_1 \beta_4 + \alpha_2 \beta_3\right) \left({\bf A} \sqrt{{\cal G}_3{\cal G}_4}\right)_r  \over 
\sqrt{{\cal G}_1 {\cal G}_3 {\cal G}_4}\left(\alpha_1 \alpha_3 - \beta_1 \beta_3\right) \left(\alpha_2 \alpha_4 - \beta_2 
\beta_4\right)}, \nonumber\\
\nd  
where ${\bf H}^{(0)}_1$ and ${\bf H}^{(0)}_2$ are given in \eqref{kingi}. Let us first consider the equation involving 
${\bf M}_1$ and ${\bf M}_4$. They appear in \eqref{planethulk} and \eqref{frogthor} from the vanishing of 
${\rm Im}~\mathbb{Z}_{41}$ and ${\rm Im}~\mathbb{Z}_{51}$ respectively as:
\bg\label{dumpling}
&&{{\bf M}_4 e^{-\phi} {\rm sinh}~\gamma \over \sigma} - {\bf M}_1 e^{-2\phi} {\rm cosh}~\gamma = 0 \nonumber\\
&& {{\bf M}_1 e^{-\phi} {\rm sinh}~\gamma \over \sigma} - {\bf M}_4 e^{-2\phi} {\rm cosh}~\gamma = 0. \nd
One solution for the above system of equations is clearly the vanishing of both ${\bf M}_1$ and ${\bf M}_4$. This will 
keep $\sigma$ undetermined. On the other hand there is also a non-trivial solution that fixes the form for $\sigma$ 
in the following way:
\bg\label{baghdadi}
{\bf M}_1 = \pm {\bf M}_4, ~~~~ \sigma = \pm e^\phi ~{\rm tanh}~\gamma. \nd     
The correct solution however may be inferred by going to the non-K\"ahler resolved conifold limit where the vielbein 
coefficients ($\alpha_i, \beta_i$) satisfy \eqref{edenlac} \cite{DEM}. Plugging in \eqref{edenlac} in \eqref{blake2} we see that the solution is the one with the minus signs in both the equations of \eqref{baghdadi} as this leads not only 
to \eqref{virgop} but also to the correct form for $\sigma$, the complex structure\footnote{Recall that $\sigma$ used here is related to the $\sigma$ used in \cite{DEM} by $-{1\over \sigma}$.}.

With this choice of the complex structure $\sigma$, we can now see that the ${\rm Im}~\mathbb{Z}_1$, 
${\rm Im}~\mathbb{Z}_2$ and ${\rm Im}~\mathbb{Z}_{31}$ in \eqref{kamran}, \eqref{shangchi} and 
\eqref{jeantalon} respectively may be easily satisfied by imposing the following conditions on ${\bf Q}_3$ and ${\bf Q}_4$ as well as on ${\bf M}_2$ and ${\bf M}_3$:
\bg\label{2la1aa2}
{\bf Q}_3  = {\bf Q}_4, ~~~~~ {\bf M}_2 =  - {\bf M}_3, \nd
which are of course trivially satisfied for the resolved conifold case in \eqref{edenlac}.  The problem however appears from the imaginary pieces in the last two equations, namely \eqref{itonya} and \eqref{bluebox}, as:
\bg\label{tax2018}
{\rm Im}~\mathbb{Z}_{33}  = - {\rm Im}~\mathbb{Z}_{32} = {e^{-\phi} {\bf Q}_3 \over 4}\left({{\rm sinh}^2\gamma - 
e^{-2\phi} {\rm cosh}^2\gamma \over {\rm sinh}~\gamma}\right), \nd
{\it do not} vanish automatically. One way it can vanish is by fixing the dilaton using $\gamma$. However this will make the warp-factor $h$ in \eqref{snowman} vanish. This is clearly not acceptable! The other way it can vanish is by making ${\bf Q}_3$ vanish, implying the vanishing of ${\bf Q}_4$ also.      
Since ${\bf Q}_3$  and ${\bf Q}_4$ are both proportional to ${\bf A}$ in \eqref{paques},  this immediately tells us that the supersymmetric constraints appearing from all the 22 equations are:
\bg\label{easter}
{\bf A} = - {\bf B} = 0, ~~~~ {\bf M}_1 = - {\bf M}_4, ~~~~ {\bf M}_2 = - {\bf M}_3. \nd     
Note that the above solution is also consistent with the ${\bf A} = {\bf B}$ constraint that we discussed in section 
\ref{a=b}, thus resolving the conundrum that appeared earlier. At this stage, 
the requirement that $\bold{A} = \bold{B}$ is worth interpreting. We note that our definition of the vielbeins 
\eqref{IT} for the deformed conifold can be rewritten in the following form:
\bg\label{khilji}
h^{-1/4}\begin{bmatrix}
\frac{1}{\sqrt{\mathcal{G}_3}}e_3 \\ \frac{1}{\sqrt{\mathcal{G}_4}}e_4 \\ \frac{1}{\sqrt{\mathcal{G}_3}}e_5 \\ \frac{1}{\sqrt{\mathcal{G}_4}}e_6
\end{bmatrix}
=
\begin{bmatrix}
\alpha_1 & 0 &\beta_3 & 0 \\
0 & -\beta_4 & 0 &\alpha_2 \\
\beta_1 & 0 & \alpha_3 & 0 \\
0 & \alpha_4 & 0 & -\beta_2
\end{bmatrix}
\begin{bmatrix}
\sigma_1 \\ \Sigma_2 \\ \Sigma_1 \\ \sigma_2
\end{bmatrix},
\nd
where $\alpha_i$ and $\beta_i$ are the coefficients of the Maurer-Cartan forms that appear in the definition of the vielbeins in \eqref{IT}, satisfying the consistency relations \eqref{lathi}. We can now 
separate the matrix  on the right hand side of \eqref{khilji} into blocks in the following way:
\bg\label{lriel}
\begin{bmatrix}
\alpha_1 & 0 &\beta_3 & 0 \\
0 & -\beta_4 & 0 &\alpha_2 \\
\beta_1 & 0 & \alpha_3 & 0 \\
0 & \alpha_4 & 0 & -\beta_2
\end{bmatrix} =
\begin{bmatrix}
\mathbb{A}_1 & \mathbb{B}_1 \\
\mathbb{A}_2 & \mathbb{B}_2
\end{bmatrix}.
\nd
Looking at the blocks, we can easily infer that the functional forms for ${\bf A}$ and ${\bf B}$, as defined in \eqref{pendant}, can be rewritten in terms $\mathbb{A}_i$ and $\mathbb{B}_i$ as: 
\bg\label{luchale}
&& \bold{A}=\det \mathbb{A}_1 + \det \mathbb{A}_2 \nonumber \\
&& \bold{B} = \det \mathbb{B}_1 + \det \mathbb{B}_2.
\nd
Each of these determinants measure the contribution of $\sigma_1 \wedge \Sigma_2$ or $\Sigma_1 \wedge \sigma_2$ to the complex structure components $e_3\wedge e_4$ or $e_5 \wedge e_6$ (up to a factor of $\sqrt{h \mathcal{G}_3 \mathcal{G}_4}$). The assumption that $\bold{A} = \bold{B}$ can then be interpreted as a sort of ``equal contribution" condition, where each of those 2-forms must contribute equally to the sum of complex structure components. Note that when we try to relax this assumption, some of the supersymmetry conditions demand that $\bold{A} = - \bold{B}$ instead. This means the ``equal contribution" condition must still hold, but now involves an orientation change, since at least one of the determinants must change sign. This change of orientation then changes the complex structure relative to the $\bold{A} = \bold{B}$ case. This changes what one means by the $(2,1)$ fluxes and it appears that the $(2,1)$ fluxes relative to this new complex structure simply can't exist on a deformed conifold background and therefore consistency imposes $\bold{A}= \pm \bold{B}=0$.


After the dust settles, we are ready to collect all our results to express the final form of the three-form flux.  The 
${\bf G}_3$ flux that solves the EOM and leads to supersymmetric configuration may now be written as:
\bg\label{tagramaiaa}
{{\bf G}_3 \over {\rm cosh}~\gamma} & = & {ie^{-2\phi} {\bf M}_1 \over 2} ~E_1 \wedge \left(E_3 \wedge {\overline E}_3
- E_2 \wedge {\overline E}_2\right) \nonumber\\
& + & {ie^{-2\phi} {\bf M}_2 \over 2} ~E_1 \wedge \left( \overline{E}_2 \wedge {E}_3
- E_2 \wedge \overline{E}_3 \right), \nd
which is expectedly a (2, 1) form and ISD. The ${\bf M}_i$ functions are defined as in \eqref{blake2} but with 
${\bf A} = 0$. If ${\bf A}$ was non-zero, we could have got  more non-trivial coefficients  for ${\bf G}_3$, although the form would have been similar to \eqref{tagramaiaa}. Thus
one might ask whether one may choose a {\it different} complex structure to get a more non-trivial result. This is the issue that we turn to next.

\subsubsection{Analysis with a different choice of complex structure \label{rebchut}}

The complex structure choice that we entertained earlier in \eqref{machine} was the simplest available, so the natural questions is to extend this to a more generic case of the form:
\bg\label{rossi}
\left(\sigma_1  + i \sigma_2, \sigma_3 + i \sigma_4, \sigma_5 + i \sigma_6\right), \nd
where we will take the $\sigma_i$ to be functions of $r$ only. This could be further generalized by making them functions of all the internal coordinates, but we will leave this as an exercise for the reader. Along the way, let us also define three other combinations:
\bg\label{bellaR}
\sigma \equiv \sigma_4 \sigma_6, ~~~~~ {1\over \alpha_{11}} \equiv 1 - {\sigma_3 \sigma_5\over \sigma},
~~~~~ {1\over \alpha_{22}} \equiv 1 + {\sigma_3 \sigma_5\over \sigma} \nd
that will be useful soon. Our goal here is to see whether one may generalize the constraints on ${\bf M}_i, {\bf Q}_3$ and ${\bf Q}_4$ from \eqref{easter} and \eqref{paques} respectively, still using the constraint \eqref{nileriver} on 
${\bf A}$ and ${\bf B}$. In this limit, as before, ${\bf N}_i, {\bf P}_i, {\bf Q}_1$ and ${\bf Q}_2$ continue to vanish. However the supersymmetry equations for the non-vanishing pieces do become more complicated. For example the 
non-zero imaginary pieces in \eqref{planethulk} and \eqref{frogthor} change to the following:
\bg\label{tagraburia}
&& {1\over \sigma_4}\left(1 - {i\sigma_1 \over \sigma_2}\right) {\bf M}_4 e^{-2\phi} ~{\rm cosh}~\gamma - 
{{\bf M}_1 e^{-\phi}~{\rm sinh}~\gamma \over \sigma_2 \sigma_4} = 0 \nonumber\\
&& {1\over \sigma_6}\left(1 - {i\sigma_1 \over \sigma_2}\right) {\bf M}_1 e^{-2\phi} ~{\rm cosh}~\gamma - 
{{\bf M}_4 e^{-\phi}~{\rm sinh}~\gamma \over \sigma_2 \sigma_6} = 0, \nd
which now have both real and imaginary pieces contributing to $\mathbb{Z}_{41}$  and $\mathbb{Z}_{51}$ in
\eqref{planethulk} and \eqref{frogthor} respectively. Solving this set of equations simultaneously we get:
\bg\label{marmichel}
{\bf M}_1 = -{\bf M}_4, ~~~~~ \sigma_1 = 0, ~~~~~ \sigma_2 = - e^\phi ~ {\rm tanh}~\gamma, \nd
 which is exactly what we had in \eqref{baghdadi} by choosing the minus sign. The fact that $\sigma_1$ vanishes implies that the complex structure cannot be made arbitrarily complicated, at least for the relevant part, if we also want to preserve supersymmetry simultaneously. 

What about the other parts of the complex structure in \eqref{rossi}? For this we will have to look at the other supersymmetry equations carefully. These equations replace the set of equations given earlier by the imaginary pieces in \eqref{kamran}, \eqref{shangchi}, \eqref{jeantalon}, \eqref{itonya} and \eqref{bluebox}. The first set of equations may be expressed as:
\vskip-.2in
{\footnotesize
\bg\label{nightgallery}
\pm e^{-2\phi}~{\rm cosh}~\gamma \left({\bf M}_2 - {\bf M}_3\right) \left({1\over \sigma_4} - {1\over \sigma_6}\right) +
{e^{-2\phi}~{\rm cosh}~\gamma\over \sigma_2}\left({{\bf Q}_3 \over \sigma} - {{\bf Q}_4 \over \alpha_{11}}\right)
+ {e^{-\phi}~{\rm sinh}~\gamma}\left({{\bf Q}_3 \over \alpha_{11}} - {{\bf Q}_4 \over \sigma}\right) = 0, \nonumber\\ \nd}
where $\alpha_{11}$ and $\sigma$ are defined in \eqref{bellaR}. The above are two equations 
that appear to have
multiple solutions. We shall discuss the explicit solutions of \eqref{nightgallery} a bit later because the solutions of \eqref{nightgallery} are influenced by the next set of equations that may be presented as:
\vskip-.2in
{\footnotesize
\bg\label{nightgallery2}
\pm e^{-2\phi}~{\rm cosh}~\gamma \left({\bf M}_2 + {\bf M}_3\right) \left({1\over \sigma_4} - {1\over \sigma_6}\right) +
{e^{-2\phi}~{\rm cosh}~\gamma\over \sigma_2}\left({{\bf Q}_3 \over \sigma} + {{\bf Q}_4 \over \alpha_{22}}\right)
+ e^{-\phi}~{\rm sinh}~\gamma\left({{\bf Q}_3 \over \alpha_{22}} + {{\bf Q}_4 \over \sigma}\right) = 0, \nonumber\\ \nd}
where $\alpha_{22}$ is defined in \eqref{bellaR}. These equations differ from \eqref{nightgallery} by relative signs between ${\bf M}_2$ and ${\bf M}_3$ as well as between 
${\bf Q}_3$ and ${\bf Q}_4$. It is easy to see that the following constraints:
\bg\label{olivieriR}
\sigma_4 =  \sigma_6, ~~~~~ {\bf Q}_3 = {\bf Q}_4 = 0, \nd
solve \eqref{nightgallery}  and \eqref{nightgallery2} simultaneously without trivializing ${\bf M}_2$ and 
${\bf M}_3$.  The condition on ${\bf M}_2$ and ${\bf M}_3$ however gets fixed from the following equation:
\vskip-.2in
{\footnotesize
\bg\label{tagraelis}
e^{-2\phi}~{\rm cosh}~\gamma \left({\bf M}_2 + {\bf M}_3\right) \left({1\over \sigma_4} + {1\over \sigma_6}\right) -
{e^{-2\phi}~{\rm cosh}~\gamma\over \sigma_2}\left({{\bf Q}_3 \over \sigma} - {{\bf Q}_4 \over \alpha_{11}}\right)
+ e^{-\phi}~{\rm sinh}~\gamma\left({{\bf Q}_3 \over \alpha_{11}} - {{\bf Q}_4 \over \sigma}\right) = 0, \nonumber\\ \nd}
which differs from both \eqref{nightgallery} as well as \eqref{nightgallery2} by  sign rearrangements of the various 
terms. Imposing \eqref{olivieriR}, we see that the only way we could solve this equation is by putting the following 
constraint on ${\bf M}_2$  and ${\bf M}_3$:
\bg\label{eliserchut} 
{\bf M}_2 = - {\bf M}_3, \nd
which when combined with \eqref{marmichel} is basically the constraint \eqref{easter} that we had earlier. Thus 
it seems taking \eqref{rossi} doesn't change the constraints in any interesting way. 

However the story is not complete as we now have additional supersymmetry equations appearing from the real parts of the complex structure \eqref{rossi}. Could they change the constraints in any meaningful way?

The new set of equations follow the same pattern as \eqref{nightgallery}, \eqref{nightgallery2} and \eqref{tagraelis}
in the sense that the first two equations take the following form:
\vskip-.2in
{\footnotesize
\bg\label{carter}
\pm {e^{-2\phi}~{\rm cosh}~\gamma\over \sigma}  \left({\bf M}_2 - {\bf M}_3\right) \left({\sigma_3} - { \sigma_5}\right) +
\left({\sigma_3\over \sigma_4} + {\sigma_5\over \sigma_6}\right)\left({{\bf Q}_4 e^{-2\phi}~{\rm cosh}~\gamma\over \sigma_2} - {\bf Q}_3 e^{-\phi} ~{\rm sinh}~\gamma\right) = 0,  \nonumber\\ \nd}
where $\sigma$ is defined in \eqref{bellaR}. 
Since ${\bf Q}_3$  and ${\bf Q}_4$ vanish, and ${\bf M}_2$ and ${\bf M}_3$ satisfy \eqref{eliserchut}, the above equations are clearly satisfied with: 
\bg\label{canarvon}
\sigma_3 = \sigma_5, \nd
or with vanishing values for $\sigma_3$ and $\sigma_5$. Either of these conclusions remain unchanged even if we look at the next three equations:
\vskip-.2in
{\footnotesize
\bg\label{kingtut}
&&~~{e^{-2\phi}~{\rm cosh}~\gamma\over \sigma}  \left({\bf M}_2 + {\bf M}_3\right) \left({\sigma_3} + { \sigma_5}\right) +
\left({\sigma_3\over \sigma_4} + {\sigma_5\over \sigma_6}\right)\left({{\bf Q}_4 e^{-2\phi}~{\rm cosh}~\gamma\over \sigma_2} + {\bf Q}_3 e^{-\phi} ~{\rm sinh}~\gamma\right) = 0 \nonumber\\
&&\pm {e^{-2\phi}~{\rm cosh}~\gamma\over \sigma}  \left({\bf M}_2 + {\bf M}_3\right) \left({\sigma_3} + { \sigma_5}\right) +
\left({\sigma_3\over \sigma_4} - {\sigma_5\over \sigma_6}\right)\left({{\bf Q}_4 e^{-2\phi}~{\rm cosh}~\gamma\over \sigma_2} + {\bf Q}_3 e^{-\phi} ~{\rm sinh}~\gamma\right) = 0,  \nonumber\\ \nd}
which are automatically consistent in the light of the constraint \eqref{eliserchut} even if \eqref{canarvon} is not imposed. On the other hand the (2, 1) ISD ${\bf G}_3$ flux remains unchanged from what we had earlier in 
\eqref{tagramaiaa}. This in itself is not surprising as \eqref{tagramaiaa} is expressed in terms of differences of three-forms. These differences {\it cancel} out any changes of the complex structure from \eqref{machine} to 
\eqref{rossi} satisfying \eqref{marmichel},  \eqref{olivieriR} and \eqref{canarvon}, thus revealing no new constraints. On the other hand the Bianchi identity should further constrain the warp-factors, but imposing the Bianchi identity is a bit subtle here if we also want to demand UV completion as new three-brane and five-brane sources may become relevant in the dual side. Additionally the fact that certain warp-factors, for example ${\cal G}_3$ and ${\cal G}_4$
in \eqref{crooked}, can be made unequal will be important for UV completion. This fact should also be derivable 
directly from the field theory construction in section \ref{rescon}, for example using {\it probe} branes.    
This and other details would be the topic of our next discussions in the following two sections, starting first with the 
probe brane dynamics.  


\section{Dynamics of probe brane and Higgsing \label{navneet}}

The gauge/gravity correspondence owes its origins to considering branes either as localized dynamic objects in a background spacetime that  carry a worldvolume gauge theory or as gravitational solutions in their own right and matching various properties of the two pictures. In this approach one constructs the two sides of the duality separately and tests the duality by looking for objects on either side that are dual to each other (e.g. objects with same quantum numbers). An interesting, though significantly more difficult task would be to try and construct the gravity dual directly starting from the field theory. 

The gravity dual can be thought of as the geometry a probe brane sees as it approaches the brane configuration that carries the gauge theory. This corresponds to augmenting the gauge theory by the degrees of freedom of the probe brane and moving in the moduli space of this larger theory. Famously, the fact that the IR behavior of the gauge theory on a stack of fractional branes at a conifold singularity is dual to a deformed conifold topology can be arrived at from precisely these kinds of considerations. However the geometry on this deformed conifold is then found entirely on the supergravity side. In this section we attempt to use the probe brane approach to determine properties of the dual geometry directly.

The general approach is as follows. We imagine probing the brane stack that carries our gauge theory of interest with an additional brane. Alternatively, one can give an expectation value to appropriate components of the scalar fields in the theory, which corresponds to splitting off a single brane from the rest of the stack and using it as the probe. Since this probe is at finite separation from the stack, the expectation value of the scalars corresponding to its relative position lead to Higgsing of the gauge group and gives mass to some subset of the fields. Integrating out these fields generates higher point vertices for the component of the gauge field strength corresponding to the probe brane worldvolume field strength as well as for the chiral multiplets corresponding to transverse oscillations of the probe brane. On the other hand, in the gravity dual, the effective action of our probe brane is the DBI action, where these higher point vertices appear at tree level and depend on the metric of the gravity dual. Matching the one-loop amplitudes to the tree level DBI couplings allows us to extract the warp factor of the dual geometry as well as information about the internal metric.

We will consider the theories arising on stacks of branes at the singular point of a conifold background. This can include fractional branes arising from D5's wrapping the vanishing but non-trivial 2-cycle, if it exists. Indeed this will be the most interesting case, otherwise we are restricted to a stack of D3 branes to have a four dimensional supersymmetric theory at all energy scales. Naturally in the latter case, we expect a conformal field theory and the gravity dual will involve AdS, while in the former case we should expect a non-conformal theory with bifundamental scalars and the gravity dual should be one of the solutions described above. We are interested in seeing which features of these dual geometries this perturbative one-loop calculation can reveal.

\subsection{DBI action as effective action after Higgsing \label{DBI}}

Consider a gauge/gravity dual pair arising from a particular brane configuration. A single probe Dp-brane in the gravity dual is governed by the sum of a DBI action and a Chern-Simons type action:
\bg\label{max1}
&& S=\int d^{p+1} x e^{-\phi}\sqrt{{\rm det}(g+\mathcal{F})} + \int d^{p+1} x~ e^\mathcal{F} \wedge C \\
&& \mathcal{F} \equiv F+B,  \nonumber
\nd
where $F$ is the worldvolume gauge field strength, $B$ is the NS-NS 2-form and $C$ contains all the R-R potentials and the exponential is to be thought of as a ``power series" involving wedge products. All the bulk fields are understood to be pulled back to the brane worldvolume.

For $p=3$, going to static gauge and expanding the square root to fourth order in the field strengths yields:
\bg \label{expandedDBI}
S_{DBI,~D3}&=&\int d^4 x \sqrt{g} e^{-\phi}\left(1 + \frac{1}{4} F^{\mu \nu}F_{\mu \nu}+\frac{1}{4} 
\partial_\mu X^i \partial^\mu X_i \right) \\
&+& \int d^4 x \sqrt{g}e^{-\phi} \left( \frac{1}{8} F^{\mu \nu} F_{\nu \mu} F^{\rho \sigma} F_{\sigma \rho } - \frac{1}{4} F^{\mu \nu} F_{\nu \rho} F^{\rho \sigma} F_{\sigma \mu} \right) \nonumber \\
&+& \int d^4 x \sqrt{g}e^{-\phi} \left(\frac{1}{8} \partial^\mu X^i \partial_\mu X_i \partial^\nu X^j \partial_\nu X_j  + \frac{1}{4} \partial^\mu X^i \partial_\nu X_i \partial^\nu X^j \partial_\mu X_j      \right) + ...\nonumber
\nd
On the field theory side, these 4-point tree-level interactions come from integrating out fields that gain a mass after Higgsing. Specifically, consider an $\mathcal{N}=1$ theory with a vector multiplet, transforming in the adjoint of the gauge group, and a set of chiral multiplets that can generally transform in any representation\footnote{We use the same symbol for trace, but which one is meant should be clear from the context. If a chiral multiplet transforms in the adjoint, or in the bi-fundamental of the same group, then this is the ${\cal N} = 2$ sector of the theory. If all the chiral multiplets transform this way, then we see the ${\cal N} = 4$ behavior.}:
\bg\label{max2}  
S=\int d^4 x d^4 \theta ~{\rm tr}~\bar{\Phi} e^{g V} \Phi + \int d^4 x d^2 \theta 
\left[\frac{1}{C_2(A) g^2} {\rm tr}~W^\alpha W_\alpha + W(\Phi) + h.c.\right],
\nd
where $V$ is the vector potential of field strength $W_\alpha$, $g$ is the gauge coupling and $W(\phi)$ is the superpotential.
Moving in the moduli space of this theory will give a mass to some of the gauge components of the chiral and vector multiplets via the Higgs mechanism. 

In the language of the brane configuration, the mass is given by the proper length of the string connecting the probe brane and the stack, while in the gravity dual, this mass serves as a radial coordinate that will directly correpond to the energy scale of the field theory. The leading order coefficient in this 4-point function should then correspond to the tree-level coupling in the DBI action, which involves the warp factors of the gravity dual. Crucially, \emph{the radial dependence of the warp factors should match the mass-dependence of the one-loop field theory result}. Demanding that this correspondence holds, we can determine the warp factors of the gravity dual from this one-loop field theory result. Note that in principle one should match \emph{all} the higher point functions that arise from the expansion of the square root in the DBI action, but we will restrict our attention to the 4-point function here.

On purely dimensional grounds, these amplitudes must go as $m^{-4}$. In the simple case of $\mathcal{N}=4$ SYM
the couplings are constant and there are $N$ massive fields running in the loops. This directly results in the warp factor for Anti-de Sitter space, i.e.:
\bg
g_{YM}^2 \left(\frac{g_{YM}^2 N}{m^4}\right) ~  \propto ~ g_s \frac{R^4}{r^4} 
\nd
precisely as dictated by the AdS/CFT dictionary with $g_s$ being the string coupling. 

We now proceed to consider the field theories that arise as low-energy limits of brane configurations on the conifold. We start with the conformal Klebanov-Witten theory \cite{klebwit} and then proceed to the non-conformal Klebanov-Strassler theory \cite{KS} in order to make contact with the geometries found in the previous sections.

\subsection{$\mathcal{N}=1 ~SU(N)\times SU(N)$ with bi-fundamental matter \label{alchut}}

This is the Klebanov-Witten theory \cite{klebwit} arising on a stack of D3 branes at the tip of a singular confold. The conifold is parametrized by four complex variables obeying a constraint. These get promoted to the chiral superfields in the field theory, denoted by $A_{1,2}$ and $B_{1,2}$ transforming in the bi-fundamental and anti-bi-fundamental representations of the $SU(N)\times SU(N)$ gauge group respectively. The action is:
\bg \label{KSaction} 
S_{KS}&=&\int d^4 x \int d^4 \theta ~{\rm tr}\left[\frac{1}{N_1 g_1^2}\left( e^{-g_1 V_1}D^\alpha e^{g_1 V_1}\right) \bar{D}^2 \left(e^{-g_1 V_1}D_\alpha e^{g_1 V_1}\right) \right] \nonumber \\
&+& \int d^4 x \int d^4 \theta ~{\rm tr}\left[\frac{1}{N_2 g_2^2} \left( e^{-g_2 V_2}D^\alpha e^{g_2 V_2}\right) \bar{D}^2 
\left(e^{-g_2 V_2}D_\alpha e^{g_2 V_2}\right)\right] \nonumber \\
&+& \int d^4 x \int d^4 \theta ~{\rm tr}\left[ \sum_{(r, i) = (1, 1)}^{(2, 2)}\bar{A}_{r} e^{g_i V_i} A_{r} 
+ \sum_{(s, i) = (1, 1)}^{(2, 2)}\bar{B}_{s} e^{ g_i V_i} B_{s} \right]\nonumber \\
&+& \int d^4x \int d^2 \theta~ h~ {\rm tr}(A_1 B_1 A_2 B_2 - A_1 B_2 A_2 B_1), 
\nd
where we split the vector multiplet into two parts, $V_{1,2}$, one for each $SU(N)$ factor and the normalization factors $N_1, N_2$ are the casimirs of the corresponding gauge group factor in the adjoint representation. The theory on the brane stack worldvolume is actually $U(N) \times U(N)$ with the extra $U(1)$ factors corresponding to an overall shift of the stack's center of mass, which decouples.

Since the tip is not a smooth point, there is no reparametrization that can render the geometry locally $R^6$, which is what allows for the enlarged gauge group at low energies. However there are deformations of this theory that transform it into the $U(N)$ case, namely smoothing out the singularity via deformation or resolution. From the brane perspective, this is obvious, since the manifold becomes smooth, the neighborhood of the brane stack becomes $R^6$ and depending on the fluxes present, the low energy theory becomes a conformal $\mathcal{N}=1~~ U(N)$ theory or in the absence of 3-form fluxes the more symmetric $\mathcal{N}=4$ SYM. More precisely, one should regard the theory as a renormalization group flow from Klebanov-Witten to an IR fixed point where the gauge group gets broken down to a single $U(N)$.

It is instructive to see how this happens from the T-dual perspective. 
The type IIA description of this system is a set of $N$ D4-branes oriented along $x_{0,1,2,3,6}$ with periodic $x_6$ in the presence of two mutually orthogonal NS5-branes along $x_{0,1,2,3,4,5}$ and $x_{0,1,2,3,8,9}$ directions respectively and separated along the $x_6$ direction. When the D4's intersect the NS5's, they effectively split into two stacks of D4's suspended on either side of the NS5 branes. This gives rise to the $U(N) \times U(N)$ gauge group. However, the D4 stack can also be placed away from the NS5's in which case there is only one $U(N)$ group, which is the diagonal subgroup of $U(N) \times U(N)$. In the IIB picture this corresponds to placing the D3 stack away from the singular point on the conifold. 

Resolving the conifold corresponds to separating the NS5 branes along the $x_7$ direction as well. Since the D4's must wrap the $x_6$ circle, they can no longer intersect both NS5's and the gauge group is once again broken down to the diagonal $U(N)$ subgroup.

\begin{figure}[t]
  \centering
    \includegraphics[width=1.0\textwidth]{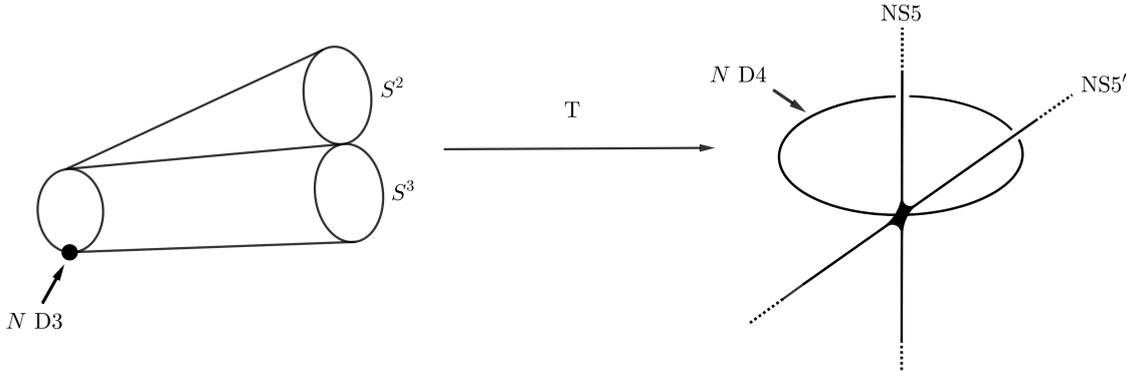}
    \vskip-3cm
  \caption{The gauge theory on $N$ D3-branes probing a K\"ahler deformed conifold can be easily ascertained from its type IIA T-dual brane configuration. The K\"ahler deformed conifold T-dualizes to two intersecting NS5-branes with the intersection point blown up to a diamond shape configuration \cite{ohta}. The D3-branes become D4-branes on a circle intersecting the diamond.}
\label{fig1MN}
\end{figure}

The deformation of the conifold is a more interesting case. The deformed conifold is T-dual to a configuration where the NS5 branes intersect and the intersection point is blown up into a diamond-shaped curved brane profile. In the conformal case, where the tension from the D4's on either side of hte NS5's is the same, the only way to have the NS5 branes intersect is to manually set their $x_{6,7}$ separation to zero. This means that the D4's intersecting this diamond no longer get split into two segments and there is still just one $U(N)$ gauge group as shown in 
{\bf fig \ref{fig1MN}}.

From the field theory side one can also see this flow as either the breaking of the gauge group down to its diagonal subgroup in the case of the resolution, or as the divergence of one of the gauge coupling, resulting in the mesons transforming into adjoint scalars for the remaining gauge group in the case of the deformation.

\subsubsection{Gauge fixing and Higgsing \label{meyechut}}

Now let us apply our probe approach to this theory. To parametrize a probe brane moving on a singular conifold one needs to give expectation values to \emph{two} of the chiral multiplets, one $A$ and one $B$ field. This corresponds to moving along the mesonic branch of the theory. Since the superpotential is quartic, this induces masses for the remaining two chiral multiplets as well as the vector multiplets of each of the two $SU(N)$ factors of the gauge group and of course the ghost fields, of which there are now two sets as well. 

For concreteness, we will give expectation values to the fields $A_2$ and $B_2$ in the following way:
\bg\label{max3}
A_2 \to a+A_2 \nonumber \\ 
B_2 \to b+B_2, 
\nd
where ($a, b$) correspond to the required expectation values. Substituting \eqref{max3} in the action \eqref{KSaction},
and  after gauge fixing, it takes the following form  \cite{Rxi}:
\bg \label{GFaction}
S&=&\int d^4 x \int d^4 \theta ~{\rm tr} \left[ \frac{1}{N_1 g_1^2}( e^{-g_1 V_1}D^\alpha e^{g_1 V_1}) \bar{D}^2 (e^{-g_1 V_1}D_\alpha e^{g_1 V_1}) \right] \nonumber \\
&+& \int d^4 x \int d^4 \theta ~{\rm tr} \left[ \frac{1}{N_2 g_2^2} ( e^{-g_2 V_2}D^\alpha e^{g_2 V_2}) \bar{D}^2 (e^{-g_2 V_2}D_\alpha e^{g_2 V_2}) \right] \nonumber \\
&+& \int d^4 x \int d^4 \theta ~{\rm tr} ~ \sum_{i=1,2}   \left[-\frac{1}{8} (\bar{D}^2V_i) (D^2 V_i) + V_i \mathcal{M}_{V_i}^2 V_i \right]\nonumber \\
&+&\int d^4 x \int d^4 \theta ~{\rm tr}~ \sum_{i=1,2} \bar{c'}_{i}\left[\left(1-\frac{\mathcal{M}_{V_i}^2}{\Box}\right)+g_i V_i+ ...\right] c_{i} \nonumber \\
&+&  \int d^4 x \int d^4 \theta ~{\rm tr} ~ \sum_{i=1,2}c'_{i}\left[\left(1-\frac{\mathcal{M}_{V_i}^2}{\Box}\right)+g_i V_i+ ...\right] \bar{c}_{i} \nonumber \\
&+& \int d^4 x \int d^4 \theta ~ {\rm tr}~\sum_{i=1,2} \bar{A}_{r}\left[\left(1-\frac{\mathcal{M}_{A_r}^2}{\Box}\right)+g_1 V_1+ g_2 V_2 +...\right] A_{r} \nonumber \\
&+& \int d^4 x \int d^4 \theta ~{\rm tr} ~\sum_{s=1,2}\bar{B}_{s}\left[\left(1-\frac{\mathcal{M}_{B_s}^2}{\Box}\right)+g_1 V_1+ g_2 V_2 +...\right] B_{s} \nonumber \\
&+& \int d^4 x \int d^2 \theta~ h~ {\rm tr}\left(A_1 B_1 a b - A_1 b a B_1 + A_1 B_1 A_2 B_2 - A_1 B_2 A_2 B_1 \right) \nonumber \\
&+&\int d^4 x \int d^2 \theta ~h~{\rm tr}\left(  A_1 B_1 a B_2 + A_1 B_1 A_2 b - A_1 B_2 a B_1 - A_1 b A_2 B_1 \right), 
\nd
where we have suppressed the gauge group indices, and the $...$ indicate higher order couplings of the chiral multiplets to the vector multiplet. The $c', c$ fields are the ghosts. They transform in the adjoint representations and we have also split them into two parts, one for each $U(N)$ factor. The mass matrices are given by:
\bg \label{masses}
&&\mathcal{M}_{V_{i}}^{2~mn}=\frac{g_i^2}{2} \left[{\rm tr}(\bar{a}(T^{(i)}_m T^{(i)}_n)a )+ {\rm tr}(\bar{b}(T^{(i)}_m T^{(i)}_n)b ) \right] \\
&&\mathcal{M}_{A_2}^{2} = \sum_{i=1,2} \sum_{m} \frac{g_i^2}{2}(T^{(i)}_m a) (\bar{a} T^{(i)}_m) \\
&&\mathcal{M}_{B_2}^{2} =\sum_{i=1,2} \sum_{m} \frac{g_i^2}{2}(T^{(i)}_m b) (\bar{b} T^{(i)}_m) \\
&&\mathcal{M}^2_{A_1, B_1} = 0, 
\nd
where $i$ denotes the copy of $U(N)$ in the gauge group and $m,n$ are adjoint indices of that copy of $U(N)$. Note that the $A_1, B_1$ fields have mass terms coming from the superpotential. 

We must now identify the components that correspond to the probe brane worldvolume gauge field and transverse fluctuations. For illustrative purposes let us consider $U(3) \times U(3)$, since all the essential features will be visible in this example. Splitting off a single brane is described, without loss of generality, by turning on the following VEV's:
\bg \label{expect}
a^{3}_{\tilde{3}}=b^{\tilde{3}}_{3} = m, 
\nd
where the untilded/tilded indices transform under the first/second copy of $U(3)$ respectively. The generators of the respective $SU(3)$ groups in the \emph{fundamental} representation are then written as:
\bg
\left[T^{(1)}_m\right]^a_b,~~~ \left[T^{(2)}_m\right]^{\tilde{a}}_{\tilde{b}},
\nd
with $m$ being the adjoint index. In the anti-fundamental representation they are of course written as $-T^*$. Now using \eqref{masses}, we can determine the components that become massive in the following way:
\bg \label{su3masses}
&\mathcal{M}_{V_{1}}^{2~mn}&=\frac{g_1^2}{2} \left[\bar{a}^{\tilde{3}}_3 (T^{(1)}_m)^3_a (T^{(1)}_n)^a_3 a^3_{\tilde{3}}+\bar{b}^3_{\tilde{3}} (T^{(1)*}_m)^a_3 (T^{(1)*}_n)^3_a b^{\tilde{3}}_3\right] \\
&&=\frac{g_1^2}{2}m^2 \left( T^{(1)}_m T^{(1)}_n+T^{(1)*}_m T^{(1)*}_n\right)^3_3 \nonumber \\
&\mathcal{M}_{V_{2}}^{2~mn}&=\frac{g_1^2}{2}m^2 \left( T^{(2)}_m T^{(2)}_n+T^{(2)*}_m T^{(2)*}_n\right)^3_3 \\
&\mathcal{M}_{A_2}^{2}&=\sum_{m} \frac{g_1^2}{2}(T^{(1)}_m)^a_3 a^3_{\tilde{3}} \bar{a}^{\tilde{3}}_3 (T^{(1)}_m)^3_b+\sum_{m} \frac{g_2^2}{2}(T^{(2)*}_m)_{\tilde{a}}^{\tilde{3}} a^3_{\tilde{3}} \bar{a}^{\tilde{3}}_3 (T^{(2)*}_m)_{\tilde{3}}^{\tilde{b}} \nonumber \\
&&= m^2 \left( \frac{g_1^2}{4} \delta^{a \tilde{3}}_{b \tilde{3}} + \frac{g_2^2}{4} \delta_{\tilde{a} 3}^{\tilde{b} 3} \right)  \\
&\mathcal{M}_{B_2}^{2}&=m^2 \left( \frac{g_1^2}{4} \delta^{a \tilde{3}}_{b \tilde{3}} + \frac{g_2^2}{4} \delta_{\tilde{a} 3}^{\tilde{b} 3} \right).
\nd
The $A_1, B_1$ also in principle gain a mass from the superpotential equal to $\frac{h}{2} m^2$ for 3 of the gauge components. Although at the field theory level the coupling $h$ is in principle independent from the gauge couplings, from the brane perspective it must be related to them. In fact, from the type IIA perspective the coupling $h$ is determined by the relative angle between the NS5-branes and for small angles can be deduced by viewing the $\mathcal{N}=1$ theory as a Higgsed $\mathcal{N}=2$ theory with massive adjoint scalars. The coupling $h$ is then suppressed relative to the gauge couplings by the mass of these adjoint scalars. In what follows we will assume that $h$ is much smaller than the gauge couplings, and so the $A_1, B_1$ fields are much lighter and we will not consider their contributions to the loops since we are not integrating down to that mass scale.

To summarize, we obtained 3 massive components for each of the 2 vector fields and ghosts, with masses related to one of the gauge coupling constants, as well as for each of the $A_2, B_2$ chiral multiplets with masses related to both gauge coupling constants. All of the masses are of course proportional to $m^2$. The $A_1, B_1$ fields also pick up some masses that are parametrically smaller.

Also note that the mass for the $3,3$ components of the chiral multiplets as well as a mass for the $T_8$ generators of the gauge groups that are twice the mass of the remaining degrees of freedom. These masses are related to the fact that by splitting off a single brane we have also shifted the center of mass of the brane configuration. In flat space, this shift is described by an overall $U(1)$ mode that we factor out to obtain the $SU(N)$ gauge group, instead of $U(N)$. We would also give traceless expectation values to the chiral multiplets, in order to maintain the overall center of mass. 

In our case, we do not want to move the rest of the stack off the singular point and the price to pay is that the $U(1)$ mode must be left in. However the degrees of freedom associated to it are heavier than the rest of the massive particles and are not enhanced by a factor of $N$, so we can safely ignore its contribution as subleading in $1/N$.

More generally, for $U(N+1)\times U(N+1)$, we use the expectation values:
\bg
a = b^\dag = {\rm diag} (0,...,m).
\nd
This breaks the gauge group down to $U(N) \times U(N) \times U(1)$, where the $U(1)$ factor is part of the diagonal subgroup of the original $U(N+1)\times U(N+1)$ theory. This is the $U(1)$ that lives on the probe brane worldvolume. Its generator can be written as the following linear combination:
\bg
T =\frac{1}{g} \left(g_1 T^{(1)} + g_2 T^{(2)} \right),
\nd
where $T^{(1)}$ and $T^{(2)}$ are represented by the same matrix when written in the fundamental representation, but are generators of the first and second $U(N+1)$ factor respectively. We also introduced a new coupling $g$. This allows us to separate out this component of the gauge field, which we'll call $v$, and write the action as:
\bg \label{diagAction}
S&=&\int d^4 x \int d^4 \theta ~{\rm tr} \left[\frac{1}{g^2}( D^\alpha v) \bar{D}^2 (D_\alpha v) \right] \nonumber \\
&+&  \int d^4 x \int d^4 \theta ~{\rm tr} ~\sum_{i=1,2} \frac{1}{g_i^2} (D^{\alpha} g_i V_i ) \,g v \, (\bar{D}^2 D_{\alpha} g_i V_i) + ... \nonumber \\
&+& \int d^4 x \int d^4 \theta ~{\rm tr} ~  \left[-\frac{1}{8} (\bar{D}^2 v) (D^2 v)\right]\nonumber \\
&+&  \int d^4 x \int d^4 \theta ~{\rm tr}  ~ c'_{i}\left[\left(1-\frac{\mathcal{M}_{V_i}^2}{\Box}\right)+g v+ ...\right] \bar{c}_{i} \nonumber \\
&+& \int d^4 x \int d^4 \theta ~{\rm tr} ~  \bar{c'}_{i}\left[\left(1-\frac{\mathcal{M}_{v}^2}{\Box}\right)+g_i V_i+ ...\right] c_{i} \nonumber \\
&+& \int d^4 x \int d^4 \theta ~{\rm tr} ~\bar{A}_{r}\left[\left(1-\frac{\mathcal{M}_{A_r}^2}{\Box}\right)+g v +...\right] A_{r} \nonumber \\
&+& \int d^4 x \int d^4 \theta ~{\rm tr} ~ \bar{B}_{s}\left[\left(1-\frac{\mathcal{M}_{B_s}^2}{\Box}\right)+g v +...\right] B_{s} \\
&+& ... ,\nonumber
\nd
where we only wrote out explicitly the kinetic and gauge fixing terms for this $U(1)$ gauge field and its linear interactions with the massive fields. From the kinetic terms we conclude that this new coupling $g$ must be related to the original gauge couplings through:
\bg
\frac{1}{g^2} = \frac{1}{g_1^2}+\frac{1}{g_2^2}.
\nd
There are also 3 massless chiral multiplet components that transform under this $U(1)$,
which correspond to the transverse fluctuations of the brane. They are related to $A_1, B_1$ and a specific linear combination of $A_2$ and $B_2$.

\subsubsection{Vector 4-point function and the warp factor \label{machli}}

We are now in a position to compute the one-loop 4-point functions of these fields. A convenient way is using supergraph techniques and essentially boils down to keeping track of the superderivatives and power counting. For an extensive review of these techniques see \cite{superspace}. The essential feature is that interaction vertices come with superderivatives acting on internal legs. These are then transfered onto other vertices or external legs through integration by parts. The final loop integral must contain $\bar{D}^2 D^2$ to give a non-vanishing result and any additional pairs of superderivatives acting on the loop get converted to internal momenta. This heavily limits the number of diagrams that actually contribute to the amplitude. The rules that we'll need are as follows:

\vskip.1in

\noindent $\bullet$ Chiral or anti-chiral 3-point vertices will contribute a $D^2$ or $\bar{D}^2$, respectively. 

\vskip.1in

\noindent $\bullet$ Vector 3-point vertex will contribute a $D^\alpha$ on one internal leg and a $\bar{D}^2 D_\alpha$ on another.

\vskip.1in

\noindent $\bullet$ Vector-chiral-antichiral vertices have a $D^2$ on the chiral leg and $\bar{D}^2$ on the anti-chiral leg. This includes the vertices involving ghosts.

\vskip.1in

\noindent $\bullet$ Propagators are given by the usual formula for the propagators in quantum field theory, namely:
\bg
\pm\frac{1}{p^2 + M^2},
\nd
with $+$ for chiral multiplets and $-$ for vector multiplets, where the vector squared mass is the $\mathcal{M}^2_{V_i}$ given in \eqref{masses}, while for the chiral multiplets the squared mass is the sum of $\mathcal{M}^2_{X}$ (where $X$ stands for the chiral field in question) and any additional mass that appears from the superpotential. In our case, all of these masses are proportional to the expectation value $m$ in that appears in \eqref{expect}.

\vskip.1in

\noindent $\bullet$ Ghost loops have an overall minus sign.

\vskip.1in

\noindent $\bullet$ Each vertex comes with the appropriate coupling constant and group theory factor.

\vskip.1in

\begin{figure}[t]
\begin{center}
\includegraphics[scale=0.3]{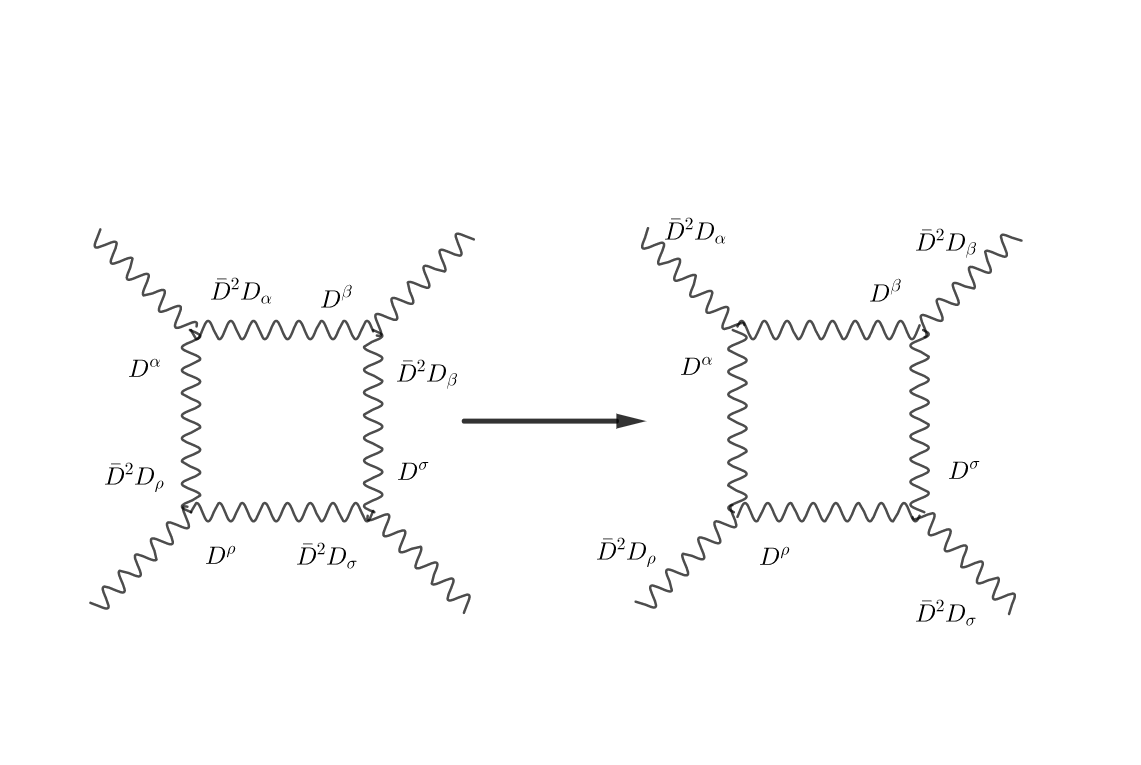}
\end{center}
\caption{A box diagram that contributes to the field strength 4-point function. Transferring the necessary derivatives onto the external legs leaves exactly 4 superderivatives in the loop, giving a non-vanishing result.}
\label{box}
\end{figure}

\begin{figure}[t]
\begin{center}
\subfloat{
\includegraphics[scale=0.4]{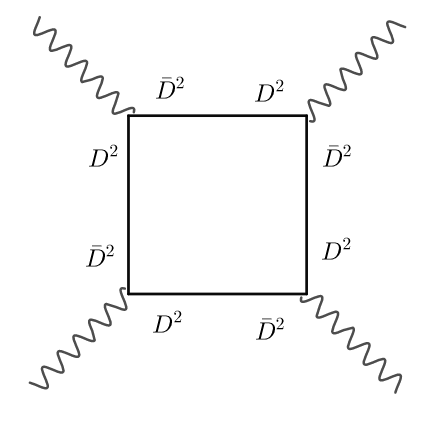}}
\subfloat{
\includegraphics[scale=0.4]{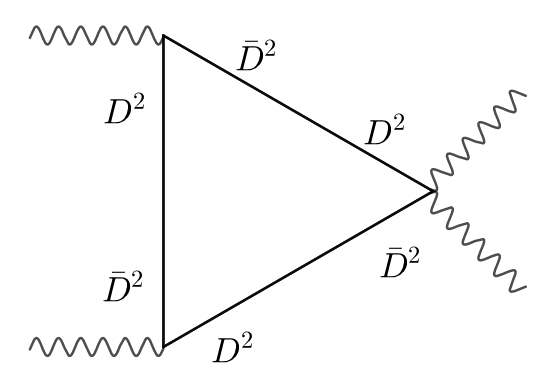}}
\subfloat{
\includegraphics[scale=0.4]{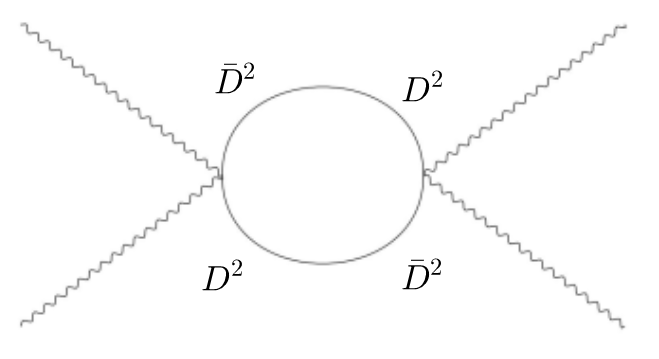}}
\end{center}
\caption{The first diagram contributes to the vector 4-point function. The other diagrams don't have enough superderivatives to give a non-vanishing gauge invariant result, which requires 3 superderivatives on each external leg and 4 in the loop.}
\label{nocontrib}
\end{figure}

\noindent Note that there are also 4-point vertices and higher present in the action \eqref{GFaction}, however they will not be important for the one-loop calculation, since they don't provide enough superderivatives to give a non-vanishing result for the quantities we're computing. 
In terms of the superfields, the two quantities we will need to compute are:
\bg
&& \langle F F F F \rangle \sim \langle W^\alpha \bar{W}^{\dot{\alpha}} W_\alpha \bar{W}_{\dot{\alpha}} \rangle  \\
&& \langle X X \partial X \partial X \rangle \sim  \langle\bar{\Phi} \Phi \bar{\Phi}\Phi \rangle .
\nd
A simple counting of the superderivatives quickly shows that the only diagrams that contribute to the field strength 4-point function are box diagrams. Indeed, applying the rules above gives box diagrams such as the one shown in {\bf fig \ref{box}}. After we transfer the necessary number of superderivatives onto the external legs, we see that the box diagrams have exactly 4 superderivatives left in the loop, which is what is necessary to give a non-vanishing answer. One could consider triangle or ``fish'' diagrams, such as in {\bf fig \ref{nocontrib}}. These use 4-point vertices that we hid in the ellipses when writing the action \eqref{GFaction}. Each vertex still contributes the same number of superderivatives, but there are fewer vertices and so the diagram vanishes.

The vector 4-point function is now very straightforward to calculate. Our $U(1)$ field couples to everything through the coupling $g$ that we introduced earlier and the loop integral is just the product of 4 propagators, so we get:
\bg
\langle F F F F \rangle &\sim& g^4 N \int d^4 k \prod_{i=1}^4 \frac{1}{q_i^2+\mu_i^2} \\
q_i &=& k+\sum_{j=1}^i p_j ,
\nd
 with $p_j$ being the external momenta and $M_i$ are the appropriate masses given by \eqref{masses}. There are also symmetry and group theory factors but they are not important for our purposes. Evaluating the integral and taking the external momenta to zero, we obtain the expected result:
\bg
\langle F F F F \rangle &\sim& \frac{g^4 N}{\mu^4}.
\nd
We now wish to compare this to the quartic term in the DBI action \eqref{expandedDBI}. We assume that the dual metric takes the following form:
\bg
ds^2 = \frac{1}{\sqrt{h}} dx^2 + \sqrt{h} \left(G_{ij} dx^i dx^j \right), 
\nd
where $h$ is the warp-factor (and not the quartic coupling we had earlier), and $G_{ij}$ is the internal metric that should be related to ${\cal G}_{ij}$ that we had in \eqref{crooked} under some approximations. The
kinetic terms in the DBI action become:
\bg \label{DBIkinetic}
S_{kin}&\sim &\int d^4 x \frac{1}{h} \left( e^{-\phi} + \frac{1}{4} e^{-\phi} h F^2+\frac{1}{4} e^{-\phi} G_{ij} h \partial X^i \partial X^j \right), 
\nd
where we omit the Lorentz indices so that we can write the warp factors explicitly. We see that if we identify our coupling $g$ with $e^{\phi/2}$, the gauge field ends up canonically normalized since the warp factor from the inverse metric cancel the warp factor from the determinant of the metric. The quartic term for the gauge field then takes the form:
\bg
S_{int} &\sim& \int d^4 x e^{-\phi} \left[\frac{1}{8} h (F F) (F F) - \frac{1}{4} h (F F F F) +... \right], 
\nd
which are the required interactions terms and the dotted terms would be going beyond the quartic order. 
We therefore demand that: 
\bg
e^{-\phi} h \propto \frac{g^4 N}{\mu^4} .
\nd
Identifying $\mu$ with a radial coordinate $r$ and $e^{\phi}$ with $g^2$ we recover the result for the warp factor of AdS, namely:
\bg
h \propto e^{2\phi} \frac{R^4}{r^4}, ~~~~R^4\sim e^\phi N. 
\nd

\subsubsection{Chiral multiplet 4-point function and resolution parameters \label{jesschut}}

We now turn to the scalars. In order to canonically normalize them, we need to express the metric $G^{ij}$ in Riemann normal coordinates. We also need to further rescale the scalar fields to cancel out the prefactors, so $X \to e^{\phi/2} X$. 

\begin{figure}[t]
\begin{center}
\subfloat{
\includegraphics[scale=0.4]{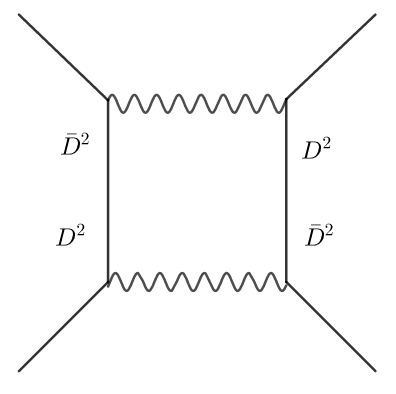}}
\subfloat{
\includegraphics[scale=0.4]{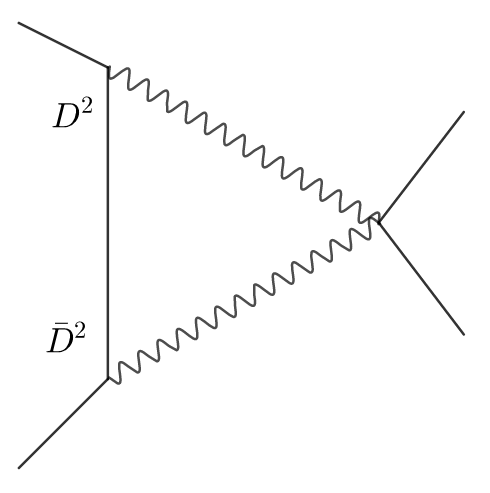}}
\subfloat{
\includegraphics[scale=0.4]{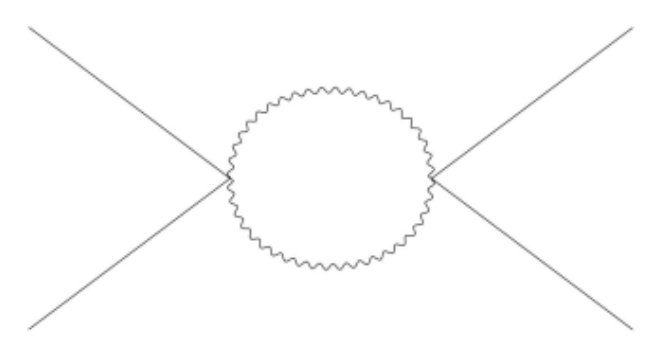}}
\end{center}
\caption{The box and triangle diagrams contribute to the 4-point function and scale as $\frac{1}{m^2}$. The fish diagram vanishes by virtue of lacking superderivatives in the loop.}
\label{scalars}
\end{figure}

To gain information about the transverse metric, we consider the next order in its the taylor expansion around the location of the brane. This is the coupling to the Riemann tensor, which with the rescaling above takes the form:
\bg
\int d^4 x \sqrt{h}~ e^{\phi} R_{ijkl} X^i X^k \partial X^j \partial X^l,
\nd
where $R_{ijkl}$ being the Reimann tensor. 
This term comes from computing the $\langle \bar{\Phi} \Phi \bar{\Phi} \Phi \rangle$ correlator without moving any superderivatives onto the external legs. {\bf Fig \ref{scalars}} shows the diagrams that one needs to consider. Note that the superpotential doesn't couple the heavy fields to the probe brane fields, so we have no purely chiral vertices. Now, box and triangle diagrams contribute, giving contributions proportional to:
\bg
\frac{g^4 N}{\mu^2}, 
\nd
where the mass dependence gets perfectly accounted for by the $\sqrt{h}~ e^{\phi}$ factor on the gravity side, indicating that the non-vanishing components of the internal Riemann tensor have no dependence on the mass scale. In other words all information on the energy scale dependence is contained in the warp factor. Note that there is still a dependence on $N$, which is no surprise, since the number of branes controls the flux and therefore the size of the near-horizon compact geometry. The exact value of the Riemann tensor would in principle correspond to the exact coefficient of this 4-point function, but the gravity dual is outside the regime of validity of our 1-loop calculation so we have no right to expect a match. However we can still study some qualitative features.

By symmetry, the warp factor and the Riemann tensor in the above calculation must be the same at every point on the $T^{1,1}$ at fixed radius, however we can break this symmetry by turning on baryon expectation values on the brane stack. If we turn on the following ``baryonic" operator:
\bg \label{a2vev}
A_2 = {\rm diag}(\epsilon,\epsilon,...,\epsilon,0), 
\nd
in addition to the probe brane degrees of freedom, certain components of the vector multiplet (and ghost) mass matrices will change by $g_i^2 \epsilon^2$. The effect of this is to break the gauge group down to the diagonal subgroup, by giving masses to the other components. However the masses of the components that couple to the probe degrees of freedom that will be running in the loop will not change. The chiral multiplets will also not pick up new masses from \eqref{masses}. However, if we instead turn on:
\bg \label{a1vev}
A_1 = {\rm diag}(\epsilon,\epsilon,...,\epsilon,0), 
\nd
we generate an additional contribution, proportional to $\epsilon^2$, to the masses both through \eqref{masses}.

This means that turning on different field expectation values transforming under the $SU(N) \times SU(N)$ subgroup will slightly alter the 4-point function of the fields transforming under the $U(1)$ subgroup, by an amount proportional to the ``baryon" expectation value and the new result depends on precisely which ``baryons" are switched on.

By dimensional analysis the 4-point functions must of course still have the same mass/radial scaling, but the exact coefficient will now change depending on which ``baryons" are switched on. We can factor out the overall radial dependence into a common warp factor but the different coefficients will now correspond to a relative warping of the internal manifold of order $\epsilon^2 / m^2$. Specifically the two 2-spheres of the $T^{1,1}$ will no longer be of the same size. Moreover, different points along the ``baryonic" branch of the theory lead to different warpings of the internal manifold. This feature will help us understand why the geometries found in the previous sections must be a warped-deformed-resolved conifold. Indeed we will argue that a very similar phenomenon can occur in the non-conformal case, but may require an extention of the theory with a larger moduli space.

\begin{figure}[h]
  \centering
    \includegraphics[width=0.6\textwidth]{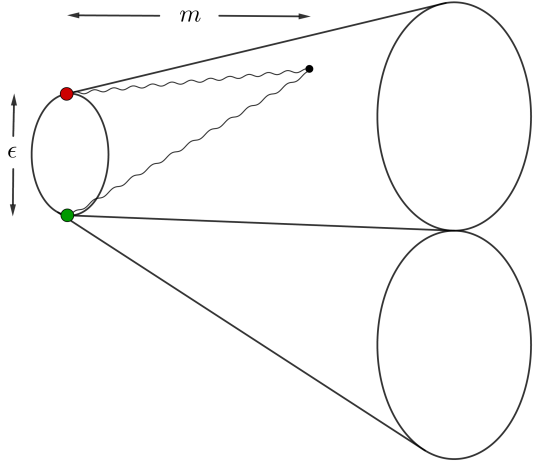}
  \caption{Effect of turning on different ``baryonic'' expectation values. The background manifold becomes a resolved conifold. Turning on the $A_2$ field places the stack at one pole of the non-vanishing 2-sphere (red dot). Turning on $A_1$ places it at the other pole (green dot). The squared masses of the stretched strings between the probe brane (black dot) and the stack changes from $m^2$ to $m^2 + \epsilon^2$.}
\label{max}
\end{figure}

The reason for this effect is easy to explain from the brane perspective as depicted in {\bf fig \ref{max}}.
As we mentioned earlier, turning on ${\rm diag} (\epsilon,\epsilon,..,\epsilon, 0)$ corresponds to a resolution of the conifold. Which field ($A_1$, $A_2$ or some linear combination thereof) takes on this value dictates where the brane stack is located on the blown up 2-cycle and thus makes a difference in the relative position of the stack and the probe. In particular, if the probe and the stack are located at the same point on the blown up 2-cycle, we have the situation described by \eqref{a2vev}, while placing the stack and the probe at opposite poles of the 2-cycle is described by \eqref{a1vev}. It's quite clear that in this case the squared masses of the stretched strings shift by an amount proportional to the resolution factor!

\subsection{$\mathcal{N}=1 ~SU(N)\times SU(N+M)$ with bi-fundamental matter \label{marmichut}}

We can now generalize the results of the previous section to the Klebanov-Strassler field theory \cite{KS}. 
This theory arises on a stack of $N$ D3-branes at the tip of the conifold combined with $M$ D5-branes wrapping the non-trivial 2-cycle. This results in additional scalar degrees of freedom corresponding to the transverse oscillations of the D5's. The fact that they can not oscillate along the wrapped 2-cycle, is ultimately responsible for the mismatch of the gauge groups, however it is much easier to see from the T-dual side, where the wrapped D5's become D4-branes suspended between the orthogonal NS5-branes (see {\bf fig \ref{laval}}(a)). 

The action is the same as \eqref{KSaction}, and the chiral multiplets still transform in the (anti-)bi-fundamental representation, but now the dimensions of the two gauge group factors are not the same which modifies our analysis in several ways. The obvious difference is that since the chiral matter is no longer representable by a square matrix, some of the expressions in the previous section need to be modified appropriately. Specifically the expectation values of the fields should be understood to still be diagonal in the $N\times N$ submatrix and zero elsewhere. The more important difference is that the theory is no longer conformal, so all the couplings in our theory now depend on the energy scale. The natural energy scale to use is the mass of the lightest massive field, since we are integrating out everything above that. 

The running of the couplings in this theory is known. For each of the gauge groups, we can treat the other as a flavor group and the running of the coupling is then given by the NSVZ beta function. Thus the $SU(N+M)$ has $2N$ ``flavors" and the $SU(N)$ has $2(N+M)$ flavors. This simply gives a logarithmic running of each (inverse squared) coupling, with the two gauge couplings running in opposite directions: 
\bg
\beta_{\frac{1}{g_1^2}} &\propto& 3(N+M)-2N(1-\gamma) \nonumber \\
\beta_{\frac{1}{g_1^2}} &\propto& 3N-2(N+M)(1-\gamma), 
\nd
where $\gamma$ is the anomalous dimension. 
To a first approximation, the anomalous dimensions of the chiral multiplets are $-1/2+ \mathcal{O}(g_i^2 M^2/N^2)$. This means that:
\bg
\frac{1}{g_1^2}+\frac{1}{g_2^2} =\frac{1}{g_0^2}\left(1- c g_0^2 \frac{M^2}{N} \log ~{\mu\over \mu_0} + ... \right), 
\nd
with some order one constant $c$, where $\mu_0$ is a scale at which both couplings are equal to $2 g_0$. It's not immediately obvious whether one should identify the entire right hand side with $1/g_s$ or just the leading term. The latter case corresponds to the constant dilaton in the gravity dual but that result is also only valid to leading order in $M/N$. We note, however that the \emph{difference} of the (inverse squared) couplings:
\bg
\frac{1}{g_1^2}-\frac{1}{g_2^2} \propto M \, \log (\mu/\mu_0) + ..., 
\nd
indicates that the $SU(N+M) \times SU(N)$ description is only perturbatively valid within a certain range of mass scales. When we exit the regime of validity of one description we can then turn to its Seiberg-dual description along the duality cascade. Within this range, however the sum of the inverse couplings doesn't really change much, so we can safely identify the dilaton with the ``bare'' coupling:
\bg
e^{-\phi} \equiv  \frac{1}{g_s} \sim \frac{1}{g_0^2}.
\nd
The warp factor in the gravity dual is once again related to the vector multiplet 4-point function, namely:
\bg
e^{-\phi} h \sim \frac{g^4 N}{r^4} = \frac{g_0^4 N}{r^4}\left(1+\,c g_0^2 \frac{M^2}{N} \,\log~{r\over r_0} + ... \right), 
\nd
where we have identified $\mu$ and $\mu_0$ with $r$ and $r_0$ respectively. 
The warp factor is then given by:
\bg
h \sim e^{2\phi}\left[\frac{e^\phi N - c\,  e^{2 \phi} M^2 \, \log(r/r_0) + ... }{r^4}\right],  
\nd
which is the form of the Klebanov-Tseytlin solution for the gravity dual of this field theory away from small $r$ \cite{KT}. To study the small $r$ behavior and the deformed conifold geometry one would need to include the non-perturbative ADS superpotential \cite{ADS}, which becomes important in the IR.

In the IR, for $N$ a multiple of $M$, the theory is forced onto the baryonic branch by the ADS superpotential. However, what is meant by ``baryons" is slightly different in this context than in the conformal case. The baryons that gain expectation values in this case can be expressed in terms of the pre-last step of the cascade, where the gauge group is $SU(2M)\times SU(M)$. They are of the form:
\bg
\epsilon^{a_1 ... a_{2M}} \epsilon_{\tilde{a}_1 ... \tilde{a}_{2M}} (A_1)_{a_1,...,a_M}^{{\tilde{a}_1,...,\tilde{a}_M}} (A_1)_{a_{M+1},...,a_{2M}}^{{\tilde{a}_M+1,...,\tilde{a}_2M}}, 
\nd
and are symmetric under exchange of the $A_1$ and the $A_2$ chiral multiplets, that is to say they are symmetric under exchange of the two spheres of the $T^{1,1}$. This means that at large $r$, we should not expect the relative warping in the internal manifold to occur, like it did on the ``baryonic" branch in the conformal case. This is of course expected since the Klebanov-Strassler supergravity solution has no such warping. It is interesting that it doesn't seem like there are any other gauge invariant ``baryon"-like operators that we can turn on in this theory due to the mismatch between the gauge group ranks at higher steps in the cascade. There are of course still mesonic operators, but these are dual to branes on the dual geometry, rather than deformations thereof.

One could then ask what the {\it resolved} warped-deformed conifold solutions with non-zero resolution parameter are dual to. One obvious possibility is that, like in the conformal case, the resolution parameter of the dual is inherited from a resolution of the original conifold background on which the brane stack is placed. However in this case, the 
D5-branes no longer wrap a vanishing cycle and the D3 charge has to be dissolved as worldvolume fluxes to maintain supersymmetry. The resulting theory is no longer four-dimensional in the UV. Another more intriguing possibility is that instead of resolving the conifold, we could instead blow up a different 2-cycle than the one wrapped by the D5-branes while maintaining the conical singularity. In this case the D5-branes remain as fractional branes but now have a moduli space of their own. Indeed, having such a branch of the moduli space is crucial to constructing a 
four-dimensional UV completion of the theory as described in \cite{mia1}, \cite{mia2}, \cite{mia3}, \cite{maxim}. 

From the field theory perspective we can note that if we add \emph{adjoint} chiral multiplets to the theory that only couple to one of the gauge group factors, giving them a diagonal expectation value would achieve a very similar effect to turning on the baryonic operators in the Klebanov-Witten case and lead to a relative warping of the spheres in the gravity dual. However, this warping would be different from the one obtained by resolving the background manifold. In the next section we will discuss in more detail this branch of the moduli space, how to enlarge the Klebanov-Strassler theory to include it, as well as its relation to the UV completion of the theory.

\section{UV completion, moduli space and supersymmetry \label{youvee}}  

In the section \ref{IR} we discussed the IR physics of a specific ${\cal N} = 1$ confining gauge theory, specifically concentrating on the issue of supersymmetry, using two dual descriptions. One, with D5-branes wrapped on the 2-cycle of a non-K\"ahler resolved conifold and two, with its gravity dual, i.e with fluxes on a non-K\"ahler deformed conifold. Either of these two pictures provide a reasonably good description of the strongly coupled IR physics, although this cannot be extended to describe the UV dynamics.  At this stage one would require a well defined 
UV behavior so that UV divergence of Wilson loop or Landau poles (in the presence of fundamental matter) do not occur. In the following we will discuss ways to UV complete the model. 

\subsection{Moduli space of a gauge theory from non-K\"ahler resolved conifold \label{dracula}}

The model that we discussed in section \ref{rescon} is related to D5-branes wrapped on a 2-cycle of a non-K\"ahler 
resolved conifold. The IR behavior of this theory is well known from the work of \cite{malnun}:  the theory confines 
in the far IR. However the UV behavior is more non-trivial. In fact the UV is no longer a four-dimensional theory, but rather 
a six-dimensional one. Such a six-dimensional theory is in general non-renormalizable, so the UV behavior
becomes questionable in this case. Additionally the dimensional reduction of this theory to four-dimensions will have 
an infinite tower of massive KK states, resulting in a complicated Lagrangian description at the UV. These and other issues prompt us to ask whether there could exist a {\it simpler} description in the UV. The simplest description in the UV is of course a conformal behavior, so the question is to construct a UV completion of the class of theories that we studied in section \ref{rescon} that are IR confining and UV conformal\footnote{For a slightly different way to UV complete Klebanov-Strassler theory the readers may want to refer to \cite{berto}. In this paper a finite UV completion is presented using orientifold planes. The difference there is that the UV completion may be done without the requirement of having infinite cascades. It will be interesting to determine the precise connections between their work and ours. We would like to thank Riccardio Argurio and Marco Bertolini for discussions on this and for bringing \cite{berto} to our attention \label{berton}.}. 

We would also want  gravity dual descriptions at all energy scales for the kind of theories that we want to construct. In fact for calculational mobility we further demand  supergravity descriptions for the duals from IR to UV.  This means we require the gauge theories to be very strongly coupled from IR to UV.     
 
 \begin{figure}[t]
        \centering
\includegraphics[height=6cm]{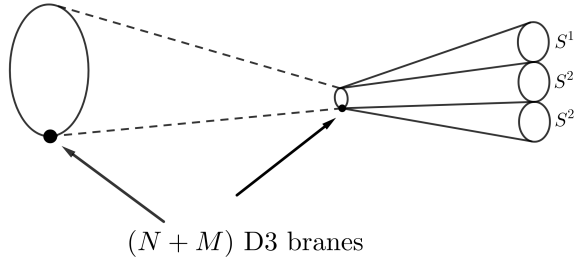}
\vskip-1cm
        \caption{$N + M$ D3-branes placed at the south pole of the resolved sphere on a non-K\"ahler warped resolved conifold.}
               \label{fig2MN}
        \end{figure}
  
  \subsubsection{Classical moduli space and UV completion \label{phantom}}
  
 Let us now consider the classical moduli space of the theory. This is of course given by the configuration in 
 section \ref{rescon}, and for our purpose it will be a non-K\"ahler resolved conifold. We will call this the brane side of the picture, to distinguish it from the gravity {\it dual} side\footnote{There are also branes that appear in the dual side 
 too, but the distinction appears clearly in the regime where the dual configuration has no branes. This is the case 
 where $p \equiv N ~{\rm mod}~ M$ vanish, where $N$ and $M$ are the number of D3 and D5-branes respectively.}.
 The world-volume gauge theory 
 is a pure glue ${\cal N} = 1$ theory and as such doesn't have a Coulomb branch. However the theory, in the presence of additional bi-fundamental matter, is known to have both Baryonic and Mesonic branches. In the 
 earlier two sections, namely \ref{rescon} and \ref{defcon}, we studied how the Baryonic branches may be understood  from both sides of the picture where the gauge group is at least:
 \bg\label{pelican}
 SU(2M) ~\times~ SU(M), \nd
 with two running couplings (for details see \cite{KS, aharony, butti, MM}). Such a configuration is {\it one} cascading step away from a $SU(M)$ confining theory \cite{KS}. 
 
 \begin{figure}[t]
  \centering
    \includegraphics[width=1.0\textwidth]{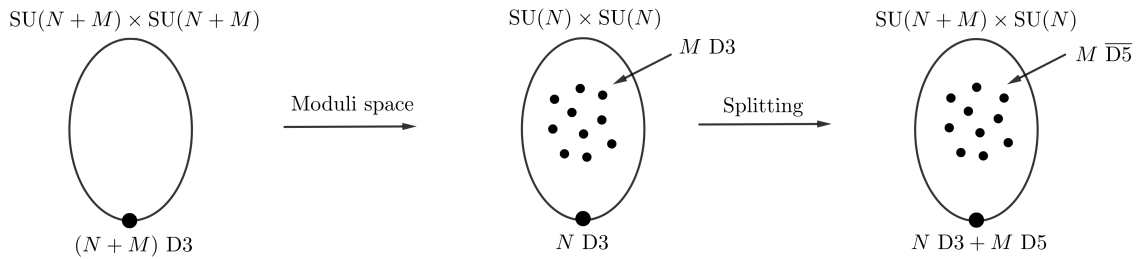}
    \vskip-3.5cm
  \caption{Moving the branes in the moduli space of almost CFTs can lead us to the UV complete Klebanov-Strassler model. In the left we have $N+M$ D3-branes localized on the south pole of the resolved sphere, out of which $M$ D3-branes may be moved away as shown in the middle figure. Finally, on the extreme right, the $M$ D3-branes spilit into $M$ D5 and  $M$ $\overline{\rm D5}$ branes to form the required UV completion of the Klebanov-Strassler model. }
\label{fig3MN}
\end{figure}

 In the present case we are looking for a different branch of the theory, where the classical moduli space still remains a resolved conifold (let us denote the manifold as $\mathbb{M}$).
 The difference lies in the choice of $N + M$ D3-branes instead of $M$ wrapped D5-branes that we considered in section \ref{rescon}. The configuration is depicted in {\bf fig \ref{fig2MN}}.  Additionally, there is  another difference from the resolved conifold configuration that we took in section \ref{rescon}. The five-branes $-$ that will appear 
 soon $ - $ will wrap a {\it vanishing} 2- cycle of $\mathbb{M}$ and therefore the D3-branes are arranged at the south pole of the resolved 2-sphere of $\mathbb{M}$. In other words, there are two 2-cycles with one of them vanishing at $r = 0$. Such a manifold 
 cannot support a K\"ahler (or Calabi-Yau) metric on it and it is therefore an example of a non-K\"ahler resolved conifold. In other words we can take:
 \bg\label{HtGlTKO}
 {\cal G}_3 = r^2 + a^2_1(r), ~~  {\cal G}_4 = r^2 + a^2_2(r), ~~  {\cal G}_5 =  {\cal G}_6 = 0, ~~
 (a_1(0), a_2(0)) \equiv (0, a_2),   \nd
 in \eqref{crooked} so that the resolved sphere\footnote{The orientations of the wrapped D5-branes can be changed from $a_1$ to $a_2$. In that case the resolved 2-cycle will have a size $a_1$ with vanishing $a_2$.} 
 at the origin has a size $a_2$.  We will also take $\gamma = 0$. This way we are essentially studying the {\it pre-dual} scenario and not the  Baryonic branch of the theory.  The constraints 
 on ${\cal G}_1, {\cal G}_2, a_1$ and $a_2$ will appear from the ${\cal N} = 1$ supersymmetry constraints 
 \eqref{baghdadi} and \eqref{easter} that we analyzed in section \ref{salonfem}, as well as from the Bianchi identity.  In the presence of $N + M$ D3-branes we then expect the gauge group to be:
 \bg\label{angrynan}
 SU(N + M) ~\times ~SU(N + M), \nd
 which is naively a CFT with bi-fundamental matter, so that the gravity dual may be given simply by taking the near horizon geometry of the D3-branes on a non-K\"ahler resolved conifold. Unfortunately this naive expectation turns out to be
 incorrect as the gauge theory fails to be a CFT because of the broken scale invariance from the 
 ${\bf F}_3$ fluxes generated by the non-K\"ahler nature of the background! 

However in the limit of constant $a_2(r)$ and almost vanishing $a_1(r)$, the manifold is closer to a K\"ahler resolved conifold and as such doesn't require a non-vanishing ${\bf F}_3$ flux to support the geometry. In this limit the gauge theory is nearly a CFT with a gauge group given by \eqref{angrynan}.  We can then move the $M$ D3-branes in the moduli space of a $SU(N) \times SU(N)$ gauge theory, or along the resolved 2-sphere as depicted in {\bf fig \ref{fig3MN}}. Once on the 2-sphere, the $M$ D3-branes split into $M$ D5 and $M$ $\overline{\rm D5}$ branes. The five-branes wrap the vanishing 2-sphere (i.e the 2-cycle parametrized by $a_1(r)$). The 
the stability of the system is a subject of detailed discussion in \cite{maxim} and also in section \ref{stabul}. The whole configuration then satisfies the minimal ${\cal N} = 1$ supersymmetry, and the forces between any pair of branes in the system vanish. 
We can then move the $M$ D5-branes at the south pole of the resolved sphere where it combines with $N$ D3-branes. At low energies, we can integrate out the massive strings between the D5 and the $\overline{\rm D5}$ branes, and so the low energy gauge group becomes:
\bg\label{nakeyoudie}
SU(N + M) ~\times ~SU(N) ~ \times ~ U(1)^M. \nd
The $U(1)$'s are  basically decoupled and therefore the IR configuration leads us to the Klebanov-Strassler model \cite{KS}. On the other hand, at very high energy, we can ignore the mass of the stretched strings, and the gauge group approaches \eqref{angrynan}, i.e a CFT. Together this leads to a UV conformal and IR confining theory, very much like the ones we had in \cite{mia1, mia2, mia3}.

How does this work from the perspective of the UV complete gauge group \eqref{angrynan}? This has already been discussed in some details in \cite{mia3}, where the UV completion uses a certain ${\cal N} = 2$ theory, but the essential framework remains the same\footnote{The following discussion, and also the underlying ${\cal N} = 2$ aspect of the story, is probably more transparent from a T-dual type IIA brane configuration. This will be discussed in section 
\ref{tdula}.}.  Let us start with the following Lagrangian:
\bg\label{rossi2}
{\cal L} & = & \int d^4\theta~{\rm tr}~\bar{\Phi} e^{g{\mathbb{V}}} \Phi+\int d^2\theta \left({\cal W}
+  {\tau \over 32\pi i} ~{\rm tr} ~\mathbb{W}^2_{\alpha}\right) + c.c.\nonumber\\
 & = & - {1\over 2} D^{\mu}\varphi_k D_{\mu}\varphi_k - {\bf V}(\varphi)
 - \sum_{i = 1}^2 {1\over 4g_i^2} {\bf F}^{a\mu\nu}_i {\bf F}^a_{i\mu\nu} + ... , \nd
where in the first line ${\cal W}$ is the superpotential; $\Phi$ is the chiral multiplet (constructed out of $A_{1, 2}, 
B_{1, 2}$ bi-fundamental matters, as well as adjoints); 
and $\mathbb{W}_{\alpha} \equiv - {1\over 4} {\overline{D}}^2 D_\alpha \mathbb{V}$ is the field strength of the vector multiplet $\mathbb{V}$. The way we have expressed the first line, all the above quantities are defined 
with respect to the product gauge group \eqref{angrynan} in the UV. A more precise representation of the Lagrangian could probably be extracted from a variant of the Lagrangian presented in \eqref{KSaction}. 
Thus $\tau$ appearing above is some complexified gauge coupling with $g$ representing the YM coupling. 
If we want to express the Lagrangian in terms of two different gauge groups 
$SU(N+M)$ and $SU(N + M)$ with couplings $g_1$ and $g_2$ respectively, then we have to perform the 
$\theta$ integration. A part of the final Lagrangian is then depicted in the second line of \eqref{rossi2} 
written in terms of the field strength of {\it two} gauge groups at UV \eqref{angrynan}. The covariant derivative 
of the scalars $\varphi_k$ from the chiral multiplet $\Phi$,  is now expressed in the following way:
\bg\label{maisoncurry}
D_\mu \varphi_k \equiv \del_\mu \varphi_k  - i g_2 {\bf A}^a_{2\mu} \left({\mathbb{T}}^a_2\right)_{kl} \varphi_l. \nd
Note that we have  required the scalar fields $\varphi_k$ to transform only under a subgroup of the {\it second} 
$SU(N + M)$ UV group. Such a requirement clearly underlies the basics of the Higgsing scenario if we demand the potential ${\bf V}(\varphi)$ to  be minimized at $\langle \varphi_k \rangle \equiv a_k$. Then the generator 
${\mathbb{T}}_2^a$ is broken if and only if:
\bg\label{punsweet}
{\mathbb{M}}^a_i \equiv ig_2 \left({\mathbb{T}}_2^a\right)_{ij} a_j \ne 0. \nd
The quantity ${\mathbb{M}}^a_i$ has a special significance in our discussion here in the sense that it will be related to the breaking of the second gauge group to a smaller subgroup. To see this, let us consider the fluctuations 
of the scalar fields $\varphi_k$ over the background value of $a_k$. Denoting the fluctuations as $\phi_k$, it is easy to see that:
\bg\label{thecommuter}
{1\over 2} D^\mu \varphi_k D_\mu\varphi_k = {1\over 2} \del^\mu \phi_k \del_\mu \phi_k + 
{1\over 2} {\mathbb{M}}_k^a {\mathbb{M}}_k^b {\bf A}_2^{a\mu} {\bf A}_{2\mu}^b + {\rm interactions}, \nd 
where the interaction terms have been given in \cite{mia3} so we don't elaborate them here. The quantity 
${\mathbb{M}}_k^a$ appears as a mass term that breaks the gauge group $SU(N + M)$. For us we only want to break it to a $SU(N)$ group, so all the ${\mathbb{M}}_k^a$ give mass to the remaining generators. 

The above analysis is rather simplified because, as one may see from {\bf fig \ref{fig3MN}}, the $\overline{\rm D5}$ 
distribution may generically be continuous. For example if we assume the strings connecting the D5-branes, localized at a point on the south pole of the resolved sphere, to a particular localized set of $\overline{\rm D5}$-branes on the resolved two-sphere, to provide an energy scale $\Lambda$, then the $\overline{\rm D5}$ distribution may be expressed as\footnote{More generic distribution is possible. Here we take the simplest one to illustrate the underlying physics.}:
\bg\label{20lisa18}
M_\Lambda \equiv {Me^{\alpha\left(\Lambda - \Lambda_0\right)}\over 
1 + e^{\alpha\left(\Lambda - \Lambda_0\right)}}, \nd
where $\Lambda_0$ is a fixed scale of the theory and $\alpha >> 1$. It is clear from above that if 
$\Lambda >> \Lambda_0$, then $M_\Lambda = M$, whereas if $\Lambda << \Lambda_0$, then 
$M_\Lambda$ vanishes. This means, at any energy scale, the gauge group may be expressed as:
\bg\label{lakeshore}
 SU(N + M_\Lambda) ~\times ~ SU(N + M) \times \left[U(1)\right]^{M - M_\Lambda }, \nd
 where again the $U(1)$ parts are decoupled from the dynamics. The above distribution makes sense because 
both $M$ and $N$ are very large numbers, and is consistent with the low energy expectation \eqref{nakeyoudie}. 

What happens in the gravity dual? The gauge theory on $N$ D3 and the $M$ D5-branes localized at the south pole of the resolved sphere can be dualized to a resolved warped-deformed conifold  \eqref{crooked} with ${\bf G}_3$ fluxes 
\eqref{tagramaiaa}. We took  constant dilaton and zero axion fields, but this can be easily generalized. However the issue of Bianchi identity was never discussed there as in the absence of localized sources both ${\bf F}_3$ and 
${\bf H}_3$ are closed. This is also true in the Baryonic branch because the D3-brane charges are dissolved in the wrapped D5-branes so there are no localized sources therein. However once we demand UV completion, as in 
{\bf fig \ref{fig3MN}}, the ${\overline{\rm D5}}$-branes reappear on the gravity dual side, distributed along the radial direction. The distribution should be similar to \eqref{20lisa18}, with $\Lambda$ and $\Lambda_0$ now replaced by $r$ and $r_0$ respectively. This means we can now demand:
\bg\label{HFcoup10}
d{\bf G}_3  = d{\bf F}_3 + i e^{-\phi} d\phi \wedge {\bf H}_3 \equiv ~{\rm sources}, ~~~ 
d{\widetilde{\bf F}}_5 = {\bf F}_3 \wedge {\bf H}_3, \nd
with ${\overline{\rm D5}}$ sources. The scenario now is very close to what we discussed as Region 2 in \cite{mia2} (except there are no fundamental matter sources), and we expect both ${\bf F}_3$  and ${\bf H}_3$ to vanish at $r \to \infty$. This in turn leads us to an $AdS_5 \times T^{1, 1}$ geometry at large $r$ with constant ${\bf F}_5$.      

\subsubsection{Stability of branes and anti-branes in a curved space \label{stabul}} 

The UV completion of the model requires the introduction of $\overline{\rm D5}$-branes. In the classical picture, these antibranes are located near the bottom of the throat but at a finite separation from the D5-branes. In the gravity dual, the $\overline{\rm D5}$ appear at a finite radius, corresponding to the energy scale at which the UV Klebanov-Witten like theory gets Higgsed to a Klebanov-Strassler type duality cascade. Neither of these configurations are stable without the introduction of world-volume fluxes on the D5 and $\overline{\rm D5}$ branes. The stabilization of a single parallel brane-antibrane pair by world-volume flux is well understood in flat space \cite{Bak} and in cases where the orthogonal directions to the branes have a curved metric \cite{mateos}. It is possible to find a combination of world-volume fluxes that induce the correct lower-dimensional brane charge to restore the force cancellation required for stability. Typically, the required flux can be expressed as having opposite space-space component (magnetic ${\cal B}$ field) and the same time-space component (electric ${\cal E}$ field) turned on on the brane and the anti-brane.
The situation becomes more subtle when the branes wrap cycles that have a curved metric. Although, the analysis of \cite{Bak} can still be carried out to determine the fluxes necessary to have the right lower-dimensional charges on the branes, these fluxes are generally no longer closed if the wrapped cycle has curvature and thus do not satisfy the equations of motion. One way around this obstacle is to consider non-abelian world-volume fluxes, since the non-abelian field strengths do not require closure, of the form: 
\bg\label{gajor}
{\cal F} = d{\cal A} + {\cal A} \wedge {\cal A}.
\nd
In \cite{maxim} the necessary conditions for restoring stability in the non-abelian case were worked out to lowest order in the field strengths and an example of a configuration of non-abelian fluxes on a ${\rm D5} - \overline{\rm D5}$ pair wrapping a 2-sphere was given. The required flux is of the form:
\bg
{\cal F}^a = {\cal B}^a ~e_{\theta} \wedge e_{\phi} + {\cal E}^a~ e_0 \wedge e_\phi = \partial_\theta {\cal A}_\phi^a 
~e_{\theta} \wedge e_{\phi} + f^{abc} {\cal A}_0^b {\cal A}_\phi^c ~e_0 \wedge e_\phi,
\nd
where ${\cal E}^a$ and ${\cal B}^a$ are now adjoint fields and on top of having the appropriate magnitudes, must also be different generators of an SU(2) subgroup of the world-volume gauge group. As can be seen, ${\cal B}$ is the closed piece of the flux, while the ${\cal E}$ component arises from the wedge product term.
A caveat to this approach, is that checking that it does indeed restore supersymmetry is a non-trivial task. Since the system is non-abelian, this requires a notion of non-abelian $\kappa$-symmetry, which is not fully understood. In 
\cite{broo} an expression for the $\kappa$-symmetry $\Gamma$ matrix was given to quadratic order in the field strength. While the fluxes used in \cite{maxim} satisfy the $\kappa$-symmetry condition to the appropriate order, the magnitudes of the fluxes are not small, so an all-order expression is desirable. Unfortunately, as pointed out in 
\cite{bilal}, the procedure used to obtain the $\Gamma$ matrix does not extend to higher orders in the field strength.

These technical challenges however do not deter us to construct indirect ways to verify the stability of the system. Switching on gauge fluxes on the D5 and the ${\overline{\rm D5}}$ branes create bound states of D3-branes on both set of branes. These D3-branes repel each other because of their RR charges, and these repulsive forces are cancelled by other attractive forces\footnote{One example would be the Coulombic attractive forces between the D5 and the ${\overline{\rm D5}}$ branes.  Another would be the gravitational attractions. There would also be repulsive forces between individual set of branes and anti-branes (for example in a given set of say D5-branes, the bound D3-branes would repel each other). Stability will happen when all the repulsive forces are balanced by the attractive forces. This is where ${\cal N} = 1$ supersymmetry will be realized.}
coming from the  closed string exchanges. Such an analysis do not rely on the 
$\kappa$-symmetry computation, but rather on the precise exchanges of closed strings. Of course again 
an exact analysis is hampered by our ignorance of string dynamics in a curved background, but at least this picture gives us a physical reason why we expect stability to happen. 
 
Yet another computation would be to actually quantize the open string between a D5-brane 
and a ${\overline{\rm D5}}$-brane. In a flat space, and when the branes approach each other, the open string develops a tachyon. This tachyon eventually 
annihilates the brane-antibrane system. In the presence of gauge fluxes on the two branes, the tachyon can be made massive and the brane anti-brane system no longer annihilates each other \cite{Bak, shmakova}. Globally the system only has a three-brane charge as the five-brane charges cancel. In a curved space however a similar computation is hard to perform because of the lack of the computational technology to quantize a string in a curved space. On the other had, if the curvature is small we expect the flat space result to carry over to this space, and the tachyon between a D5-brane and a ${\overline{\rm D5}}$-brane should become massive as before in the presence of world-volume fluxes. A generalization of the above computation for $M$ localized D5-branes and $M$ distributed ${\overline{\rm D5}}$-branes as in {\bf fig \ref{fig3MN}} should not be too difficult.  

\begin{figure}[t]
  \centering
    \includegraphics[width=1.0\textwidth]{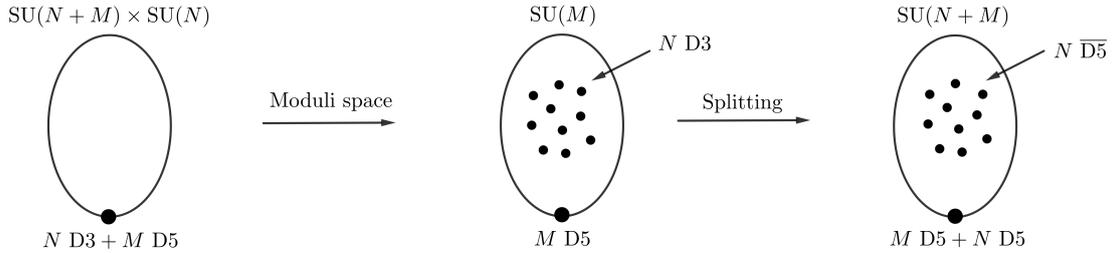}
    \vskip-3.5cm
  \caption{Another branch of the non-conformal cascading theory where the moduli space may be studied either from $N$ D3-branes in a $SU(M)$ theory, as shown in the middle figure, or from $N$ ${\overline{\rm D5}}$-branes in a $SU(N + M)$ theory, as shown in the right figure.}
\label{fig4MN}
\end{figure}

\subsection{Moduli space of a non-conformal cascading theory \label{cvetic}}

The Higgsing that we discussed above breaks the UV gauge group to a cascading one, along with a decoupled sector  that plays no role here. In fact if we integrate out the massive strings, all we see is the cascading gauge group of $SU(N+M) \times SU(N)$ given by $N$ D3 and $M$ D5-branes localized at the south pole of the resolved sphere. 
If the D5-branes wrap a non-vanishing 2-cycle, then the parallel configuration with $N$ D3-branes would break supersymmetry. However supersymmetry may be preserved if all the D3-branes are dissolved inside the wrapped D5-branes as gauge fluxes. When $N = M$, this is of course the Baryonic branch that we discussed earlier. For 
$N < M$, the situation is tricky as Seiberg dualities cannot be performed. In the Klebanov-Strassler case, one may instead go to the Mesonic branch. The story has been discussed elsewhere, see for example \cite{chetan}, so we will not dwell on this here anymore. 
Instead we want to study the case $N > M$ from $SU(M)$ moduli space by moving the $N$ 
D3-branes  on the resolved sphere as shown in the middle figure of {\bf fig \ref{fig4MN}}. 
This then brings us to the configuration studied in section \ref{rescon} with the following choice for ${\cal G}_i$ in \eqref{crooked}: 
\bg\label{getty}   
&& {\cal G}_2 = {\cal G}_2(r), ~~~~~ {\cal G}_5 = {\cal G}_6 = 0, ~~~~ \phi=\phi_0  \nonumber\\
&&{\cal G}_4= \left[1 - {2 + {\cal G}_2 \over {\cal G}_2^{1/3} (3 + 2{\cal G}_2)^{2/3}}\right]{{\cal G}^{2/3}_2\over (3 + 2{\cal G}_2)^{2/3}}  \nonumber\\
&& {\cal G}_1 = {4 {\cal G}_{2r}^2 \over {\cal G}_2^{5/3} (3 + 2{\cal G}_2)^{10/3}}, ~~~~~
{\cal G}_3 = 1 - {2+ {\cal G}_2 \over {\cal G}_2^{1/3} (3 + 2{\cal G}_2)^{2/3}},  \nd
expressed completely in terms of ${\cal G}_2(r)$ whose functional form may be determined by solving the Bianchi identity with sources \cite{DEM}. The left-most configuration in {\bf fig \ref{fig4MN}} is the one that we discussed earlier: it is the configuration that appears after Higgsing of a UV conformal theory. In the middle configuration, we require the D5-branes to wrap a vanishing cycle. The constraint on ${\cal G}_2$ is that it has to be bigger than 2 in units of length used in expressing the metric \eqref{crooked}. Let us then take the following  parametrization 
of ${\cal G}_2$:
\bg\label{paulgetty}
{\cal G}_2  = 2\left(1 + \sum_{n = 0}^\infty c_n r^n\right), \nd
which is described for $r \to 0$, as we are more concerned about the behavior at the tip of the cone. The constant terms $c_0$ could be made zero, and the other $c_n$ may be fixed by solving the Bianchi identity as given in \cite{DEM}. The other two relevant warp factors, namely ${\cal G}_3$ and ${\cal G}_4$ may be approximated by:
\bg\label{romanduris}
{\cal G}_3 \approx 0.132\left(1 + 1.409 \sum_{n = 1}^\infty c_n r^n\right), ~~~~
{\cal G}_4 \approx 0.057\left(1 + 1.695 \sum_{n = 1}^\infty c_n r^n\right), \nd
with the D5-branes wrapping the 2-cycle whose warp factor is  ${\cal G}_4$ (expressed in the same units as in the metric \eqref{crooked}). This doesn't vanish, at least for the choice \eqref{getty}, so the supersymmetric configuration
would be the right-most configuration in {\bf fig \ref{fig4MN}}, provided of course the D5 and the 
${\overline{\rm D5}}$-branes are stabilized accordingly as in section \ref{stabul}. 

\subsection{Moduli space descriptions from Type IIA  brane configurations \label{tdula}}

The above discussion tells us that the branch of the moduli space where the Klebanov-Strassler model may be viewed as $N$ ${\overline{\rm D5}}$-branes moving on the resolved sphere could in principle be connected to the known branches of the theory, although we don't make this precise here. The fact that this is identified to be in the same moduli space as of a $SU(N + M)$ theory, shouldn't come as any surprise. In the following we will provide T-dual type IIA brane configurations that will hopefully shed a much clearer light on all the moduli space discussions we have presented here. 

\subsubsection{Type IIA dual of the UV complete type IIB model \label{bhadush}}

\begin{figure}[t]       
  \centering
    \includegraphics[width=1.0\textwidth]{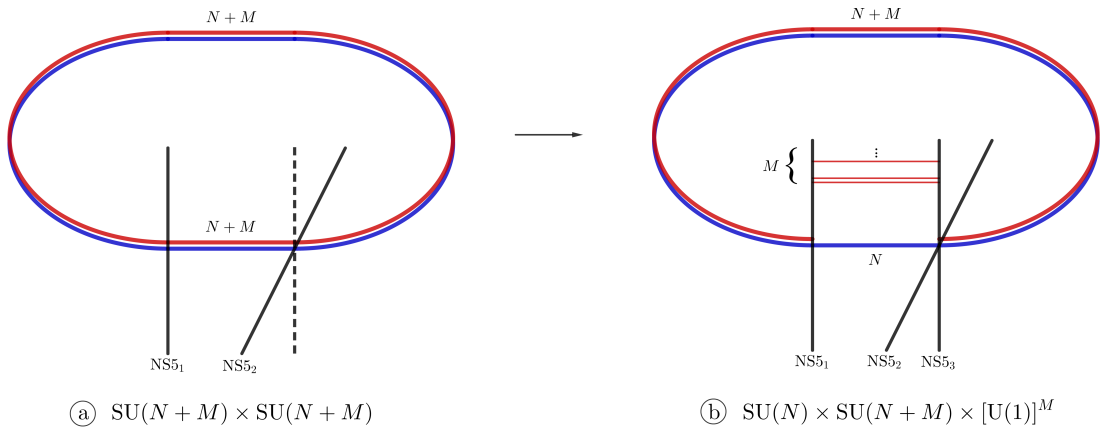}
    \vskip-2.5cm
  \caption{Type IIA brane dual for the UV complete model discussed in a IIB setting of the previous section. On the left is the usual brane construction for the Klebanov-Witten type model with orthogonal NS5-branes. The dotted line is the possibility of adding an extra NS5-brane without breaking the underlying ${\cal N} = 1$ supersymmetry. On the right is the new branch along which we can move the NS5-branes to generate, at low energies, the Klebanov-Strassler type model. Therefore at high energies we have a CFT, whereas at low energies the theory becomes confining.}
\label{laval}
\end{figure}

The type IIA dual of $N + M$ D3-branes at the tip of a conifold is now well known \cite{angel, DM1, DM2}. 
The T-dual is given by two orthogonal NS5-branes oriented along directions ($x_0, x_1, x_2, x_3, x_4, x_5$) and ($x_0, x_1, x_2, x_3, x_8, x_9$) respectively with $N + M$ D4-branes wrapping the compact $x_6$ circle and oriented along ($x_0, x_1, x_2, x_3$) directions. The configuration is shown in {\bf fig \ref{laval}}(a). 
The minimal 
${\cal N} = 1$ supersymmetry prohibits a Coulomb branch in the model, so as such looks like we cannot move the 
$N + M$ D4-branes straddling between the two orthogonal NS5-branes. 

The question then is whether we can add another NS5-brane to the existing configuration to change the behavior of the system. The answer turns out to be in the affirmative, and is shown by the dotted line in {\bf fig \ref{laval}}(a): 
we can add another parallel NS5-brane {\it intersecting} the orthogonal NS5-brane over a $3+1$ dimensional spacetime oriented along 
($x_0, x_1, x_2, x_3$). In the absence of the additional NS5-brane the gauge group is 
$SU(N + M)  \times  SU(N + M)$. Once we add an additional parallel NS5-brane the configuration is given in 
{\bf fig \ref{laval}}(b).
Clearly a branch opens up along which we can move some of the D4-branes. If we move $M$ number of 
D4-branes, the strings connecting the $M$ D4-branes and the $N$ remaining D4-branes become heavy. At the energy scale smaller than the lowest mass of the extended strings, the gauge group becomes:
\bg\label{nightmare}
SU(N) ~\times ~ SU(N + M) ~\times~ {\mathbb{G}}, \nd
where $\mathbb{G}$ is a subgroup of $SU(M)$ and is decoupled at the energy scale that we are interested in. Note that the brane construction in {\bf fig \ref{laval}}(b) makes many subtle points, that we discussed earlier in section \ref{dracula}, transparent. For example:

\vskip.1in

\noindent $\bullet$ A T-duality on the brane configuration of {\bf fig \ref{laval}}(b) along the compact direction $x_6$, first in the absence of the straddling D4-branes and following similar techniques as in \cite{DM1}, leads to the metric configuration similar to \eqref{crooked} with vanishing ${\cal G}_5$ and ${\cal G}_6$ but with 
${\cal G}_3 \ne {\cal G}_4$. The latter is implemented by the difference in the harmonic functions of the parallel and the orthogonal NS5-branes. The parallel NS5-branes are separated by a distance along the $x_6$ direction, and this information is captured by the harmonic function, thus making it different from the single orthogonal NS5-brane\footnote{Of course this requires using {\it localized} NS5-branes in the T-duality rules.}. Once we insert back the D4-branes, the T-duality rules are more subtle, but the eventual answer should be related to \eqref{crooked} again 
with vanishing ${\cal G}_5$ and ${\cal G}_6$ and 
${\cal G}_3 \ne {\cal G}_4$.

\vskip.1in

\noindent $\bullet$ The Higgsing process studied via the Lagrangian \eqref{rossi} and the mass term 
\eqref{thecommuter}, now simply becomes the motion of the $M$ D4-branes along the ($x_4, x_5$) directions. It should also be clear why, as in \eqref{maisoncurry}, a class of scalar fields transform only under the second gauge group $\mathbb{A}_{2\mu}^a$ to facilitate Higgsing. 

\vskip.1in

\noindent $\bullet$ The UV gauge theory is clearly a CFT with gauge group  $SU(N+M) \times SU(N+M)$. This should be apparent from either {\bf fig \ref{laval}}(a) or {\bf fig \ref{laval}}(b). 
At very high energy, we can ignore the mass of the stretched strings and 
 therefore effectively the separation between the $N$ D4 and $M$ D4-branes do not matter. In the IR, after integrating out the massive strings and removing the decoupled sector (represented here by the gauge group $\mathbb{G}$), we recover the Klebanov-Strassler system. Cascading now can be easily explained by moving the set of intersecting NS5-branes across the parallel NS5-brane multiple times \cite{DM1, DOT}. 

\vskip.1in

\noindent $\bullet$ Both in {\bf fig \ref{fig2MN}} and {\bf fig \ref{fig3MN}}, and also in \cite{mia1, mia2, mia3}, the existence of a 
 blown-up 2-sphere at $r = 0$ is an important ingredient for UV completion. In fact this requirement also makes the underlying manifold a 
 non-K\"ahler resolved conifold. Here we see that, as depicted in {\bf fig \ref{laval}}(b), the blown-up 2-sphere is precisely represented by the additional parallel NS5-brane. Of course the brane configuration does not automatically 
 guarantee the compactness of the $(x_4, x_5$) directions, but this can be easily implemented\footnote{Note that the issue of    
the compactness of ($x_4, x_5$) as well as of ($x_8, x_9$) directions is already present in the original brane constructions of \cite{angel, DM1, DM2}.}.

\vskip.1in

\noindent $\bullet$ The UV complete type IIB configuration shown in {\bf fig \ref{fig3MN}}, is realized using 
$M$ D5  and $M$ ${\overline{\rm D5}}$-branes  as well as $N$ D3-branes. In the type IIA brane 
construction of {\bf fig \ref{laval}}(b), the 
$N$ D4-branes that go all the way along the $x_6$ circle are T-duals of the $N$ D3-branes. 
On the other hand the two sets of $M$ D4-branes:  
one, that are moved along the ($x_4, x_5$) directions along parallel NS5-branes and the other,
that lie coincident with the $N$ D4-branes, are precisely T-duals of the two sets of $D5$ and 
${\overline{\rm D5}}$-branes. Both these set of five-branes carry fractional D3-brane bound states, and their global five-brane charges cancel. What we see on the type IIA side are exactly the T-duals of these fractional D3-branes. 

\vskip.1in 

\noindent $\bullet$ Once we move away from the intersection surface of the two orthogonal NS5-branes, say along directions ($x_4, x_5$) in {\bf fig \ref{laval}}(b), we are basically considering the dynamics of two parallel NS5-branes. 
These lead to an 
effective ${\cal N} = 2$ theory. It was speculated in the literature that certain aspects of ${\cal N} = 2$ theory will become necessary to UV complete ${\cal N} = 1$ cascading models. Here we justify why this should be the case. In fact we see that the ${\cal N} = 2$ Coulomb branch of the theory helps us to properly Higgs the extra states, despite the fact that we don't expect Coulomb branches in ${\cal N} = 1$ theories.

\vskip.1in

\noindent $\bullet$ The configuration of the separated $M$ D4-branes in {\bf fig \ref{laval}}(b) looks particularly stable, and therefore one might ask where the issue of instability, discussed in section \ref{stabul}, may arise. There are two ways to 
answer this question: one, by invoking the bending of the NS5-branes, and two, by taking the compactness of the 
($x_4, x_5$) directions. The bendings of the NS5-branes happen when there are unequal numbers of D4-branes on either sides. These are most prominent for ${\cal N} = 2$ theories  \cite{bending}, where they are related to the RG flows on the straddling D4-branes. The parallel NS5-branes are then affected\footnote{The orthogonal NS5-branes should also be affected.}, so the D4-branes are no longer parallel as in {\bf fig \ref{laval}}(b) leading to possible un-cancelled forces. 
On the other hand, the compactness of the ($x_4, x_5$) directions might also play a role here, although that is significantly more difficult to compute. In any case, we can even turn the argument around and use this to justify the stability of the IIB model as discussed in section \ref{stabul}. 

\vskip.1in

\noindent With the above set of points, we see that the type IIA brane configuration as given by {\bf fig \ref{laval}}(b) is powerful enough to not only provide a consistent UV completion of the cascading Klebanov-Strassler model, but also allows us to study many properties of the system that are difficult to approach from the type IIB 
perspective\footnote{See also \cite{berto} for an alternative scenario and the discussions in footnote \ref{berton}.}. 
In fact many other branches of the moduli space opens up, and in the following section we will speculate on some of them. 

\subsubsection{Other branches of the moduli space and Higgsing \label{branches}}

\begin{figure}[t]
  \centering
    \includegraphics[width=1.0\textwidth]{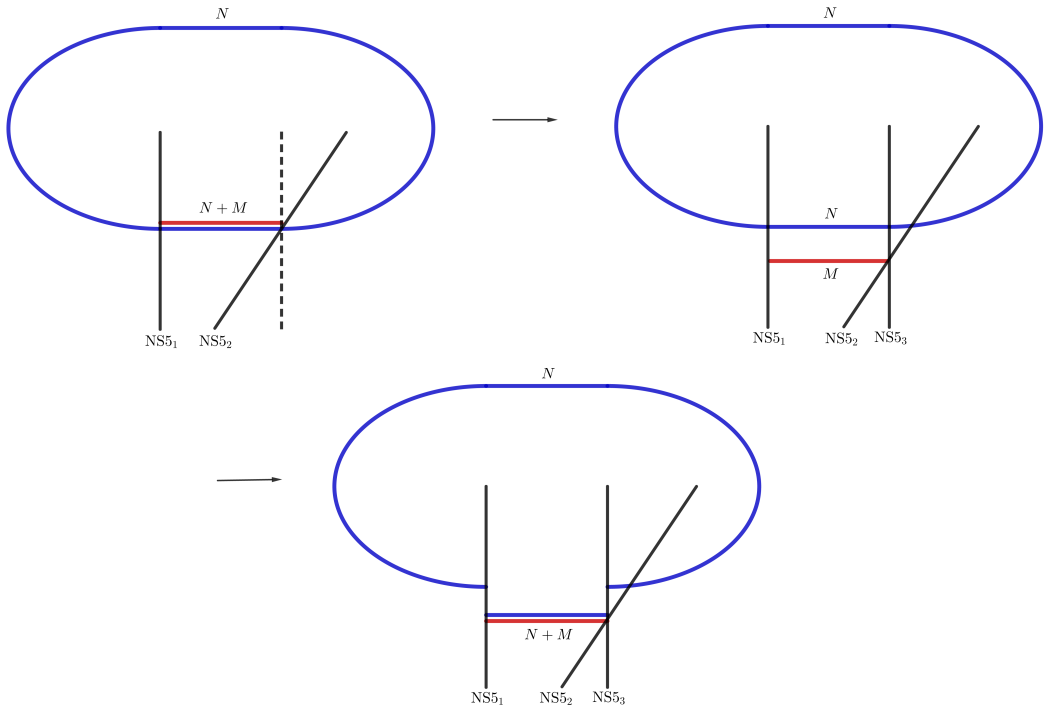}
    \vskip-0.5cm
  \caption{Another branch of the Klebanov-Strassler model generated by T-dualizing the configuration studied in 
  the corresponding type IIB setting. On the left is the standard brane dual of the Klebanov-Strassler 
  model with the dotted line representing the possibility of adding an extra NS5-brane. The middle figure represents the motion of $N$ D4-branes along  directions ($x_4, x_5$). These $N$ D4-branes can now be split and rejoined with the $M$ remaining D4-branes as shown in the bottom figure.}
\label{batman}
\end{figure}

One of the curious thing about the type IIA brane construction of {\bf fig \ref{laval}}(b) is not it's sheer simplicity, but the possibilities of moving the pieces of D4-branes in multiple ways leading to interesting dynamics of the underlying gauge theory. A possibility is depicted in {\bf fig \ref{batman}}. In the left-most figure, we have $N + M$ D4-branes between two orthogonal NS5-branes and $N$ D4-branes the other way around. The configuration lies at the zero of the Coulomb branch in the usual way. The dotted line represents the possibility of adding an orthogonal NS5-brane without breaking any extra supersymmetry just like we had before.

Once we add the NS5-brane, we can move $N$ D4-branes along directions ($x_4, x_5$) as shown in the middle 
figure in {\bf fig \ref{batman}}. This configuration is T-dual to the middle figure in {\bf fig \ref{fig4MN}}. There are however a few subtleties 
that need to be addressed before we can make the precise identification. First, note that near the vicinity of the $N$ D4-branes, the NS5-branes have no additional bendings, and the only bendings 
appear from the remaining $M$ D4-branes. In fact the theory on the $M$ D4-branes is exactly a $SU(M)$ confining gauge theory. Of course all of these are seemingly consistent with what we expect from the type IIB side as depicted in the middle figure of {\bf fig \ref{fig4MN}}. Secondly, the subtlety alluded to above however has more to do with the issue of supersymmetry than the mere identifications of the states. This is because, if we follow the warp factor ansatze \eqref{getty} or more appropriately \eqref{paulgetty} and \eqref{romanduris}, we are in principle dealing with parallel sets of wrapped D5-branes and D3-branes in the type IIB set-up. Supersymmetry is broken when the D5-branes wrap non-vanishing 2-cycle, as in the case of \eqref{romanduris}, unless the D3-branes are dissolved on them. The latter is of course the Baryonic branch, which was discussed in much details in sections \ref{rescon} and 
\ref{defcon}. The question is whether we can move the D3-branes away from the wrapped D5-branes without breaking any additional supersymmetry. 

\begin{figure}[t]
  \centering
    \includegraphics[width=1.0\textwidth]{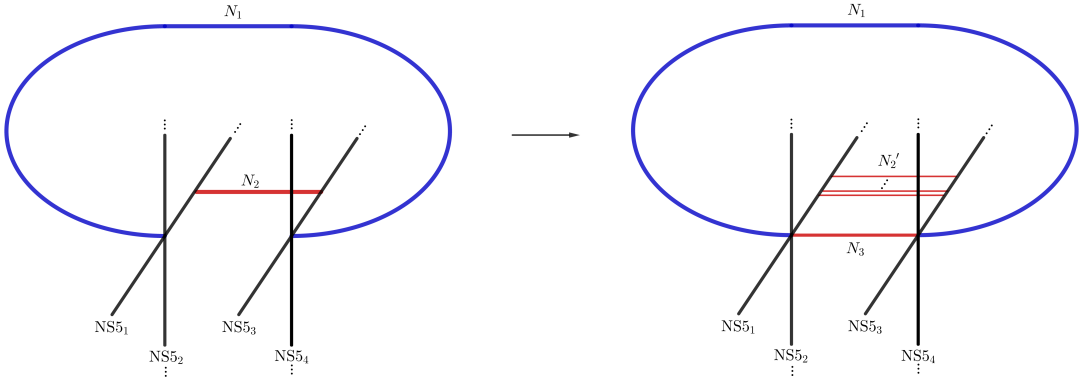}
    \vskip-3.0cm
  \caption{Inserting two set of orthogonal NS5-branes can lead to additional branches in the moduli space of the underlying gauge theory. The figure on the left is from moving $N_2$ D4-branes along ($x_8, x_9$) directions, whereas on the right is the further splitting of the $N_2$ D4-branes. Both supersymmetric and non-supersymmetric possibilities may be realized.}
\label{pizza}
\end{figure}

The answer lies in going beyond the ansatze \eqref{getty} by allowing a non-trivial dilaton. This way we can still allow the D5-branes to wrap vanishing 2-cycle and then move the D3-branes on the resolved sphere. An example of this can be given directly in the type IIB side by allowing the following choice of the warp factors ${\cal G}_i$:
\bg\label{phantomthread}
{\cal G}_1 = {e^{-\phi}\over 2 {\cal G}}, ~~~ {\cal G}_2 = {r^2 {\cal G} e^{-\phi}\over 2}, ~~~ 
{\cal G}_3 = {r^2 e^{-\phi}\over 4}, ~~~ {\cal G}_4 = {r^2 e^{-\phi}\over 4} + {\cal O}(a^2), \nd
with vanishing ${\cal G}_5$ and ${\cal G}_6$ in \eqref{crooked}. The choice \eqref{phantomthread} satisfies EOM 
and supersymmetry once the resolution $a^2$ is incorporated. However 
somewhat different from \eqref{getty}, we now have a  
new function ${\cal G}(r)$ that needs to be fixed
from the Bianchi identity with an appropriate dilaton\footnote{It is possible that both are fixed, but we haven't checked the details.} input $\phi(r)$. Since ${\cal G}_3$ vanishes when $r \to 0$, the D5-branes wrap vanishing cycle in the type IIB side, and therefore the middle figure in {\bf fig \ref{batman}} precisely corresponds to a supersymmetric setting. This can be further verified from the fact that any additional bendings of the NS5-branes near the vicinity of the $N$ D4-branes are exactly zero, as mentioned above. Thus the $M$ D4 and the $N$ D4-branes  are mutually parallel, despite the fact that there are some bendings near the $M$ D4-branes. 

We can now split the $N$ D4-branes along the ($x_4, x_5$) directions and this correspond to the splitting of $N$ D3-branes to D5 and ${\overline{\rm D5}}$-branes in the type IIB set-up as depicted in the rightmost figure of
{\bf fig \ref{fig4MN}}. Un-cancelled forces in the type IIB set-up, or bendings of the NS5-branes in the 
type IIA dual side, will now reappear but stability/supersymmetry can be restored along the lines of the discussion in section \ref{stabul}. After the dust settles, the type IIA configuration is depicted in the bottom figure of 
{\bf fig \ref{batman}}. This is supersymmetric and matches well with the corresponding type IIB expectations. 

The story however is richer as may be seen from the configurations depicted in {\bf fig \ref{pizza}}. Inserting 
yet another NS5-brane orthogonal to the left NS5-brane, as shown in left of {\bf fig \ref{pizza}}, we can move the D4-branes along ($x_8, x_9$) directions instead. The $N_2$ D4-branes can be further split along ($x_8, x_9$) directions to the one shown on the right of {\bf fig \ref{pizza}}. It is easy to see that both supersymmetric and non-supersymmetric scenarios may be constructed this way, although the type IIB duals of any of these constructions from {\bf fig \ref{pizza}} may be more non-trivial\footnote{Needless to say, we haven't tried to find the type IIB duals of the configurations of {\bf fig \ref{pizza}} and their possible generalizations. These are left for future works.}. 
One can even get more ambitious and insert additional parallel and/or orthogonal NS5-branes to the existing configurations. The straddling D4-branes may be moved along these directions leading to generalizations of the models studied in \cite{francok}. 

Note however that, as mentioned above, not all configurations can be supersymmetric. Restoring supersymmetry, as discussed in section \ref{stabul} for the type IIB side, is a delicate affair and various subtle cancellations need to happen for this to work out consistently. On the type IIA side, it is the bendings and the compactness of the NS5-branes that need to be carefully taken into account while studying supersymmetric configurations in the moduli space of the vacua. As we saw, there do exist ``Goldilocks zones" in the moduli spaces of gauge theories where minimal supersymmetry can be realized, and these are also the regions where well defined UV completions appear hand in hand.

\section{Discussions and conclusions \label{konoklota}}

In this paper three related topics are studied, all aiming towards understanding the UV completion of a Klebanov-Strassler type model. The Klebanov-Strassler model \cite{KS} is a ${\cal N} = 1$ supersymmetric gauge theory with a product gauge group that breaks the conformal invariance in the theory leading, in the far IR, to confinement, although other IR behaviors are also possible by appropriately choosing the colors and the bi-fundamental flavors. 
These theories have UV issues, and so UV completions are called for. In this paper we analyzed the possibility of having an asymptotically conformal behavior at UV. Such a possibility has been discussed earlier in \cite{mia1, mia2, mia3} (see also \cite{berto}), and here we make this more precise.  

The issue that presents itself while UV completing such theories appears directly from the brane constructions of these theories. The IR physics is generated by allowing a certain number of D3 and D5 branes at the conifold point. Typically we take $N$ D3-branes at the conifold point and $M$ D5-branes wrapping a vanishing cycle of the conifold. For large $M$ and
$N \equiv k M$, with integer $k$, the theory confines in the far IR and simultaneously has a dual gravity description \cite{KS}. However to UV complete the theory, we require no wrapped D5-branes. One way to get this is to allow $M$ 
${\overline{\rm D5}}$-branes with world-volume fluxes so that the D5 and the $\overline{\rm D5}$-branes together give rise to $M$ D3-branes. This means, to have a UV conformal and IR confining theory, we will require the 
$\overline{\rm D5}$-branes  to be arranged on a resolved 2-cycle,  separated from the other coincident D5 and D3 system, while still wrapping the vanishing cycle.  

This in turn means we are looking for non-K\"ahler resolved and deformed conifolds for describing the gauge theory and its gravity dual respectively. The generic metric for such manifolds is given by \eqref{crooked} with appropriate choices for the warp-factors ${\cal G}_i$. In section \ref{IR} we show that such a background with fluxes {\it and} with possibles sources can be supersymmetric if and only if they satisfy \eqref{easter}. The allowed three-form flux turns out to be \eqref{tagramaiaa}, 
alongwith a complex structure given in \eqref{baghdadi} with a minus sign.  

The crucial issue of resolutions in these manifolds may in turn be shown directly by using a probe brane analysis in the dual geometry by turning on ``baryonic" operators. This is detailed in section \ref{navneet}. Finally, in section \ref{youvee} we show how to study the UV completions, both from type IIB as well as the T-dual type IIA descriptions. In fact the IIA descriptions turn out to be richer as many new branches of the moduli space open up from the possibilities of moving the D4-branes on the NS5-branes.   

\vskip.15in

\centerline{\bf Acknowledgements}

\vskip.1in

\noindent We would like to thank  Simon Caron-Huot, Veronica Errasti Diez, Juan Maldacena, and Matt Strassler for many helpful discussions. The work of K. D, J. E, M. E and A-K. T, is supported in part by the Natural Sciences and Engineering 
Research Council of Canada (NSERC) grant.

{}

\newpage

\begin{thebibliography}{}


\bibitem{candelas90} 
  P.~Candelas and X.~C.~de la Ossa,
  ``Comments on Conifolds,''
  Nucl.\ Phys.\ B {\bf 342}, 246 (1990);
  P.~Candelas, P.~S.~Green and T.~Hubsch,
  ``Rolling Among Calabi-Yau Vacua,''
  Nucl.\ Phys.\ B {\bf 330}, 49 (1990).

\bibitem{DRS} 
  K.~Dasgupta, G.~Rajesh and S.~Sethi,
  ``M theory, orientifolds and G - flux,''
  JHEP {\bf 9908}, 023 (1999)
    [hep-th/9908088].

 \bibitem{anke} 
  M.~Becker, K.~Dasgupta, A.~Knauf and R.~Tatar,
  ``Geometric transitions, flops and non-K\"ahler manifolds. I.,''
  Nucl.\ Phys.\ B {\bf 702}, 207 (2004)
    [hep-th/0403288];
  S.~Alexander, K.~Becker, M.~Becker, K.~Dasgupta, A.~Knauf and R.~Tatar,
  ``In the realm of the geometric transitions,''
  Nucl.\ Phys.\ B {\bf 704}, 231 (2005)
    [hep-th/0408192]; 
  M.~Becker, K.~Dasgupta, S.~H.~Katz, A.~Knauf and R.~Tatar,
  ``Geometric transitions, flops and non-K\"ahler manifolds. II.,''
  Nucl.\ Phys.\ B {\bf 738}, 124 (2006)
    [hep-th/0511099].

\bibitem{CHSW} 
  P.~Candelas, G.~T.~Horowitz, A.~Strominger and E.~Witten,
  ``Vacuum Configurations for Superstrings,''
  Nucl.\ Phys.\ B {\bf 258}, 46 (1985).

\bibitem{torsion} 
  G.~Lopes Cardoso, G.~Curio, G.~Dall'Agata, D.~Lust, P.~Manousselis and G.~Zoupanos,
  ``NonKahler string backgrounds and their five torsion classes,''
  Nucl.\ Phys.\ B {\bf 652}, 5 (2003)
    [hep-th/0211118]; 
    S.~Gurrieri, J.~Louis, A.~Micu and D.~Waldram,
  ``Mirror symmetry in generalized Calabi-Yau compactifications,''
  Nucl.\ Phys.\ B {\bf 654}, 61 (2003)
    [hep-th/0211102].

\bibitem{GKP} 
  S.~B.~Giddings, S.~Kachru and J.~Polchinski,
  ``Hierarchies from fluxes in string compactifications,''
  Phys.\ Rev.\ D {\bf 66}, 106006 (2002)
    [hep-th/0105097].

\bibitem{angel} 
  A.~M.~Uranga,
  ``Brane configurations for branes at conifolds,''
  JHEP {\bf 9901}, 022 (1999)
    [hep-th/9811004].
  
  \bibitem{DM1} 
  K.~Dasgupta and S.~Mukhi,
  ``Brane constructions, conifolds and M theory,''
  Nucl.\ Phys.\ B {\bf 551}, 204 (1999)
    [hep-th/9811139].
  
  \bibitem{DM2} 
  K.~Dasgupta and S.~Mukhi,
  ``Brane constructions, fractional branes and Anti-de Sitter domain walls,''
  JHEP {\bf 9907}, 008 (1999)
    [hep-th/9904131].



\bibitem{klebwit} 
  I.~R.~Klebanov and E.~Witten,
  ``Superconformal field theory on three-branes at a Calabi-Yau singularity,''
  Nucl.\ Phys.\ B {\bf 536}, 199 (1998)
    [hep-th/9807080].

\bibitem{KS} 
  I.~R.~Klebanov and M.~J.~Strassler,
  ``Supergravity and a confining gauge theory: Duality cascades and chi SB resolution of naked singularities,''
  JHEP {\bf 0008}, 052 (2000)
    [hep-th/0007191].

\bibitem{mia1} 
  M.~Mia, K.~Dasgupta, C.~Gale and S.~Jeon,
  ``Five Easy Pieces: The Dynamics of Quarks in Strongly Coupled Plasmas,''
  Nucl.\ Phys.\ B {\bf 839}, 187 (2010)
    [arXiv:0902.1540 [hep-th]].

\bibitem{mia2} 
  M.~Mia, K.~Dasgupta, C.~Gale and S.~Jeon,
  ``Toward Large N Thermal QCD from Dual Gravity: The Heavy Quarkonium Potential,''
  Phys.\ Rev.\ D {\bf 82}, 026004 (2010)
    [arXiv:1004.0387 [hep-th]].
  
  \bibitem{mia3} 
  F.~Chen, L.~Chen, K.~Dasgupta, M.~Mia and O.~Trottier,
  ``Ultraviolet complete model of large N thermal QCD,''
  Phys.\ Rev.\ D {\bf 87}, no. 4, 041901 (2013)
    [arXiv:1209.6061 [hep-th]];
    M.~Dhuria and A.~Misra,
  ``Towards MQGP,''
  JHEP {\bf 1311}, 001 (2013)
    [arXiv:1306.4339 [hep-th]]; 
    K.~Sil,
  ``Application of top-down holographic thermal QCD at finite coupling,''
  arXiv:1804.01290 [hep-th].
  
  \bibitem{maxim} 
  K.~Dasgupta, M.~Emelin, C.~Gale and M.~Richard,
  ``Renormalization Group Flow, Stability, and Bulk Viscosity in a Large N Thermal QCD Model,''
  Phys.\ Rev.\ D {\bf 95}, no. 8, 086018 (2017)
    [arXiv:1611.07998 [hep-th]].
  
 
  
  \bibitem{DOT} 
  K.~Dasgupta, K.~Oh and R.~Tatar,
  ``Geometric transition, large N dualities and MQCD dynamics,''
  Nucl.\ Phys.\ B {\bf 610}, 331 (2001)
    [hep-th/0105066];
  ``Open / closed string dualities and Seiberg duality from geometric transitions in M theory,''
  JHEP {\bf 0208}, 026 (2002)
    [hep-th/0106040];
  K.~Dasgupta, K.~h.~Oh, J.~Park and R.~Tatar,
  ``Geometric transition versus cascading solution,''
  JHEP {\bf 0201}, 031 (2002)
    [hep-th/0110050].
  
  \bibitem{vafaGT} 
  C.~Vafa,
  ``Superstrings and topological strings at large N,''
  J.\ Math.\ Phys.\  {\bf 42}, 2798 (2001)
    [hep-th/0008142];
  F.~Cachazo, K.~A.~Intriligator and C.~Vafa,
  ``A Large N duality via a geometric transition,''
  Nucl.\ Phys.\ B {\bf 603}, 3 (2001)
    [hep-th/0103067].
  
    
  \bibitem{DEM} 
  K.~Dasgupta, M.~Emelin and E.~McDonough,
  ``Non-K\"ahler resolved conifold, localized fluxes in M-theory and supersymmetry,''
  JHEP {\bf 1502}, 179 (2015)
    [arXiv:1412.3123 [hep-th]].
  
  \bibitem{buscher} 
  E.~Bergshoeff, C.~M.~Hull and T.~Ortin,
  ``Duality in the type II superstring effective action,''
  Nucl.\ Phys.\ B {\bf 451}, 547 (1995)
    [hep-th/9504081].

\bibitem{MM} 
  J.~Maldacena and D.~Martelli,
  ``The Unwarped, resolved, deformed conifold: Fivebranes and the baryonic branch of the Klebanov-Strassler theory,''
  JHEP {\bf 1001}, 104 (2010)
    [arXiv:0906.0591 [hep-th]].


\bibitem{malnun} 
  J.~M.~Maldacena and C.~Nunez,
  ``Towards the large N limit of pure N=1 superYang-Mills,''
  Phys.\ Rev.\ Lett.\  {\bf 86}, 588 (2001)
    [hep-th/0008001].

\bibitem{butti} 
  A.~Butti, M.~Grana, R.~Minasian, M.~Petrini and A.~Zaffaroni,
  ``The Baryonic branch of Klebanov-Strassler solution: A supersymmetric family of SU(3) structure backgrounds,''
  JHEP {\bf 0503}, 069 (2005)
    [hep-th/0412187].

\bibitem{cvetic} 
  M.~Cvetic, G.~W.~Gibbons, H.~Lu and C.~N.~Pope,
  ``A G(2) unification of the deformed and resolved conifolds,''
  Phys.\ Lett.\ B {\bf 534}, 172 (2002)
    [hep-th/0112138].

\bibitem{ankegwyn} 
  R.~Gwyn and A.~Knauf,
  ``Conifolds and geometric transitions,''
  Rev.\ Mod.\ Phys.\  {\bf 8012}, 1419 (2008)
  [hep-th/0703289].

\bibitem{minpis} 
  R.~Minasian and D.~Tsimpis,
  ``On the geometry of nontrivially embedded branes,''
  Nucl.\ Phys.\ B {\bf 572}, 499 (2000)
    [hep-th/9911042].


\bibitem{ADS}
 I. Affleck, M. Dine and N. Seiberg,
``Dynamical supersymmetry breaking in supersymmetric QCD",
 Nucl.\ Phys.\ B {\bf241} 49 (1984) 

\bibitem{superspace}
S.J. Gates, Marcus T. Grisaru, M. Rocek, W. Siegel
``Superspace Or One Thousand and One Lessons in Supersymmetry''
Front.\ Phys. \  {\bf58} (1983) 

\bibitem{Rxi}
Burt A. Ovrut and Julius Wess
``Supersymmetric $R_\xi$ gauge and radiative symmetry breaking"
Phys. Rev. D {\bf25} 409 (1982)

\bibitem{KT} 
  I.~R.~Klebanov and A.~A.~Tseytlin,
  ``Gravity duals of supersymmetric SU(N) x SU(N+M) gauge theories,''
  Nucl.\ Phys.\ B {\bf 578}, 123 (2000)
    [hep-th/0002159].

\bibitem{berto} 
  R.~Argurio and M.~Bertolini,
  ``Orientifolds and duality cascades: confinement before the wall,''
  JHEP {\bf 1802}, 149 (2018)
    [arXiv:1711.08983 [hep-th]].

\bibitem{aharony} 
  O.~Aharony,
  ``A Note on the holographic interpretation of string theory backgrounds with varying flux,''
  JHEP {\bf 0103}, 012 (2001)
    [hep-th/0101013];
    ``The non-AdS / non-CFT correspondence, or three different paths to QCD,''
  hep-th/0212193.




\bibitem{Bak} 
  D.~s.~Bak and A.~Karch,
  ``Supersymmetric brane anti-brane configurations,''
  Nucl.\ Phys.\ B {\bf 626}, 165 (2002)
    [hep-th/0110039];
    D.~Bak and N.~Ohta,
  ``Supersymmetric D2 anti-D2 strings,''
  Phys.\ Lett.\ B {\bf 527} (2002) 131  [hep-th/0112034];
Y.~Hyakutake and N.~Ohta,
  ``Supertubes and supercurves from M ribbons,''
  Phys.\ Lett.\ B {\bf 539} (2002) 153  [hep-th/0204161];
D.~s.~Bak, N.~Ohta and M.~M.~Sheikh-Jabbari,
  ``Supersymmetric brane - anti-brane systems: Matrix model description, stability and decoupling limits,''
  JHEP {\bf 0209} (2002) 048  [hep-th/0205265];
D.~Bak, N.~Ohta and P.~K.~Townsend,
  ``The D2 Susy zoo,''
  JHEP {\bf 0703} (2007) 013  [hep-th/0612101].    

\bibitem{mateos} 
  D.~Mateos, S.~Ng and P.~K.~Townsend,
  ``Tachyons, supertubes and brane / anti-brane systems,''
  JHEP {\bf 0203}, 016 (2002)
    [hep-th/0112054].

\bibitem{broo} 
  E.~A.~Bergshoeff, M.~de Roo and A.~Sevrin,
  ``NonAbelian Born-Infeld and kappa symmetry,''
  J.\ Math.\ Phys.\  {\bf 42}, 2872 (2001)
    [hep-th/0011018].

\bibitem{bilal} 
  E.~A.~Bergshoeff, A.~Bilal, M.~de Roo and A.~Sevrin,
  ``Supersymmetric nonAbelian Born-Infeld revisited,''
  JHEP {\bf 0107}, 029 (2001)
    [hep-th/0105274].

\bibitem{shmakova} 
  K.~Dasgupta and M.~Shmakova,
  ``On branes and oriented B fields,''
  Nucl.\ Phys.\ B {\bf 675}, 205 (2003)
    [hep-th/0306030].

\bibitem{chetan} 
  C.~Krishnan and S.~Kuperstein,
  ``The Mesonic Branch of the Deformed Conifold,''
  JHEP {\bf 0805}, 072 (2008)
    [arXiv:0802.3674 [hep-th]].

\bibitem{ohta} 
  K.~Ohta and T.~Yokono,
  ``Deformation of conifold and intersecting branes,''
  JHEP {\bf 0002}, 023 (2000)
    [hep-th/9912266].

\bibitem{bending} 
  E.~Witten,
  ``Solutions of four-dimensional field theories via M theory,''
  Nucl.\ Phys.\ B {\bf 500}, 3 (1997)
    [hep-th/9703166].

\bibitem{francok} 
  R.~Argurio, M.~Bertolini, S.~Franco and S.~Kachru,
  ``Gauge/gravity duality and meta-stable dynamical supersymmetry breaking,''
  JHEP {\bf 0701}, 083 (2007)
    [hep-th/0610212].


\end{thebibliography}
 \end{document}